\documentclass[aps,prd,onecolumn,groupedaddress,nofootinbib,amssymb]{revtex4}
\usepackage[T1]{fontenc}
\usepackage[latin1]{inputenc}
\usepackage{graphicx}
\usepackage[english]{babel}

\def\ba{\begin{eqnarray}}
\def\ea{\end{eqnarray}}
\def\beq{\begin{equation}}
\def\eeq{\end{equation}}

\def\pr{{\it Phys. Rev.}\ }
\def\prl{{\it Phys. Rev. Lett.}\ }
\def\pl{{\it Phys. Lett.}\ }

\def\ijmp{{\it Int. Journ. Mod. Phys.}\ }

\def\cqg{{\it Class. Quantum Grav.}\ }

\def\grg{{\it Gen. Relativ. Grav.}\ }

\def\apj{{\it Ap. J.}\ }

\def\mnras{{\it Mon. Not. R. Ast. Soc.}\ }

\def\araa{{\it Ann. Rev. Astr. Ap.}\ }

\def\ie{{\it i.e. }}

\def\pr{{\it Phys. Rev.}\ }
\def\pl{{\it Phys. Lett.}\ }

\def\apj{{\it Ap. J.}\ }
\def\mnras{{\it Mon. Not. R. Ast. Soc.}\ }
\def\araa{{\it Ann. Rev. Astr. Ap.}\ }

\def\vol{\int d^4x\,\sqrt{-g}}

\def\half{\frac{1}{2}}
\def\gu{g^{\mu\nu}}
\def\gd{g_{\mu\nu}}

\def\umunu{^{\mu\nu}}
\def\dmunu{_{\mu\nu}}
\def\ua{^{\alpha}}

\def\dab{_{\alpha\beta}}

\def\ddeab{_{;\alpha\beta}}
\def\ddemunu{_{;\mu\nu}}

\def\ddemu{_{;\mu}}  
\def\ddenu{_{;\nu}}  
\def\ddea{_{;\alpha}}  \def\udea{^{;\alpha}}
\def\ddeb{_{;\beta}}

\def\pa{\partial}

\let\alp=\alpha

\renewcommand{\epsilon}{\varepsilon}

\def\beqt{\begin{tabular}}
\def\eet{\end{tabular}}
\def\beqf{\begin{figure}}
\def\eef{\end{figure}}
\def\beqa{\begin{eqnarray}}
\def\eeqa{\end{eqnarray}}
\def\vol{\int d^4x\,\sqrt{-g}}

\def\half{\frac{1}{2}}

\def\gu{g^{\mu\nu}}
\def\gd{g_{\mu\nu}}

\def\umunu{^{\mu\nu}}
\def\dmunu{_{\mu\nu}}
\def\ua{^{\alpha}}

\def\dab{_{\alpha\beta}}

\def\ddeab{_{;\alpha\beta}}
\def\ddemunu{_{;\mu\nu}}

\def\ddemu{_{;\mu}}

\def\ddenu{_{;\nu}}

\def\ddea{_{;\alpha}}
\def\udea{^{;\alpha}}
\def\ddeb{_{;\beta}}

\def\pa{\partial}

\def\ie{{\it i.e. }}

\def\p{\phi}

\def\v{V(\phi)}

\def\l{\cal L}

\let\alp=\alpha

\renewcommand{\epsilon}{\varepsilon}

\def\disp{\displaystyle}

\def\vol{d^4x\,\sqrt{-g}}

\def\half{\frac{1}{2}}
\def\gu{g^{\mu\nu}}
\def\gd{g_{\mu\nu}}

\def\pa{\partial}

\def\ome{{\omega}}

\def\ua{^{\alpha}}

\def\umunu{^{\mu\nu}}
\def\dmunu{_{\mu\nu}}

\def\dab{_{\alpha\beta}}

\def\udea{_{;}^{\alpha}}

\def\ddemu{_{;\mu}}
\def\ddenu{_{;\nu}}
\def\ddea{_{;\alpha}}
\def\ddeb{_{;\beta}}
\def\ddemunu{_{;\mu\nu}}

\def\ddeab{_{;\alpha\beta}}

\def\f{{\phi}}

\def\VDEF{{V_{\phi}}}

\def\FDEF{{F_{\phi}}}

\def\ie{{\it i.e. }}

\def\pr{{\it Phys. Rev.}\ }
\def\prl{{\it Phys. Rev. Lett.}\ }
\def\pl{{\it Phys. Lett.}\ }

\def\ijmp{{\it Int. Journ. Mod. Phys.}\ }

\def\cqg{{\it Class. Quantum Grav.}\ }

\def\grg{{\it Gen. Relativ. Grav.}\ }

\def\apj{{\it Ap. J.}\ }

\def\mnras{{\it Mon. Not. R. Ast. Soc.}\ }

\def\araa{{\it Ann. Rev. Astr. Ap.}\ }

\let\alp=\alpha

\renewcommand{\epsilon}{\varepsilon}

\def\pr{{\it Phys. Rev.}\ }
\def\prl{{\it Phys. Rev. Lett.}\ }
\def\pl{{\it Phys. Lett.}\ }

\def\ijmp{{\it Int. Journ. Mod. Phys.}\ }

\def\cqg{{\it Class. Quantum Grav.}\ }

\def\grg{{\it Gen. Relativ. Grav.}\ }

\def\apj{{\it Ap. J.}\ }

\def\mnras{{\it Mon. Not. R. Ast. Soc.}\ }

\def\araa{{\it Ann. Rev. Astr. Ap.}\ }

\def\ie{{\it i.e. }}

\def\p{\phi}

\def\v{V(\phi)}

\def\l{\cal L}

\begin{document}

\input{epsf.sty}

\title{  Extended Theories of Gravity and their Cosmological and Astrophysical Applications}

\author{Salvatore Capozziello$^1$, Mauro Francaviglia$^2$}

\affiliation{\it $^1$Dipartimento di Scienze fisiche, Universit\`a
 di Napoli {}`` Federico II'', and INFN Sez. di Napoli, Compl.
Univ. di
Monte S. Angelo, Edificio G, Via Cinthia, I-80126, Napoli, Italy\\
 $^2$Dipartimento di Matematica, Universit\`a di Torino, and INFN
Sez. di Torino,Via Carlo Alberto 10, 10123 Torino, Italy.}

\begin{abstract}
Astrophysical observations are pointing out huge amounts of "dark
matter" and "dark energy" needed to explain the observed large
scale structure and  cosmic dynamics.  The emerging picture is
 a spatially flat, homogeneous Universe undergoing the today
observed accelerated phase. Despite of the good quality of
astrophysical surveys, commonly addressed as {\it Precision
Cosmology}, the nature and the nurture of dark energy and dark
matter, which should constitute the bulk of cosmological
matter-energy,  are still unknown. Furthermore, up to now, no
experimental evidence has been found, at fundamental level, to
explain such mysterious components.

The problem could be completely reversed considering dark matter
and dark energy as "shortcomings" of General Relativity in its
simplest formulation (a linear theory in the Ricci scalar $R$,
minimally coupled to the standard perfect fluid matter) and
claiming for the "correct" theory of gravity as that derived by
matching the largest number of observational data, without
imposing any theory {\it a priori}. As a working hypothesis,
accelerating behavior of cosmic fluid, large scale structure,
potential of galaxy clusters, rotation curves of spiral galaxies
could be reproduced by means of
 {\it extending} the standard theory of General Relativity. In other
words, gravity could acts in different ways at different scales
and the above "shortcomings" could be due  to  incorrect
extrapolations of the Einstein gravity, actually tested at short
scales and low energy regimes.

After a  survey of what is intended for {\it Extended Theories of
Gravity} in the so called "metric" and "Palatini" approaches, we
discuss some cosmological and astrophysical applications where the
issues related to the dark components are addressed  by enlarging
the Einstein theory to more general $f(R)$ Lagrangians, where
$f(R)$ is a generic function of Ricci scalar $R$, not assumed
simply linear. Obviously, this  is not the final answer to the
problem of "dark-components" but it can be considered as an
operative scheme whose aim is to avoid the addition of unknown
exotic  ingredients to the cosmic pie.
\end{abstract}


\maketitle

\section{Introduction}

General Relativity (GR)  is a comprehensive theory of spacetime,
gravity and matter. Its formulation implies that space and time
are not "absolute" entities, as in Classical Mechanics, but
dynamical quantities strictly related to the distribution of
matter and energy. As a consequence, this approach gave rise to a
new conception of the Universe itself which, for the first time,
was considered as a  dynamical system. In other words, Cosmology
has been enclosed in the realm of Science and not only of
Philosophy, as before the Einstein work. On the other hand, the
possibility of a scientific investigation of the Universe has led
to the formulation of the  Standard Cosmological Model
\cite{weinberg} which, quite nicely, has matched with
observations.

Despite of these results, in the last thirty years, several
shortcomings came out in the Einstein theory and people began to
investigate whether GR is the only fundamental theory capable of
explaining the gravitational interaction. Such issues come,
essentially, from cosmology and quantum field theory. In the first
case, the presence of the Big Bang singularity, the flatness and
horizon problems \cite{guth} led to the statement that
Cosmological Standard Model, based on  the GR and the Standard
Model of Particle Physics, is inadequate to describe the Universe
at extreme regimes. On the other hand, GR is a {\it classical}
theory which does not work as a fundamental theory, when one wants
to achieve a full quantum description of spacetime (and then of
gravity).

Due to these  facts and, first of all, to the lack of a definitive
quantum gravity theory, alternative theories  have been considered
in order to attempt, at least, a semi-classical scheme where GR
and its positive results could be recovered. One of the most
fruitful approaches has been that of {\it Extended Theories of
Gravity} (ETG)  which have become a sort of paradigm in the study
of gravitational interaction. They are based on corrections and
enlargements of the  Einstein theory. The paradigm consists,
essentially,  in adding higher-order curvature invariants and
minimally or non-minimally coupled scalar fields into dynamics
which come out from  the effective action of quantum gravity
\cite{odintsov}.

Other motivations to modify GR come from the issue of a full
recovering of the Mach principle which leads to assume a varying
gravitational coupling. The principle states that the local
inertial frame is determined by some average of the motion of
distant astronomical objects \cite{bondi}. This fact implies that
the gravitational coupling can be scale-dependent and related to
some scalar field. As a consequence,  the concept of ``inertia''
and the Equivalence Principle have to be revised. For example, the
Brans-Dicke theory \cite{brans} is a serious attempt to define an
alternative theory to the Einstein gravity: it takes into account
a variable Newton gravitational coupling, whose dynamics is
governed by a scalar field non-minimally coupled to the geometry.
In such a way, Mach's principle is better implemented
\cite{brans,cimento,sciama}.

Besides, every unification scheme as Superstrings, Supergravity or
Grand Unified Theories, takes into account effective actions where
non-minimal couplings to the geometry or higher-order terms in the
curvature invariants are present. Such contributions are due to
one-loop or higher-loop corrections in the high-curvature regimes
near the full (not yet available) quantum gravity regime
\cite{odintsov}. Specifically, this scheme was adopted in order to
deal with the quantization on curved spacetimes and the result was
that the interactions among quantum scalar fields and background
geometry or the gravitational self-interactions yield corrective
terms in the Hilbert-Einstein Lagrangian \cite{birrell}. Moreover,
it has been realized that such corrective terms are inescapable in
order to obtain the effective action of quantum gravity at scales
closed to the Planck one \cite{vilkovisky}. All these approaches
are not the ``{\it full quantum gravity}" but are needed as
working schemes toward it.

In summary, higher-order terms in curvature invariants (such as
$R^{2}$, $R\umunu R\dmunu$,
$R^{\mu\nu\alp\beta}R_{\mu\nu\alp\beta}$, $R \,\Box R$, or $R
\,\Box^{k}R$) or non-minimally coupled terms between scalar fields
and geometry (such as $\p^{2}R$) have to be added to the effective
Lagrangian of gravitational field when quantum corrections are
considered. For instance, one can notice that such terms occur in
the effective Lagrangian of strings or in Kaluza-Klein theories,
when the mechanism of dimensional reduction is used
\cite{veneziano}.

On the other hand, from a conceptual viewpoint, there are no {\it
a priori} reason to restrict the gravitational Lagrangian to a
linear function of the Ricci scalar $R$, minimally coupled with
matter \cite{francaviglia}. Furthermore, the idea that there are
no ``exact'' laws of physics could be taken into serious account:
in such a case, the effective Lagrangians of physical interactions
are ``stochastic'' functions. This feature means that the local
gauge invariances (\ie conservation laws) are well approximated
only in the low energy limit and the fundamental physical
constants can vary \cite{ottewill}.

Besides fundamental physics motivations, all these theories have
acquired a huge interest in cosmology due to the fact that they
``naturally" exhibit inflationary behaviors able to overcome the
shortcomings of  Cosmological Standard Model (based on GR). The
related cosmological models seem  realistic and capable of
matching with the CMBR observations \cite{starobinsky,kerner,la}.
Furthermore, it is possible to show that, via conformal
transformations, the higher-order and non-minimally coupled terms
always correspond to the Einstein gravity plus one or more than
one minimally coupled scalar fields
\cite{teyssandier,maeda,wands1,wands,gottloeber}.

More precisely, higher-order terms appear always  as contributions
of order two in the field equations. For example, a term like
$R^{2}$ gives fourth order equations \cite{ruzmaikin}, $R \ \Box
R$ gives sixth order equations \cite{gottloeber,sixth}, $R
\,\Box^{2}R$ gives eighth order equations \cite{eight} and so on.
By a conformal transformation, any 2nd-order  derivative term
corresponds to a scalar field\footnote{The dynamics of such scalar
fields is usually given by the corresponding Klein-Gordon
Equation, which is  second order.}: for example, fourth-order
gravity gives Einstein plus one scalar field, sixth-order gravity
gives Einstein plus two scalar fields and so on
\cite{gottloeber,schmidt1}.

Furthermore, it is possible to show that the $f(R)$-gravity is
equivalent not only to a scalar-tensor one but also to the
Einstein theory plus an ideal fluid \cite{cno}. This feature
results very interesting if we want to obtain multiple
inflationary events since an early stage could select ``very''
large-scale structures (clusters of galaxies today), while a late
stage could select ``small'' large-scale structures (galaxies
today) \cite{sixth}. The philosophy is that each inflationary era
is related to the dynamics of a scalar field. Finally, these
extended schemes  could naturally solve the problem of ``graceful
exit" bypassing the shortcomings of former inflationary models
\cite{la,aclo}.

In addition to the  revision of Standard Cosmology at early epochs
(leading to the Inflation),  a new approach is necessary also at
late epochs.   ETGs could play a fundamental role also in this
context. In fact,  the increasing bulk of data that have been
accumulated in the last few years have paved the way to the
emergence of a new cosmological model usually referred to as the
{\it Concordance Model}.

The Hubble diagram of Type Ia Supernovae (hereafter SNeIa),
measured by both the Supernova Cosmology Project \cite{SCP} and
the High\,-\,z Team \cite{HZT} up to redshift $z \sim 1$, has been
the first evidence  that the Universe is undergoing a phase of
accelerated expansion. On the other hand, balloon born
experiments, such as BOOMERanG \cite{Boomerang} and MAXIMA
\cite{Maxima}, determined the location of the first and second
peak in the anisotropy spectrum of the cosmic microwave background
radiation (CMBR) strongly pointing out that the geometry of the
Universe is spatially flat. If combined with constraints coming
from galaxy clusters on the matter density parameter $\Omega_M$,
these data indicate that the Universe is dominated by a
non-clustered fluid with negative pressure, generically dubbed
{\it dark energy}, which is able to drive the accelerated
expansion. This picture has been further strengthened by the
recent precise measurements of the CMBR spectrum, due to the WMAP
experiment \cite{WMAP,hinshaw,hinshaw1}, and by the extension of
the SNeIa Hubble diagram to redshifts higher than 1
\cite{Riess04}.

After these observational evidences, an  overwhelming flood of
papers has appeared: they present a great variety of models trying
to explain this phenomenon. In any case,  the simplest explanation
is claiming for the well known cosmological constant $\Lambda$
\cite{LCDMrev}. Although it is the best fit to most of the
available astrophysical data \cite{WMAP}, the $\Lambda$CDM model
fails in explaining why the inferred value of $\Lambda$ is so tiny
(120 orders of magnitude lower!) if compared with the typical
vacuum energy values predicted by particle physics and why its
energy density is today comparable to the matter density  (the so
called {\it coincidence problem}).

As a tentative solution, many authors have replaced the
cosmological constant with a scalar field rolling down its
potential and giving rise to the model now referred to as {\it
quintessence} \cite{QuintRev,tsu1}. Even if successful in fitting
the data, the quintessence approach to dark energy is still
plagued by the coincidence problem since the dark energy and
matter densities evolve differently and reach comparable values
for a very limited portion of the Universe evolution  coinciding
at present era. To be more precise, the quintessence dark energy
is tracking matter and evolves in the same way for a long time.
But then, at late time, somehow it has to change its behavior into
no longer tracking the dark matter but starting to dominate as a
cosmological constant. This is the coincidence problem of
quintessence.

Moreover, it is not clear where this scalar field originates from,
thus leaving a great uncertainty on the choice of the scalar field
potential. The subtle and elusive nature of  dark energy has led
many authors to  look for completely different scenarios able to
give a quintessential behavior without the need of exotic
components. To this aim, it is worth stressing that the
acceleration of the Universe only claims for a negative pressure
dominant component, but does not tell anything about the nature
and the number of cosmic fluids filling the Universe.

This consideration suggests that it could be possible to explain
the accelerated expansion by introducing a single cosmic fluid
with an equation of state causing it to act like dark matter at
high densities and dark energy at low densities. An attractive
feature of these models, usually referred to as {\it Unified Dark
Energy} (UDE) or {\it Unified Dark Matter} (UDM) models, is that
such an approach naturally solves, al least phenomenologically,
the coincidence problem. Some interesting examples are the
generalized Chaplygin gas \cite{Chaplygin}, the tachyon field
\cite{tachyon} and the condensate cosmology \cite{Bassett}. A
different class of UDE models has been proposed \cite{Hobbit}
where a single fluid is considered: its energy density scales with
the redshift in such a way that the radiation dominated era, the
matter  era and the accelerating phase can be naturally achieved.
It is worth noticing that such class of models are extremely
versatile since they can be interpreted both in the framework of
UDE models and as a two-fluid scenario with dark matter and scalar
field dark energy. The main ingredient of the approach is that a
generalized equation of state can be always obtained and
observational data can be fitted.

Actually, there is still a different way to face the problem of
cosmic acceleration. As stressed in \cite{LSS03}, it is possible
that the observed acceleration is not the manifestation of another
ingredient in the cosmic pie, but rather the first signal of a
breakdown of our understanding of the laws of gravitation (in the
infra-red limit).

From this point of view, it is thus tempting to modify the
Friedmann equations to see whether it is possible to fit the
astrophysical data with  models comprising only the standard
matter. Interesting examples of this kind are the Cardassian
expansion \cite{Cardassian} and the DGP gravity \cite{DGP}. Moving
in this same framework, it is possible to  find alternative
schemes where a quintessential behavior is obtained by taking into
account effective models coming from fundamental physics giving
rise to generalized or higher-order gravity actions
\cite{curvature} (for a comprehensive review see \cite{odinoj}).

For instance, a cosmological constant term may be recovered as a
consequence of a non\,-\,vanishing torsion field thus leading to a
model which is consistent with both SNeIa Hubble diagram and
Sunyaev\,-\,Zel'dovich data coming from clusters of galaxies
\cite{torsion}. SNeIa data could also be efficiently fitted
including higher-order curvature invariants in the gravity
Lagrangian \cite{curvfit,camfr,odgauss,camfrgauss}. It is worth
noticing that these alternative models provide naturally a
cosmological component with negative pressure whose origin is
related to the geometry of the Universe thus overcoming the
problems linked to the physical significance of the scalar field.

It is evident, from this short overview, the high number of
cosmological models which are viable candidates to explain the
observed accelerated expansion. This abundance of models is, from
one hand, the signal of the fact that we have a  limited number of
cosmological tests to discriminate among rival theories, and, from
the other hand, that a urgent degeneracy problem has to be faced.
To this aim, it is useful to remark that both the SNeIa Hubble
diagram and the angular size\,-\,redshift relation of compact
radio sources \cite{AngTest} are distance based methods to probe
cosmological models so then systematic errors and biases could be
iterated. From this point of view, it is interesting to search for
tests based on time-dependent observables.

For example, one can take into account the {\it lookback time} to
distant objects since this quantity can discriminate among
different cosmological models. The lookback time is
observationally estimated as the difference between the present
day age of the Universe and the age of a given object at redshift
$z$. Such an estimate is possible if the object is a galaxy
observed in more than one photometric band since its color is
determined by its age as a consequence of stellar evolution. It is
thus possible to get an estimate of the galaxy age by measuring
its magnitude in different bands and then using stellar
evolutionary codes to choose the model that reproduces the
observed colors at best.

Coming to the weak-field-limit approximation, which essentially
means considering Solar System scales,  ETGs are expected to
reproduce GR which, in any case, is firmly tested only in this
limit \cite{will}. This fact is matter of debate since several
relativistic theories {\it do not} reproduce exactly the Einstein
results in the Newtonian approximation but, in some sense,
generalize them. As it was firstly noticed by Stelle
\cite{stelle}, a $R^2$-theory gives rise to Yukawa-like
corrections in the Newtonian potential. Such a feature could have
interesting physical consequences. For example, some authors claim
to explain the flat rotation curves of galaxies by using such
terms \cite{sanders}. Others \cite{mannheim} have shown that a
conformal theory of gravity is nothing else but a fourth-order
theory containing such terms in the Newtonian limit. Besides,
indications of an apparent, anomalous, long-range acceleration
revealed from the data analysis of Pioneer 10/11, Galileo, and
Ulysses spacecrafts could be framed in a general theoretical
scheme by taking   corrections to the Newtonian potential into
account \cite{anderson,bertolami}.

In general, any relativistic theory of gravitation  yields
corrections to the Newton potential (see for example
\cite{schmidt}) which, in the post-Newtonian (PPN) formalism,
could be a test for the same theory \cite{will}. Furthermore the
newborn {\it gravitational lensing astronomy} \cite{ehlers} is
giving rise to additional tests of gravity over small, large, and
very large scales which  soon will provide direct measurements for
the variation of the Newton coupling \cite{krauss}, the potential
of galaxies, clusters of galaxies and several other features of
self-gravitating systems.

Such data will be, very likely, capable of confirming or ruling
out the physical consistency of GR or of any ETG. In summary, the
general features of ETGs are that the Einstein field equations
result to be modified in two senses: $i)$ geometry can be
non-minimally coupled to some scalar field, and/or $ii)$ higher
than second order derivative terms in the metric come out. In the
former case, we generically deal with scalar-tensor theories of
gravity; in the latter, we deal with higher-order theories.
However  combinations of non-minimally coupled and higher-order
terms can emerge as contributions in effective Lagrangians. In
this case, we deal with higher-order-scalar-tensor theories of
gravity.

Considering a mathematical viewpoint, the problem of reducing more
general theories to Einstein standard form has been extensively
treated; one can see that, through a ``Legendre'' transformation
on the metric, higher-order theories, under suitable regularity
conditions on the Lagrangian, take the form of the Einstein one in
which a scalar field (or more than one) is the source of the
gravitational field (see for example
\cite{francaviglia,sokolowski,ordsup,magnano-soko}); on the other
side, as discussed above, it has been studied the mathematical
equivalence between models with variable gravitational coupling
with the Einstein standard gravity through suitable conformal
transformations (see \cite{dicke,nmc}).

In any case, the debate on the physical meaning of conformal
transformations is far to be solved [see \cite{faraoni} and
references therein for a comprehensive review]. Several authors
claim for a true physical difference between Jordan frame
(higher-order theories and/or variable gravitational coupling)
since there are experimental and observational evidences which
point out that the Jordan frame could be suitable to better match
solutions with data. Others state that the true physical frame is
the Einstein one according to the energy theorems
\cite{magnano-soko}. However, the discussion is open and no
definitive statement has been formulated up to now.

The problem should be faced from a more general viewpoint and the
Palatini approach to gravity could be useful to this goal. The
Palatini approach in gravitational theories was firstly introduced
and analyzed by Einstein himself \cite{palaeinstein}. It was,
however, called the Palatini approach as a consequence of an
historical misunderstanding \cite{buchdahl,frafe}.

The fundamental idea  of the Palatini formalism is to consider the
(usually torsion-less) connection $\Gamma$, entering the
definition of the Ricci tensor, to  be independent of the metric
$g$ defined on the spacetime ${\cal M}$. The Palatini formulation
for the standard Hilbert-Einstein  theory results to be equivalent
to the purely metric theory: this follows from the fact that the
field equations for the connection  $\Gamma$, firstly considered
to be independent of the metric,  give the Levi-Civita connection
of the metric $g$. As a consequence, there is  no reason to impose
the Palatini variational principle in the standard
Hilbert-Einstein theory instead of the metric  variational
principle.

However, the situation completely changes if we consider the ETGs,
depending on functions of  curvature invariants, as $f(R)$, or
non-minimally coupled  to some scalar field. In these cases, the
Palatini and the metric variational principle provide different
field equations and the theories thus derived differ
\cite{magnano-soko,FFV}. The relevance of  Palatini approach, in
this framework, has been recently proven in relation to
cosmological applications
\cite{curvature,odinoj,palatinifR,palatinicam1,palatinicam2}.

It has also been studied the crucial problem of the Newtonian
potential in alternative theories of Gravity and its relations
with the conformal factor \cite{meng_rev}. From a physical
viewpoint, considering the metric $g$ and the connection $\Gamma$
as independent fields means to decouple the metric structure of
spacetime and its geodesic structure (being, in general, the
connection $\Gamma$ not the Levi-Civita connection of $g$). The
chronological structure of spacetime is governed by $g$ while the
trajectories of particles, moving in the spacetime, are governed
by $\Gamma$.

This decoupling enriches the geometric structure of spacetime and
generalizes the purely metric formalism. This metric-affine
structure of spacetime  is naturally translated, by means of the
same (Palatini) field equations, into a bi-metric structure of
spacetime. Beside the \textit{physical} metric $g$, another metric
$h$ is involved. This new metric is related, in the case of
$f(R)$-gravity, to the connection. As a matter of fact, the
connection $\Gamma$ results to be the Levi-Civita connection of
$h$ and thus provides the geodesic structure of spacetime.

If we consider the case of non-minimally coupled interaction in
the gravitational Lagrangian (scalar-tensor theories), the new
metric $h$ is  related to the non-minimal coupling. The new metric
$h$ can be thus related to a different geometric and physical
aspect of the gravitational theory. Thanks to the Palatini
formalism, the non-minimal coupling and the scalar field, entering
the evolution of the gravitational fields, are separated from the
metric structure of spacetime. The situation mixes when we
consider the case of higher-order-scalar-tensor theories. Due to
these features, the Palatini approach could greatly contribute to
clarify the physical meaning of conformal transformation
\cite{ACCF}.

\vspace{3.mm}

In this review paper,  without claiming for completeness, we want
to give a survey on the formal and physical aspects of ETGs in
metric and Palatini approaches, considering the cosmological and
astrophysical applications of some ETG models.

The layout is the following. Sect.II is a rapid overview of GR. We
summarize what a good theory of gravity is requested to do and
what  the  foundations of the Einstein theory are. The goal is to
demonstrate that ETGs  have the same theoretical bases but, in
principle, could avoid some shortcomings of GR which is nothing
else but a particular case of ETG, $f(R)=R$.

The field equations for generic ETGs are derived in Sec.III.
Specifically, we discuss two interesting cases: $f(R)$ and
scalar-tensor theories considering their relations with GR by
conformal transformations.

The Palatini approach and its intrinsic conformal structure is
discussed in Sec.IV giving some peculiar examples.

Cosmological applications are considered in Sec.V. After a short
summary of $\Lambda$CDM model, we show that dark energy and
quintessence issues can be addressed as "curvature effects", if
ETGs (in particular $f(R)$ theories) are considered. We work out
some cosmological models  comparing the solutions with data coming
from observational surveys. As further result, we show that also
the stochastic cosmological background of gravitational waves
could be "tuned" by ETGs. This fact could open new perspective
also in the issues of detection and production of gravitational
waves which should be investigated not only in the standard
framework of GR.

Sec.VI is devoted to the galactic dynamics under the standard of
ETGs. Also in this case, we show that flat rotation curves and
haloes of spiral galaxies could be explained as curvature effects
which give rise to corrections to the Newton potential without
taking into account huge amounts of dark matter.  Discussion and
conclusions are drawn in Sec.VII.

\section{What a good theory of Gravity has to do:  General Relativity and its extensions}

From a phenomenological point of view, there are some minimal
requirements that any relativistic theory of gravity has to match.
First of all, it has to explain the astrophysical observations
(e.g. the orbits of planets, the potential of self-gravitating
structures).

This means that it has to reproduce the Newtonian dynamics in the
weak-energy limit. Besides, it has to pass the classical Solar
System tests which are all experimentally well founded
\cite{will}.

As second step, it should reproduce galactic dynamics considering
the observed baryonic constituents (e.g. luminous components as
stars, sub-luminous components as planets, dust and gas),
radiation and Newtonian potential which is, by assumption,
extrapolated to galactic scales.

Thirdly, it should address the problem of large scale structure
(e.g. clustering of galaxies) and finally cosmological dynamics,
which means to reproduce, in a self-consistent way, the
cosmological parameters as the expansion rate, the Hubble
constant, the density parameter and so on. Observations and
experiments, essentially, probe the standard baryonic matter, the
radiation and an attractive overall interaction, acting at all
scales and depending on distance: the gravity.

The simplest theory which try to satisfies the above requirements
was formulated by Albert Einstein in the years 1915-1916
\cite{einstein} and it is known as the Theory of General
Relativity. It is firstly based on the assumption that space and
time have to be entangled into a single spacetime structure,
which, in the limit of no gravitational forces, has to reproduce
the Minkowski spacetime structure. Einstein profitted also of
ideas earlier put forward by Riemann, who stated that the Universe
should be a curved manifold and that its curvature should be
established on the basis of astronomical observations
\cite{riemann}.

In other words, the distribution of matter has to influence point
by point the local curvature of the spacetime structure. The
theory, eventually formulated by Einstein in 1915, was strongly
based on three assumptions that the Physics of Gravitation has to
satisfy.

The "{\it Principle of Relativity}", that amounts to require all
frames to be good frames for Physics, so that no preferred
inertial frame should be chosen a priori (if any exist).

The "{\it Principle of Equivalence}", that amounts to require
inertial effects to be locally indistinguishable from
gravitational effects (in a sense, the equivalence between the
inertial and the gravitational mass).

The "{\it Principle of General Covariance}", that requires field
equations to be "generally covariant" (today, we would better say
to be invariant under the action of the group of all spacetime
diffeomorphisms) \cite{schroedinger}.

And - on the top of these three principles - the requirement that
causality has to be preserved (the "{\it Principle of Causality}",
i.e. that each point of spacetime should admit a universally valid
notion of past, present and future).

Let us also recall that the older Newtonian theory of spacetime
and gravitation - that Einstein wanted to reproduce at least in
the limit of small gravitational forces (what is called today the
"post-Newtonian approximation") - required space and time to be
absolute entities, particles moving in a preferred inertial frame
following curved trajectories, the curvature of which (i.e., the
acceleration) had to be determined as a function of the sources
(i.e., the "forces").

On these bases, Einstein was  led to postulate that the
gravitational forces have to be expressed by the curvature of a
metric tensor field  $ds^2 = g_{\mu\nu}dx^{\mu}dx^{\nu}$ on a
four-dimensional spacetime manifold, having the same signature of
Minkowski metric, i.e., the so-called "Lorentzian signature",
herewith assumed to be $(+,-,-,-)$. He also postulated that
spacetime is curved in itself and that its curvature is locally
determined by the distribution of the sources, i.e. - being
spacetime a continuum - by the four-dimensional generalization of
what in Continuum Mechanics is called the "matter stress-energy
tensor", i.e. a rank-two (symmetric) tensor  $T^m_{\mu\nu}$.

Once a metric $g_{\mu\nu}$ is given, its curvature is expressed by
the Riemann (curvature) tensor

\begin{equation}
R^{\alpha}\hspace{0.01pt}_{\beta\mu\nu}=\Gamma_{\beta\nu}^{\alpha}\hspace{0.01pt}_{,\mu}-
\Gamma_{\beta\mu}^{\alpha}\hspace{0.01pt}_{,\nu}+\Gamma_{\beta\nu}^{\sigma}
\Gamma_{\sigma\mu}^{\alpha}
-\Gamma_{\beta\mu}^{\sigma}\Gamma_{\sigma\nu}^{\alpha}\,
\end{equation}
where the comas are partial derivatives. Its contraction
\begin{equation}
R^{\alpha}\hspace{0.01pt}_{\mu\alpha\nu}=R_{\mu\nu}\,,
\end{equation}
is the "Ricci tensor" and the scalar
\begin{equation}
R=R^{\mu}\hspace{0.01pt}_{\mu}=g^{\mu\nu}R_{\mu\nu}
\end{equation}
is called the "scalar curvature" of $g_{\mu\nu}$. Einstein was led
to postulate the following equations for the dynamics of
gravitational forces
\begin{equation}\label{wrong}
R_{\mu\nu} = \frac{\kappa}{2} T^m_{\mu\nu}
\end{equation}
where $\kappa=8\pi G$, with $c=1$ is a coupling constant. These
equations turned out to be physically and mathematically
unsatisfactory.

As Hilbert pointed out \cite{schroedinger}, they were not of a
variational origin, i.e. there was no Lagrangian able to reproduce
them exactly (this is  slightly wrong, but this remark is
unessential here). Einstein replied that he knew that the
equations were physically unsatisfactory, since they were
contrasting with the continuity equation of any reasonable kind of
matter. Assuming that matter is given as a perfect fluid, that is
\begin{equation}
T^m_{\mu\nu} = (p + \rho)u_{\mu}u_{\nu} - pg_{\mu\nu}
\end{equation}
where $u_{\mu}u_{\nu}$ is a comoving observer, $p$ is the pressure
and $\rho$ the density of the fluid, then the continuity equation
requires $T^m_{\mu\nu}$ to be covariantly constant, i.e. to
satisfy the conservation law
\begin{equation}\label{conservation}\nabla^{\mu} T^m_{\mu\nu} = 0\,,\end{equation}
where  $\nabla^{\mu}$ denotes the covariant derivative with
respect to the metric.

In fact, it is not true that $\nabla^{\mu} R_{\mu\nu}$ vanishes
(unless $R = 0$). Einstein and Hilbert reached independently the
conclusion that the wrong field equations (\ref{wrong}) had to be
replaced by the correct ones
\begin{equation}\label{field}
G_{\mu\nu} = \kappa T^m_{\mu\nu}
\end{equation}
where
\begin{equation}
G_{\mu\nu} = R_{\mu\nu} - \frac{1}{2}  g_{\mu\nu}R
\end{equation}
that is currently called the "Einstein tensor" of $g_{\mu\nu}$.
These equations are both variational and satisfy the conservation
laws (\ref{conservation}) since the following relation holds \beq
\nabla^{\mu} G_{\mu\nu} = 0\,,\eeq as a byproduct of the so-called
"Bianchi identities" that the curvature tensor of $g_{\mu\nu}$ has
to satisfy \cite{weinberg}.

The Lagrangian that allows to obtain the field equations
(\ref{field}) is the sum of a "matter Lagrangian"  ${\l}_m$, the
variational derivative of which is exactly $T^m_{\mu\nu}$, i.e.
\beq T^m_{\mu\nu} = \frac{\delta {\cal L}_m}{\delta
g^{\mu\nu}}\eeq and of a "gravitational Lagrangian", currently
called the Hilbert-Einstein Lagrangian\beq \label{HE} L_{HE} =
g^{\mu\nu} R_{\mu\nu} \sqrt{-g}=R \sqrt{-g}\,,\eeq where
$\sqrt{-g}$  denotes the square root of the value of the
determinant of the metric $g_{\mu\nu}$.

The choice of Hilbert and Einstein was completely arbitrary  (as
it became clear a few years later), but it was certainly the
simplest one both from the mathematical and the physical
viewpoint. As it was later clarified by Levi-Civita in 1919,
curvature is  not a "purely metric notion" but, rather, a notion
related to the   "linear connection"  to which "parallel
transport" and "covariant derivation" refer \cite{levicivita}.

In a sense, this is the precursor idea of what in the sequel would
be called a "gauge theoretical framework" \cite{gauge}, after the
pioneering work by Cartan in 1925 \cite{cartan}. But at the time
of Einstein, only metric concepts were at hands and his solution
was the only viable.

It was later clarified  that the three principles of relativity,
equivalence and covariance, together with causality, just require
that the spacetime structure has to be determined by either one or
both of two fields, a Lorentzian metric  $g$  and a linear
connection  $\Gamma$, assumed to be torsionless for the sake of
simplicity.

The metric  $g$  fixes the causal structure of spacetime (the
light cones) as well as its metric relations (clocks and rods);
the connection  $\Gamma$  fixes the free-fall, i.e. the locally
inertial observers. They have, of course, to satisfy a number of
compatibility relations which amount to require that photons
follow null geodesics of $\Gamma$, so that $\Gamma$ and $g$ can be
independent, {\it a priori}, but constrained, {\it a posteriori},
by some physical restrictions. These, however, do not impose that
$\Gamma$ has necessarily to be the Levi-Civita connection of  $g$
\cite{palatiniorigin}.

This justifies - at least on a purely theoretical basis - the fact
that one can envisage the so-called "alternative theories of
gravitation", that we prefer  to call "{\it Extended Theories of
Gravitation}" since their starting points are exactly those
considered by Einstein and Hilbert: theories in which gravitation
is described by either a metric (the so-called "purely metric
theories"), or by a linear connection (the so-called "purely
affine theories") or by both fields (the so-called "metric-affine
theories", also known as "first order formalism theories"). In
these theories,  the Lagrangian is a scalar density  of the
curvature invariants constructed out of both $g$ and $\Gamma$.

The choice (\ref{HE})  is by no means unique and it turns out that
the Hilbert-Einstein Lagrangian is in fact the only choice that
produces an invariant that is linear in second derivatives of the
metric (or first derivatives of the connection). A Lagrangian
that, unfortunately, is rather singular from the Hamiltonian
viewpoint, in much than same way as Lagrangians, linear in
canonical momenta, are rather singular in Classical Mechanics (see
e.g. \cite{arnold}).

A number of attempts to generalize GR (and unify it to
Electromagnetism) along these lines were  followed by Einstein
himself and many others (Eddington, Weyl, Schrodinger, just to
quote the main contributors; see, e.g., \cite{unification}) but
they were eventually given up in the fifties of XX Century, mainly
because of a number of difficulties related to the definitely more
complicated structure of a non-linear theory (where by
"non-linear" we mean here a theory that is based on non-linear
invariants of the curvature tensor), and also because of the new
understanding of Physics that is currently based on four
fundamental forces and requires the more general "gauge framework"
to be adopted (see \cite{unification2}).

Still a number of sporadic investigations about "alternative
theories" continued even after 1960 (see \cite{will} and refs.
quoted therein for a short history). The search of a coherent
quantum theory of gravitation or the belief that gravity has to be
considered as a sort of low-energy limit of string theories (see,
e.g., \cite{green}) - something that we are not willing to enter
here in detail - has more or less recently revitalized the idea
that there is no reason to follow the simple prescription of
Einstein and Hilbert and to assume that gravity should be
classically governed by a Lagrangian linear in the curvature.

Further curvature invariants or non-linear functions of them
should be also considered, especially in view of the fact that
they have to be included in both the semi-classical expansion of a
quantum Lagrangian or in the low-energy limit of a string
Lagrangian.

Moreover, it is clear from the recent astrophysical observations
and from the current cosmological hypotheses that Einstein
equations are no longer a good test for gravitation  at Solar
System, galactic, extra-galactic and cosmic scale, unless one does
not admit that the matter side of Eqs.(\ref{field}) contains some
kind of exotic matter-energy which is the "dark matter" and "dark
energy" side of the Universe.

The idea which we propose here is much simpler. Instead of
changing the matter side of Einstein Equations (\ref{field}) in
order to fit the "missing matter-energy" content of the currently
observed Universe (up to the $95\%$ of the total amount!), by
adding any sort of inexplicable and strangely behaving matter and
energy, we claim that it is simpler and more convenient to change
the gravitational side of the equations, admitting corrections
coming from non-linearities in the Lagrangian. However, this is
nothing else but a matter of taste and, since it is possible, such
an approach should be explored. Of course, provided that the
Lagrangian can be conveniently tuned up (i.e., chosen in a huge
family of allowed Lagrangians) on the basis of its best fit with
all possible observational tests, at all scales (solar, galactic,
extragalactic and cosmic).

Something that - in spite of some commonly accepted but disguised
opinion - can and should be done before rejecting a priori a
non-linear theory of gravitation (based on a non-singular
Lagrangian) and insisting that the Universe has to be necessarily
described by a rather singular gravitational Lagrangian (one that
does not allow a coherent perturbation  theory from a good
Hamiltonian viewpoint) accompanied by matter that does not follow
the behavior that standard baryonic matter, probed in our
laboratories, usually satisfies.

\section{ The Extended Theories of Gravity}

With the above considerations in mind, let us start with a general
class of higher-order-scalar-tensor theories in four dimensions
 \footnote{For the aims of this review, we do
not need more complicated invariants like $R_{\mu\nu}R^{\mu\nu}$,
$R_{\mu\nu\alpha\beta}R^{\mu\nu\alpha\beta}$,
$C_{\mu\nu\alpha\beta}C^{\mu\nu\alpha\beta}$ which are also
possible. } given by the action \beq \label{3.1} {\cal A}=\int
d^{4}x\sqrt{-g}\left[F(R,\Box R,\Box^{2}R,..\Box^kR,\p)
 -\frac{\epsilon}{2}
g\umunu \phi\ddemu \phi\ddenu+ {\l}_{m}\right], \eeq where $F$ is
an unspecified function of curvature invariants and of a scalar
field $\p$. The term ${\l}_{m}$, as above, is the minimally
coupled ordinary matter contribution. We shall use physical units
$8\pi G=c=\hbar=1$;
 $\epsilon$ is a constant which specifies the theory. Actually its
 values can be $\epsilon =\pm 1,0$ fixing the nature and the
 dynamics of the scalar field which can be a standard scalar
 field, a phantom field or a field without dynamics (see
 \cite{valerio,CP} for details).

In the metric approach, the field equations are obtained by
varying (\ref{3.1}) with respect to  $\gd$.  We get \beqa
\label{3.2} G\umunu&=&\frac{1}{{\cal
G}}\left[T\umunu+\frac{1}{2}\gu (F-{\cal G}R)+
(g^{\mu\lambda}g^{\nu\sigma}-\gu g^{\lambda\sigma})
{\cal G}_{;\lambda\sigma}\right.\nonumber\\
& & +\frac{1}{2}\sum_{i=1}^{k}\sum_{j=1}^{i}(\gu
g^{\lambda\sigma}+
  g^{\mu\lambda} g^{\nu\sigma})(\Box^{j-i})_{;\sigma}
\left(\Box^{i-j}\frac{\pa F}{\pa \Box^{i}R}\right)_{;\lambda}\nonumber\\
& &\left.-\gu g^{\lambda\sigma}\left((\Box^{j-1}R)_{;\sigma}
\Box^{i-j}\frac{\pa F}{\pa \Box^{i}R}\right)_{;\lambda}\right]\,,
\eeqa where $G\umunu$ is the above Einstein tensor and \beq
\label{3.4}
  {\cal G}\equiv\sum_{j=0}^{n}\Box^{j}\left(\frac{\pa F}{\pa \Box^{j} R}
\right)\;. \eeq The differential Eqs.(\ref{3.2}) are of order
$(2k+4)$. The stress-energy tensor is due to the kinetic part of
the scalar field and to the ordinary matter: \beq \label{3.5}
T\dmunu=T^{m}\dmunu+\frac{\epsilon}{2}[\p\ddemu\p\ddenu-
\frac{1}{2}\p\udea\p\ddea]\;. \eeq The (eventual) contribution of
a potential $\v$ is contained in the definition of $F$. From now
on, we shall indicate by a capital $F$ a Lagrangian density
containing also the contribution of a potential $\v$ and by
$F(\p)$, $f(R)$, or $f(R,\Box R)$ a function of such fields
 without potential.

By varying with respect to the scalar field $\p$, we obtain the
Klein-Gordon equation \beq \label{3.6} \epsilon\Box\p=-\frac{\pa
F}{\pa\p}\,. \eeq Several approaches can be used to deal with such
equations. For example, as we said, by a conformal transformation,
it is possible to reduce an ETG to a (multi) scalar-tensor theory
of gravity \cite{schmidt,wands1,wands,gottloeber,damour}.

The simplest extension of GR is achieved assuming \beq\label{fr}
F=f(R)\,,\qquad \epsilon=0\,,\eeq in the action (\ref{3.1});
 $f(R)$ is an arbitrary (analytic) function of the Ricci
curvature scalar $R$. We are  considering here the simplest case
of fourth-order gravity but we could
 construct such kind of theories also using other invariants
in $R\dmunu$ or $R\ua_{\beta\mu\nu}$.  The standard
Hilbert-Einstein action is, of course, recovered for $f(R)=R$.
Varying with respect to $g\dab$, we get the field equations
\begin{equation}\label{h2}
f'(R)R\dab-\frac{1}{2}f(R)g\dab=f'(R)^{;\umunu}\left(
g_{\alpha\mu}g_{\beta\nu}-g\dab\gd\right)\,, \end{equation} which
are fourth-order equations due to the term $f'(R)^{;\mu\nu}$; the
prime  indicates the derivative with respect to  $R$.
Eq.(\ref{h2}) is also the equation for $T\dmunu=0$ when the matter
term is absent.

By a suitable manipulation, the above equation can be rewritten
as:
 \begin{equation}\label{h3}
G\dab=\frac{1}{f'(R)}\left\{\frac{1}{2}g\dab\left[f(R)-Rf'(R)\right]
+f'(R)\ddeab -g\dab\Box f'(R)\right\}\,,
\end{equation}
where the gravitational contribution due to higher-order terms can
be simply reinterpreted as a stress-energy tensor contribution.
This means that additional and higher-order terms in the
gravitational action act, in principle, as a stress-energy tensor,
related to the  form of $f(R)$.  Considering also the standard
perfect-fluid matter contribution, we have
 \begin{equation}\label{h4}
G\dab=\frac{1}{f'(R)}\left\{\frac{1}{2}g\dab\left[f(R)-Rf'(R)\right]
+f'(R)\ddeab -g\dab\Box f'(R)\right\}+ \frac{T^{m}_{\alpha
\beta}}{f'(R)}=T^{curv}_{\alpha\beta}+\frac{T^{m}_{\alpha
\beta}}{f'(R)}\,,
\end{equation}
where $T^{curv}_{\alpha\beta}$ is an effective stress-energy
tensor constructed by the extra curvature terms.  In the case of
GR,   $T^{curv}_{\alpha\beta}$ identically vanishes while the
standard, minimal coupling is recovered for the matter
contribution. The peculiar behavior of $f(R)=R$ is  due to the
particular form of the Lagrangian itself which, even though it is
a second order Lagrangian, can be non-covariantly rewritten as the
sum of a first order  Lagrangian plus a pure divergence term. The
Hilbert-Einstein Lagrangian can be in fact recast as follows:
\begin{equation}
L_{HE}= {\cal L}_{HE} \sqrt{-g}  =\Big[ p^{\alpha \beta}
(\Gamma^{\rho}_{\alpha \sigma} \Gamma^{\sigma}_{\rho
\beta}-\Gamma^{\rho}_{\rho \sigma} \Gamma^{\sigma}_{\alpha
\beta})+ \nabla_\sigma (p^{\alpha \beta} {u^{\sigma}}_{\alpha
\beta}) \Big]
\end{equation}
\noindent where:
\begin{equation}
 p^{\alpha \beta} =\sqrt{-g}  g^{\alpha \beta} = \frac{\pa {\cal{L}}}{\pa R_{\alpha \beta}}
\end{equation}
$\Gamma$ is the Levi-Civita connection of $g$ and
$u^{\sigma}_{\alpha \beta}$ is a quantity constructed out with the
variation of $\Gamma$ \cite{weinberg}. Since $u^{\sigma}_{\alpha
\beta}$ is not a tensor, the above expression is not covariant;
however a standard procedure has been studied to recast covariance
in the first order theories \cite{FF1}. This clearly shows that
the field equations should consequently be second order  and the
Hilbert-Einstein Lagrangian is thus degenerate.

From the action (\ref{3.1}), it is possible to obtain another
interesting case by choosing \beq F=F(\f)R-V(\f)\,,\qquad \epsilon
=-1\,.\eeq In this case, we get
\begin{equation} \label{s1} {\cal A}= \int \vol \left[F(\f) R+ \half \gu
\f\ddemu \f\ddenu- V(\f) \right] \end{equation} $V(\f)$ and
$F(\f)$ are generic functions describing respectively the
potential and the coupling of a scalar field $\f$.  The
Brans-Dicke theory of gravity is a particular case of the action
(\ref{s1}) for $V(\f)$=0 \cite{noibrans}. The variation with
respect to $\gd$ gives the second-order field equations
\begin{equation} \label{s2} F(\f) G\dmunu= F(\f)\left[R\dmunu-
\half R \gd \right]= -\half T^{\f}\dmunu- g\dmunu \Box_{g} F(\f)+
F(\f)\ddemunu\,, \end{equation}  here $\Box_{g}$ is the d'Alembert
operator with respect to the metric $g$  The energy-momentum
tensor relative to the scalar field is
\begin{equation} \label{s4} T^{\f}\dmunu= \f\ddemu \f\ddenu- \half g\dmunu \f\ddea
\f\udea+ g\dmunu V(\f) \end{equation} The variation with respect
to $\f$ provides the Klein - Gordon equation, i.e. the field
equation for the scalar field:  \begin{equation} \label{s5}
\Box_{g} \f- R \FDEF(\f)+ \VDEF(\f)= 0
\end{equation}
where $\FDEF= dF(\f)/d\f$, $\VDEF= dV(\f)/d\f$. This last equation
is equivalent to the Bianchi contracted identity \cite{cqg}.
Standard fluid matter can be treated as above.

\subsection{Conformal transformations}

Let us now introduce conformal transformations to show that any
higher-order or scalar-tensor theory, in absence of ordinary
matter, e.g. a perfect fluid,  is conformally equivalent to an
Einstein theory plus minimally coupled scalar fields. If standard
matter is present, conformal transformations allow to transfer
non-minimal coupling to the matter component \cite{magnano-soko}.
The conformal transformation on the metric $\gd$ is
\begin{equation} \label{s6}
\tilde{g}\dmunu= e^{2 \ome} \gd
\end{equation}
in which $e^{2 \ome}$ is the conformal factor. Under this
transformation,  the Lagrangian  in (\ref{s1}) becomes
\begin{equation} \label{s7}
\begin{array}{ll}
\disp{ \sqrt{-g} \left(F R+ \half \gu \f\ddemu \f\ddenu- V\right
)}
 &= \sqrt{-\tilde{g}} e^{-2 \ome} \left(F \tilde{R}- 6 F \Box_{\tilde{g}}
\ome+\right. \\
~ & ~ \\
~ & \disp{ \left. -6 F \ome\ddea \ome\udea+ \half \tilde{g}\umunu
\f\ddemu \f\ddenu- e^{-2 \ome} V\right)}
\end{array}
\end{equation}
in which $\tilde{R}$ and $\Box_{\tilde{g}}$ are the Ricci scalar
and the d'Alembert operator  relative to the metric $\tilde{g}$.
Requiring the theory in the metric $\tilde{g}\dmunu$ to appear as
a standard Einstein theory  \cite{conf1},  the conformal factor
has to be related to $F$, that is
\begin{equation} \label{s8} e^{2 \ome}= -2
F. \end{equation} where $F$ must be negative in order to restore
physical coupling. Using this relation and
 introducing a new scalar field $\tilde{\f}$ and a new potential $\tilde{V}$,
defined respectively by
\begin{equation} \label{s10} \tilde{\f}\ddea=
\sqrt{\frac{3\FDEF^2- F}{2 F^2}}\, \f\ddea,~~~
\tilde{V}(\tilde{\f}(\f))= \frac{V(\f)}{4 F^2(\f)},
\end{equation} we see that the Lagrangian  (\ref{s7})
becomes
\begin{equation} \label{s11} \sqrt{-g}
\left( F R+ \half \gu \f\ddemu \f\ddenu- V\right) =
\sqrt{-\tilde{g}} \left( -\half \tilde{R}+ \half \tilde{\f}\ddea
\tilde{\f}\udea- \tilde{V}\right) \nonumber \end{equation} which
is the usual Hilbert-Einstein Lagrangian plus the standard
Lagrangian  relative to the scalar field $\tilde{\f}$. Therefore,
every non-minimally coupled scalar-tensor theory, in absence of
ordinary matter, e.g. perfect fluid,  is conformally equivalent to
an Einstein theory, being the conformal transformation and the
potential suitably defined by (\ref{s8}) and (\ref{s10}). The
converse is also true: for a given $F(\f)$, such that $3 \FDEF^2-
F> 0$, we can transform a standard Einstein theory into a
non-minimally coupled scalar-tensor theory. This means that, in
principle, if we are able to solve the field equations in the
framework of the Einstein theory in presence of a scalar field
with a given potential, we should be able to get the solutions for
the scalar-tensor theories, assigned by the coupling $F(\f)$, via
the conformal transformation  (\ref{s8}) with the constraints
given by (\ref{s10}). Following the standard terminology, the
``Einstein frame'' is the framework of the Einstein theory with
the minimal coupling and the
 ``Jordan frame'' is the framework of the non-minimally coupled theory
\cite{magnano-soko}.\\
In the context of alternative theories of gravity, as previously
discussed, the gravitational contribution to the stress-energy
tensor of the theory can be reinterpreted by means of a conformal
transformation as the stress-energy tensor of a suitable scalar
field and then as ``matter" like terms. Performing the conformal
transformation (\ref{s6}) in the field equations (\ref{h3}), we
get:
\begin{equation}\label{hl4}
\tilde{G}\dab=\frac{1}{f'(R)}\left\{\frac{1}{2}g\dab\left[f(R)-
Rf'(R)\right] +f'(R)\ddeab- g\dab\Box f'(R)\right\}+
\end{equation}
$$ +2\left(\omega_{;\alpha ;\beta}
+g\dab\Box \omega -\omega_{;\alpha}\omega_{;\beta}+
\frac{1}{2}g\dab\omega_{;\gamma}\omega^{;\gamma}\right)\,.$$ We
can then choose the conformal factor to be
 \begin{equation}\label{h5}
\omega=\frac{1}{2}\ln |f'(R)|\,, \end{equation} which has now to
be substituted into (\ref{h4}). Rescaling $\omega$ in such a way
that
\begin{equation}\label{h6} k\phi =\omega\,, \end{equation} and $k=\sqrt{1/6}$, we obtain
the Lagrangian equivalence \begin{equation} \label{h7} \sqrt{-g}
f(R)= \sqrt{-\tilde{g}} \left( -\half \tilde{R}+ \half
\tilde{\f}\ddea \tilde{\f}\udea- \tilde{V}\right) \end{equation}
and the Einstein equations in standard form
\begin{equation}\label{h8} \tilde{G}\dab=
\phi\ddea\phi\ddeb-\frac{1}{2}\tilde{g}\dab\phi_{;\gamma}\phi^{;\gamma}
+\tilde{g}\dab V(\phi)\,, \end{equation}  with the potential
\begin{equation}\label{h9}
V(\phi)=\frac{e^{-4k\phi}}{2}\left[{\cal P}(\phi)-{\cal
N}\left(e^{2k\phi}\right)e^{2k\phi}\right]
=\frac{1}{2}\frac{f(R)-Rf'(R)}{f'(R)^{2}}\,.
\end{equation}
Here ${\cal N}$ is the inverse function of
 ${\cal P}'(\phi)$ and ${\cal P}(\phi)=\int \exp (2k\phi) d{\cal N}$. However, the
problem is
 completely solved if
${\cal P}'(\phi)$ can be analytically inverted. In summary, a
fourth-order theory is conformally equivalent to the standard
second-order Einstein theory plus a scalar field (see also
\cite{francaviglia,ordsup}).\\
This procedure can be extended to more general theories. If the
theory is assumed to be higher than fourth order, we may have
Lagrangian densities of the form \cite{buchdahl,gottloeber},
\begin{equation}\label{h10} {\cal L}={\cal L}(R,\Box R,...\Box^{k} R)\,. \end{equation}
Every $\Box$ operator introduces two further terms of derivation
into the field equations. For example a theory like
\begin{equation}\label{h11} {\cal L}=R\Box R\,, \end{equation} is a sixth-order theory
and the above approach can be pursued by considering a conformal
factor of the form
 \begin{equation}
\label{h12} \omega=\frac{1}{2}\ln \left|\frac{\pa {\cal L}}{\pa R}
+\Box\frac{\pa {\cal L}}{\pa \Box R}\right|\,.
\end{equation}
In general,  increasing two orders of derivation in the field
equations (\ie for every term $\Box R$), corresponds to adding a
scalar field in the conformally transformed frame
\cite{gottloeber}. A sixth-order theory can be reduced to an
Einstein theory with two minimally coupled scalar fields; a
$2n$-order theory can be, in principle, reduced to an Einstein
theory plus $(n-1)$-scalar fields. On the other hand, these
considerations can be directly generalized to
higher-order-scalar-tensor theories in any number of dimensions as
shown in \cite{maeda}.

As concluding remarks, we can say that conformal transformations
work at three levels: $i)$ on the Lagrangian of the given theory;
$ii)$ on the field equations; $iii)$ on the solutions. The table
below summarizes the situation for fourth-order gravity (FOG),
non-minimally coupled scalar-tensor theories (NMC) and standard
Hilbert-Einstein (HE) theory. Clearly, direct and inverse
transformations correlate all the steps of the table but no
absolute criterion, at this point of the discussion, is able to
select which is the ``physical" framework since, at least from a
mathematical point of view, all the frames are equivalent
\cite{magnano-soko}. This point is up to now unsolved even if wide
discussions are present in literature \cite{faraoni}.
\begin{center}
\begin{tabular}{|ccccc|} \hline
  ${\cal L}_{FOG}$ & $\longleftrightarrow$ & ${\cal L}_{NMC}$ &
$\longleftrightarrow$ &
     ${\cal L}_{HE}$ \\
  $\updownarrow$ &  & $\updownarrow$ &  & $\updownarrow$ \\
  FOG Eqs. & $\longleftrightarrow$ & NMC Eqs. & $\longleftrightarrow$ &
Einstein Eqs. \\
  $\updownarrow$ &  & $\updownarrow$ &  & $\updownarrow$ \\
  FOG Solutions & $\longleftrightarrow$ & NMC Solutions &
$\longleftrightarrow$ &
  Einstein Solutions \\ \hline
\end{tabular}
\end{center}

\section{The Palatini Approach and the Intrinsic Conformal Structure}

As we said, the Palatini approach, considering $g$ and $\Gamma$ as
independent fields, is ``intrinsically" bi-metric and capable of
disentangling  the geodesic structure from the chronological
structure of a given manifold. Starting from these considerations,
conformal transformations assume a fundamental role in defining
the affine connection which is merely ``Levi-Civita" only for the
Hilbert-Einstein theory.

In this section, we work out examples showing how conformal
transformations assume a fundamental physical role in relation to
the Palatini approach in ETGs \cite{ACCF}.

Let us start from the case of fourth-order gravity where Palatini
variational principle is straightforward in showing the
differences with Hilbert-Einstein variational principle, involving
only metric. Besides, cosmological applications of $f(R)$ gravity
have shown the relevance of  Palatini formalism, giving physically
interesting results with singularity - free solutions
\cite{palatinifR}. This last nice feature is not present in the
standard metric approach.

An important remark is in order at this point. The Ricci scalar
entering in $f(R)$ is $R\equiv R( g,\Gamma)
=g^{\alpha\beta}R_{\alpha \beta}(\Gamma )$ that is a
\textit{generalized Ricci scalar} and $ R_{\mu \nu }(\Gamma )$ is
the Ricci tensor of a torsion-less connection $\Gamma$, which,
{\it a priori}, has no relations with the metric $g$ of spacetime.
The gravitational part of the Lagrangian is controlled by a given
real analytical function of one real variable $f(R)$, while
$\sqrt{-g}$ denotes a related scalar density of weight $1$. Field
equations, deriving from the Palatini variational principle are:
\begin{equation}
f^{\prime }(R)R_{(\mu\nu)}(\Gamma)-\frac{1}{2}f(R)g_{\mu \nu }=
T^{m}_{\mu\nu}\label{ffv1}
\end{equation}
\begin{equation}
\nabla _{\alpha }^{\Gamma }(\sqrt{-g}f^{\prime}(R)g^{\mu \nu })=0
\label{ffv2}
\end{equation}
where  $\nabla^{\Gamma}$ is the covariant derivative with respect
to $\Gamma$. As above, we assume $8\pi G=1$.  We shall use the
standard notation denoting by $R_{(\mu\nu)}$ the symmetric part of
$R_{\mu\nu}$, \ie $R_{(\mu\nu)}\equiv
\frac{1}{2}(R_{\mu\nu}+R_{\nu\mu})$.

In order to get (\ref{ffv2}), one has to additionally assume that
${\l}_m$ is functionally independent of $\Gamma$; however it may
contain metric  covariant derivatives $\stackrel{g}{\nabla}$ of
fields. This means that the matter stress-energy tensor
$T^m_{\mu\nu}=T^m_{\mu\nu}(g,\Psi)$ depends on the metric $g$ and
some matter fields denoted here by $\Psi$, together with their
derivatives (covariant derivatives with respect to the Levi-Civita
connection of $g$). From (\ref{ffv2}) one sees that
$\sqrt{-g}f^{\prime }(R)g^{\mu \nu }$ is a symmetric twice
contravariant tensor density of weight $1$. As previously
discussed in \cite{FFV,ACCF}, this naturally leads to define a new
metric $h_{\mu \nu}$, such that the following relation holds:
\begin{equation}\label{h_met}
\sqrt{-g}f^{\prime }(R)g^{\mu \nu}=\sqrt{-h}h^{\mu \nu }\,.
\end{equation}
This \textit{ansatz} is suitably made in order to impose $\Gamma$
to be the Levi-Civita connection of $h$ and the only restriction
is that $\sqrt{-g}f^{\prime }(R)g^{\mu \nu}$ should be
non-degenerate. In the case of Hilbert-Einstein Lagrangian, it is
$f^{\prime}(R)=1$ and the statement is trivial.

The above Eq.(\ref{h_met}) imposes that the two metrics $h$ and
$g$ are conformally equivalent. The corresponding conformal factor
can be easily found to be $f^{\prime}(R)$ (in dim ${\cal M}=4)$
and the conformal transformation results to be ruled by:
\begin{equation}\label{h_met1}
h_{\mu \nu }=f^{\prime}(R)g_{\mu \nu }
\end{equation}
Therefore, as it is well known, Eq.(\ref{ffv2}) implies that
$\Gamma =\Gamma _{LC}(h)$ and $R_{(\mu\nu)}(\Gamma)=R_{\mu \nu
}(h)\equiv R_{\mu\nu}$. Field equations can be supplemented by the
scalar-valued equation obtained by taking the trace of
(\ref{ffv1}), (we define $\tau=\mathrm{tr}\hat T$)
\begin{equation} \label{structR}
f^{\prime }(R)R-2f(R)=g^{\alpha\beta}T^m_{\alpha\beta}\equiv
\tau^m
\end{equation}
which controls solutions of (\ref{ffv2}). We shall refer to this
scalar-valued equation as the \textit{structural equation} of the
spacetime. In the vacuum case (and  spacetimes filled with
radiation, such that $\tau^m=0$) this scalar-valued equation
admits  constant solutions, which are different from zero only if
one add a cosmological constant. This means that the universality
of Einstein field equations holds \cite{FFV}, corresponding to a
theory with cosmological constant \cite{cosmconst}.

In the case of interaction with matter fields, the structural
equation (\ref{h_met1}), if explicitly solvable, provides an
expression of $R=F(\tau)$, where $F$ is a generic function, and
consequently both $f(R)$ and $f^\prime (R)$ can be expressed in
terms of $\tau$. The matter content of spacetime thus rules the
bi-metric structure of spacetime and, consequently, both the
geodesic and metric structures which are intrinsically different.
This behavior generalizes the vacuum case and corresponds to the
case of a time-varying cosmological constant. In other words, due
to these features, conformal transformations, which allow to pass
from a metric structure to another one, acquire an intrinsic
physical meaning since ``select" metric and geodesic structures
which, for a given ETG, in principle, {\it do not} coincide.

Let us now try to extend the above formalism to  the case of
non-minimally coupled scalar-tensor theories. The effort is to
understand if and how the bi-metric structure of spacetime behaves
in this cases and which could be its geometric and physical
interpretation.

We start by considering scalar-tensor theories in the Palatini
formalism, calling $A_1$ the action functional. After, we take
into account the case of decoupled non-minimal interaction between
a scalar-tensor theory and a $f(R)$ theory, calling $A_2$ this
action functional. We finally consider the case of
non-minimal-coupled interaction between the scalar field $\phi$
and the gravitational fields $(g, \Gamma)$, calling $A_3$ the
corresponding action functional. Particularly significant is, in
this case, the limit  of low curvature $R$. This resembles the
physical relevant case of present values of curvatures of the
Universe and it is important for cosmological applications.

The  action (\ref{s1}) for scalar-tensor gravity  can be
generalized, in order to better develop the Palatini approach, as:
\begin{equation} \label{lagrfR1}
A_1=\int \sqrt{-g} \; \; [ F(\phi) R+{\epsilon \over 2}
\stackrel{g}{\nabla}_\mu \phi  \stackrel{g}{\nabla}^{ \mu} \phi
-V(\phi)+ {\l}_{m}(\Psi, \stackrel{g}{\nabla} \Psi) ] d^{4}x\,.
\end{equation}
As above, the values of $\epsilon=\pm 1$ selects between standard
scalar field theories and quintessence (phantom) field theories.
The relative ``signature" can be selected by conformal
transformations. Field equations for the gravitational part of the
action are, respectively for the metric $g$ and the connection
$\Gamma$:
\begin{equation}\label{40}
\cases{ F(\phi) [R_{(\mu\nu)}-{1 \over 2} R g_{\mu \nu} ] =
T^\phi_{\mu \nu }+T^{m}_{\mu \nu }  \cr \nabla _{\alpha }^{\Gamma
}(\sqrt{-g} F (\phi)g^{\mu \nu })=0 }
\end{equation}
 $R_{(\mu\nu)}$ is the same defined in (\ref{ffv1}). For matter
fields we have the following field equations:
 \begin{equation}\label{42}
\cases{ \epsilon \Box \phi= - V_{\f} (\phi) + F_{\f} (\phi) R \cr
\frac{\delta {\l}_{m}}{\delta \Psi}=0 }\,.
\end{equation}
In this case, the structural equation of spacetime implies that:
 \begin{equation} \label{stru1a}
R=-\frac{\tau^\phi+\tau^\mathrm{m}}{F(\phi)}
\end{equation}
which expresses the value of the Ricci scalar curvature in terms
of the traces of the stress-energy tensors of standard matter and
scalar field (we have to require $F(\phi) \ne 0$). The bi-metric
structure of spacetime is thus defined by the ansatz:
 \begin{equation}\label{41}
\sqrt{- g} F (\phi) g^{\mu \nu }=\sqrt{- h} h^{\mu \nu }
 \end{equation}
such that $g$ and $h$ result to be conformally related
 \begin{equation} \label{bimetri1}
h_{\mu \nu }=  F (\phi) g_{\mu \nu }\,.
 \end{equation}
The conformal factor is exactly the  interaction factor. From
(\ref{stru1a}), it follows that in the vacuum case $\tau^\phi=0$
and $\tau^{m}=0$: this theory is equivalent to the standard
Einstein one without matter. On the other hand, for $F(\phi)=F_0$
we recover the Einstein theory plus a minimally coupled scalar
field: this means that the Palatini approach intrinsically gives
rise to the conformal structure (\ref{bimetri1}) of the theory
which is trivial in the Einstein, minimally coupled case.

As a further step, let us generalize the previous results
considering the case of a non-minimal coupling in the framework of
$f(R)$ theories. The action functional can be written as:
\begin{equation} \label{lagrfR2}
A_2=\int \sqrt{ -g} \; \; [ F(\phi) f(R)+{\epsilon \over 2}
\stackrel{g}{\nabla}_\mu \phi  \stackrel{g}{\nabla}^{ \mu} \phi
-V(\phi)+ {\l}_{m}(\Psi, \stackrel{g}{\nabla} \Psi) ] d^{4}x
\end{equation}
where $f(R)$ is, as usual, any analytical function of the  Ricci
scalar $R$. Field equations (in the Palatini formalism) for the
gravitational part of the action are:
\begin{equation}
\cases{ F(\phi) [f^\prime (R ) R_{(\mu\nu)}-{1 \over 2} f(R)
g_{\mu \nu} ] = T^\phi_{\mu \nu }+T^{m}_{\mu \nu } \cr \nabla
_{\alpha }^{\Gamma }(\sqrt{ -g} F(\phi) f^{\prime}(R) g^{\mu \nu
})=0 \,.}
\end{equation}
For scalar and matter fields we have, otherwise, the following
field equations:
 \begin{equation}
\cases{ \epsilon \Box \phi= - V_{\f} (\phi) + \sqrt{ -g} F_{\f}
(\phi) f(R) \cr \frac{\delta {\l}_{m}}{\delta \Psi}=0 }
\end{equation}
where the non-minimal interaction term enters into the modified
Klein-Gordon equations. In this case the structural equation of
spacetime implies that:
 \begin{equation} \label{stru1}
f^\prime (R) R-2 f(R)=\frac{\tau^\phi+\tau^{m}}{F(\phi)}\,.
\end{equation}
We remark again that this equation, if solved,  expresses the
value of the Ricci scalar curvature in terms of traces of the
stress-energy tensors of standard matter and scalar field (we have
to require again that $F(\phi) \ne 0$). The bi-metric structure of
spacetime is thus defined by the ansatz:
 \begin{equation}\label{47}
\sqrt{- g} F (\phi) f^\prime (R) g^{\mu \nu }=\sqrt{- h} h^{\mu
\nu }
 \end{equation}
such that $g$ and $h$ result to be conformally related by:
 \begin{equation}
h_{\mu \nu }=  F (\phi) f^\prime (R) g_{\mu \nu }\,.
 \end{equation}
Once the structural equation is solved, the conformal factor
depends  on the values of the matter fields ($\phi, \Psi$) or,
more precisely, on the traces of the stress-energy tensors and the
value of $\phi$. From equation (\ref{stru1}), it follows that in
the vacuum case, \ie  both $\tau^\phi=0$ and $\tau^{m}=0$, the
universality of Einstein field equations still holds as in the
case of minimally interacting $f(R)$ theories \cite{FFV}. The
validity of this property is related to the decoupling of the
scalar field and the gravitational field.

Let us finally consider the case where the gravitational
Lagrangian is a general function of $\phi$ and $R$. The action
functional can thus be written as:
\begin{equation} \label{lagrfR3}
A_3=\int \sqrt{ -g} \; \; [ K(\phi,R)+{\epsilon \over 2}
\stackrel{g}{\nabla}_\mu \phi  \stackrel{g}{\nabla}^{ \mu} \phi
-V(\phi)+  {\l}_{m}(\Psi, \stackrel{g}{\nabla} \Psi) ] d^{4}x
\end{equation}
Field equations for the gravitational part of the action are:
\begin{equation}
\cases{ \left[ \frac{\partial \; K(\phi,R)}{\partial R}  \right]
R_{(\mu\nu)}-{1 \over 2} K(\phi, R)  g_{\mu \nu}  = T^\phi_{\mu
\nu }+T^{m}_{\mu \nu } \cr \nabla _{\alpha }^{\Gamma } \left(
\sqrt{ -g} \left[ \frac{\partial \; K(\phi,R)}{\partial R} \right]
g^{\mu \nu } \right) =0\,. }
\end{equation}
For matter fields, we have:
 \begin{equation}
\cases{ \epsilon \Box \phi= - V_{\f} (\phi) + \left[
\frac{\partial \; K(\phi,R)}{\partial \phi }  \right] \cr
\frac{\delta L_{\mathrm{mat}}}{\delta \Psi}=0\,. }
\end{equation}
The structural equation of spacetime can be expressed as:
 \begin{equation} \label{stru2}
\frac{\partial K(\phi, R)}{\partial R} R-2 K(\phi, R) =
\tau^\phi+\tau^{m}
\end{equation}
This equation, if  solved,  expresses again the form of the Ricci
scalar curvature in terms of traces of the stress-energy tensors
of matter and scalar field (we have to impose regularity
conditions and, for example, $K(\phi,R) \ne 0$). The bi-metric
structure of spacetime is thus defined by the ansatz:
 \begin{equation}
\sqrt{- g} \frac{\partial K(\phi, R)}{\partial R}   g^{\mu \nu
}=\sqrt{- h} h^{\mu \nu }
 \end{equation}
such that $g$ and $h$ result to be conformally related by
 \begin{equation} \label{conf3}
h_{\mu \nu }=  \frac{\partial K(\phi, R)}{\partial R}  g_{\mu \nu
}
 \end{equation}
Again, once the structural equation is solved, the conformal
factor depends just on the values of the matter fields and (the
trace of) their stress energy tensors. In other words, the
evolution, the definition of the conformal factor and the
bi-metric structure is ruled by the values of traces of the
stress-energy tensors and by the value of the scalar field $\phi$.
In this case, the universality of Einstein field equations does
not hold anymore in general. This is evident from (\ref{stru2})
where the strong coupling between $R$ and $\phi$ avoids the
possibility, also in the vacuum case, to achieve  simple constant
solutions.

We consider, furthermore, the case of small values of $R$,
corresponding to  small curvature spacetimes. This limit
represents, as a good approximation, the present epoch of the
observed Universe under suitably regularity conditions. A Taylor
expansion of the analytical function $K(\phi, R)$ can be
performed:
 \begin{equation}
 K(\phi, R)=K_0 (\phi)+K_1 (\phi) R+o(R^2)
 \end{equation}
where only the first leading term in $R$ is considered and we have
defined:
 \begin{equation}
\cases{ K_0 (\phi)=  K(\phi, R)_{R=0}\cr K_1 (\phi)=\left(
\frac{\partial K(\phi, R)}{ \partial R } \right)_{R=0}}\,.
 \end{equation}
Substituting this expression in (\ref{stru2}) and (\ref{conf3}) we
get (neglecting higher order approximations in $R$) the structural
equation and the bi-metric structure in this particular case. From
the structural equation, we get:
  \begin{equation} \label{Rgenapr}
 R=\frac{1}{K_1 (\phi) } [ -(\tau^\phi+\tau^{m})- 2
K_0 (\phi)]
 \end{equation}
such that the value of the Ricci scalar is always determined, in
this first order approximation, in terms of
$\tau^\phi,\tau^\mathrm{m},\phi$. The bi-metric structure is,
otherwise, simply defined by means of the first term of the Taylor
expansion, which is
\begin{equation}
h_{\mu \nu }=  K_1 (\phi) g_{\mu \nu }\,.
\end{equation}
It reproduces, as expected, the scalar-tensor case
(\ref{bimetri1}).  In other words, scalar-tensor theories  can be
recovered in a first order approximation of a general theory where
gravity and non-minimal couplings are any (compare (\ref{Rgenapr})
with (\ref{stru1})). This fact agrees with the above
considerations where Lagrangians of physical interactions can be
considered as stochastic functions with local gauge invariance
properties \cite{ottewill}.

Finally we have to say that there are also bi-metric theories
which cannot be conformally related (see for example the summary
of alternative theories given in \cite{will}) and torsion field
should be taken into account, if one wants to consider the most
general viewpoint \cite{hehl,classtor}. We will not take into
account these general theories in this review.

After this short review of ETGs in metric and Palatini approach,
we are going to face some  remarkable applications to cosmology
and astrophysics. In particular, we deal with the straightforward
generalization of GR, the  $f(R)$ gravity, showing that, in
principle,  no further ingredient, a part a generalized gravity,
could be necessary to address issues as missing matter (dark
matter) and cosmic acceleration (dark energy).  However what we
are going to consider here are nothing else but toy models which
are not able to fit the whole expansion history, the structure
growth law and the CMB anisotropy  and polarization. These issues
require more detailed  theories which, up to now, are not
available but what we are discussing could be a useful working
paradigm as soon as  refined experimental tests to probe such
theories will be proposed and pursued. In particular, we will
outline an independent test, based on the stochastic background of
gravitational waves, which could be extremely useful to
discriminate between ETGs and GR or among the ETGs themselves. In
this latter case, the data delivered from  ground-based
interferometers, like VIRGO and LIGO, or the forthcoming space
interferometer LISA, could be of extreme relevance in such a
discrimination.

Finally, we do not take into account the well known inflationary
models based on ETGs (e.g. \cite{starobinsky,la}) since we want to
show that also the last cosmological epochs, directly related to
the so called {\it Precision Cosmology}, can be framed in such a
new "economic" scheme.

\section{ Applications to Cosmology}

As discussed in the Introduction, many rival theories  have been
advocated to fit the observed accelerated expansion and to solve
the puzzle of  dark energy.

As a simple classification scheme, we may divide the different
cosmological models in three wide classes. According to the models
of the first class, the dark energy is a new ingredient of the
cosmic Hubble flow, the simplest case being the $\Lambda$CDM
scenario and its quintessential generalization (the QCDM models).

This is in sharp contrast with the assumption of UDE models (the
second class) where there is a single fluid described by an
equation of state comprehensive of all regimes of cosmic evolution
\cite{Hobbit} (the {\it parametric density models} or generalized
{\it EoS}\footnote{EoS for Equation of State.} models).

Finally, according to the third class of models, accelerated
expansion is the first evidence of a breakdown of the Einstein GR
(and thus of the Friedmann equations) which has to be considered
as a particular case of a more general theory of gravity. As an
example of this kind of models, we will consider the
$f(R)$\,-\,gravity \cite{curvature,odinoj,curvfit,camfr}.

Far from being exhaustive, considering these three classes of
models allow to explore very different scenarios proposed to
explain the observed cosmic acceleration
\cite{lookback,mimick,IJGMP}. However, from the above
considerations, it is possible to show that one of the simplest
extension of GR, the $f(R)$ gravity can, in principle, address the
dark energy issues both in metric and Palatini approach. In this
section, without claiming for completeness, we sketch some $f(R)$
models matching solutions against some sets of data. The goal is
to show that the dark energy issue could be addressed as a
curvature effect in ETGs.

\subsection{ The $\Lambda$CDM model: the  paradigm}

\begin{figure}
\centering \resizebox{10.0cm}{!}{\includegraphics{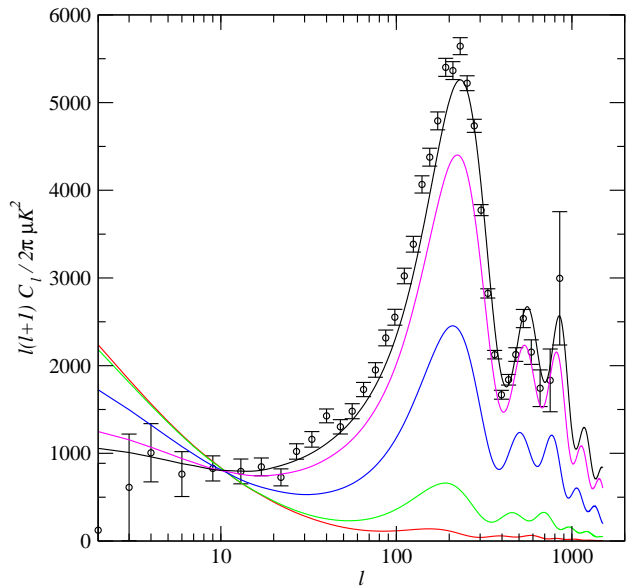}}
\caption{The CMBR anisotropy spectrum for different values of $w$.
Data points are the WMAP measurements and the best fit is obtained
for $w\simeq -1$. If $w\neq -1$ the clustering of dark energy has
been considered in this plot. } \label{fig: zqcl}
\end{figure}

Cosmological constant $\Lambda$ has  become  a textbook candidate
to drive the accelerated expansion of the spatially flat Universe.
Despite its {\it conceptual} problems, the $\Lambda$CDM model
turns out to be the best fit to a combined analysis of completely
different astrophysical data ranging from SNeIa to CMBR anisotropy
spectrum and galaxy clustering \cite{WMAP,SDSS03}. As a simple
generalization, one may consider the QCDM scenario in which the
barotropic factor $w \equiv p/\rho$ takes at a certain epoch a
negative value with $w = -1$ corresponding to the standard
cosmological constant. Testing whether  such a barotropic factor
deviate or not from $-1$ is one of the main issue of modern
observational cosmology. How such a negative pressure fluid drives
the cosmic acceleration may be easily understood by looking at the
Friedmann equations\,:

\begin{equation}
H^2 \equiv \left ( \frac{\dot{a}}{a} \right )^2 = \frac{1}{3}
(\rho_{m} + \rho_{\Lambda}) \ , \label{eq: fried1}
\end{equation}

\begin{equation}
2 \frac{\ddot{a}}{a} + H^2 = -  p_{\Lambda} = -  w \rho_{\Lambda}
\ , \label{eq: fried2}
\end{equation}
where $a(t)$ is the scale factor of the Universe, the dot denotes
differentiation with respect to cosmic time $t$, $H$ is the Hubble
parameter and the Universe is assumed spatially flat as suggested
by the position of the first peak in the CMBR anisotropy spectrum
(see also Fig.\ref{fig: zqcl}). )\cite{Boomerang,Maxima,WMAP}.

From the continuity equation, $\dot{\rho} + 3 H (\rho + p) = 0$,
we get for the $i$\,-\,th fluid with $p_i = w_i \rho_i$\,:

\begin{equation}
\Omega_i = \Omega_{i,0} a^{-3 (1 + w_i)} = \Omega_{i,0} (1 + z)^{3 (1 + w_i)} \ ,
\label{eq: omegavsz}
\end{equation}
where $z \equiv 1/a - 1$ is the redshift, $\Omega_i =
\rho_i/\rho_{crit}$  is the density parameter for the $i$\,-\,th
fluid in terms of the critical density which, defined in standard
units, is $\rho_{crit} = 3H_0^2/8\pi G$ and, hereafter, we label
all the quantities evaluated today with a subscript $0$. It is
important to stress that Eq.(\ref{eq: omegavsz}) works only for
$w_i=$ constant. Inserting this result into Eq.(\ref{eq: fried1}),
one gets\,:

\begin{equation}
H(z) = H_0 \sqrt{\Omega_{M,0} (1 + z)^3 + \Omega_{\Lambda,0} (1 +
z)^{3 (1 + w)}} \ . \label{eq: hvsz}
\end{equation}
The subscript $M$ means all the matter content, inclusive of dark
and baryonic components.  Using Eqs.(\ref{eq: fried1}), (\ref{eq:
fried2}) and the definition of the deceleration parameter $q
\equiv - a \ddot{a}/\dot{a}^2$, one finds\,:

\begin{equation}
q_0 = \frac{1}{2} + \frac{3}{2} w (1 - \Omega_{M,0}) \ .
\label{eq: qlambda}
\end{equation}
The SNeIa Hubble diagram, the large scale galaxy clustering  and
the CMBR anisotropy spectrum can all be fitted by the $\Lambda$CDM
model with $(\Omega_{M,0}, \Omega_{\Lambda}) \simeq (0.3, 0.7)$
thus giving $q_0 \simeq -0.55$, i.e. the Universe turns out to be
in an accelerated expansion phase. The simplicity of the model and
its capability of fitting the most of the data are the reasons why
the $\Lambda$CDM scenario is the leading candidate to explain the
dark energy cosmology. Nonetheless,  its generalization,  QCDM
models, i.e.  mechanisms allowing the evolution of $\Lambda$ from
the past are invoked  to remove the $\Lambda$-problem and the
$coincidence$ problem.

Here, we want to show that  assuming  $f(R)$ gravity,  not
strictly linear in $R$ as  GR, it is possible to give rise to the
evolution of the barotropic factor $w=p/\rho$, today leading to
the value $w=-1$, and to obtain models capable of matching with
the observations. However,  also if the paradigm could result
valid,  it is very difficult to address, in the same comprehensive
$f(R)$ model, different issues as structure formation,
nucleosynthesis, Hubble diagram, radiation and matter dominated
behaviors as we shall discuss below.

Before considering specific $f(R)$ theories, let us discuss
methods to constrain models by samples of data.

\subsection{Methods to constrain models by distance and time indicators}

In principle, cosmological models can be constrained using
suitable distance and/or time indicators. As a general remark,
solutions coming from cosmological models have to be matched with
observations by using the redshift $z$ as the natural time
variable for the Hubble parameter, i.e.

\begin{equation} H(z)=-\frac{\dot{z}}{z+1}\,. \end{equation}
Data can be obtained for various values of redshift $z$: for
example, CMB probes recombination at  $z\simeq 1100$ and $z\simeq
1$ via the late integrated Sachs-Wolfe effect; for $10 < z < 100$
the planned 21cm observations could give  detailed information
 \cite{white}; futuristic LSS surveys and SNe could probe the Universe up to $z\simeq
4$. The method consists in building up a reasonable patchwork of
data coming from different epochs and then matching them with the
same cosmological solution ranging, in principle, from inflation
to present accelerated era.

In order to constrain the parameters characterizing the
cosmological solution, a reasonable approach is to maximize the
following likelihood function\,:

\begin{equation}
{\cal{L}} \propto \exp{\left [ - \frac{\chi^2({\bf p})}{2} \right
]} \label{eq: deflike1}
\end{equation}
where {\bf p} are the parameters characterizing  the cosmological
solution. The $\chi^2$ merit function can be defined as\,:

\begin{equation}
\chi^2({\bf p})  =  \sum_{i = 1}^{N}{\left [ \frac{y^{th}(z_i,
{\bf p}) - y_i^{obs}}{\sigma_i} \right ]^2}  +
\displaystyle{\left [ \frac{{\cal{R}}({\bf p}) - 1.716}{0.062}
\right ]^2} + \displaystyle{\left [ \frac{{\cal{A}}({\bf p}) -
0.469}{0.017} \right ]^2}  \ . \label{eq: defchi1}\
\end{equation}
Terms entering Eq.(\ref{eq: defchi1}) can be characterized as
follows. For example, the dimensionless coordinate distances $y$
to objects at redshifts $z$ are considered in the first term. They
are defined as\,:

\begin{equation}
y(z) = \int_{0}^{z}{\frac{dz'}{E(z')}} \label{eq: defy}
\end{equation}
where $E(z)=H(z)/H_0$ is the normalized Hubble parameter. This is
the main quantity which allows to compare the theoretical results
with data. The function $y$ is related to the luminosity distance
$D_L = (1 + z) y(z)$.

A sample of data at $y(z)$ for the 157 SNeIa is discussed in the
Riess et al. \cite{Riess04} Gold dataset and 20 radio-galaxies are
in \cite{DD04}. These authors fit with good accuracy the linear
Hubble law at low redshift ($z < 0.1$) obtaining the Hubble
dimensionless parameter $h = 0.664 {\pm} 0.008\,$, with $h$ the
Hubble constant in units of $100 \ {\rm km \ s^{-1} \ Mpc^{-1}}$.
Such a number can be consistently taken into account at low
redshift.  This value  is in agreement with $H_0 = 72 {\pm} 8 \
{\rm km \ s^{-1} \ Mpc^{-1}}$ given by the HST Key project
\cite{Freedman} based on the local distance ladder and estimates
coming from  time delays in multiply imaged quasars \cite{H0lens}
and Sunyaev\,-\,Zel'dovich effect in X\,-\,ray emitting clusters
\cite{H0SZ}. The second term in Eq.(\ref{eq: defchi1}) allows to
extend the $z$-range to probe $y(z)$ up to the last scattering
surface $(z\geq 1000)$.  The {\it shift parameter}
\cite{WM04,WT04} $ {\cal R} \equiv \sqrt{\Omega_M} y(z_{ls}) $ can
be determined from the CMBR anisotropy spectrum, where $z_{ls}$ is
the redshift of the last scattering surface which can be
approximated as  $ z_{ls} = 1048 \left ( 1 + 0.00124
\omega_b^{-0.738} \right ) \left ( 1 + g_1 \omega_M^{g_2} \right )
$ with $\omega_i = \Omega_i h^2$ (with $i = b, M$ for baryons and
total matter respectively) and $(g_1, g_2)$ given in \cite{HS96}.
The parameter $\omega_b$ is constrained by the baryogenesis
calculations contrasted to the observed abundances of primordial
elements. Using this method, the value $ \omega_b = 0.0214 {\pm}
0.0020  $  is found \cite{Kirk}.

In any case, it is worth noticing that the exact value of $z_{ls}$
has a negligible impact on the results and setting $z_{ls} = 1100$
does not change constraints and priors on the other  parameters of
the given model. The third term in the function $\chi^2$ takes
into account  the {\it acoustic peak} of the large scale
correlation function at $100 \ h^{-1} \ {\rm Mpc}$ separation,
detected by using  46748 luminous red galaxies (LRG) selected from
the SDSS Main Sample \cite{Eis05,SDSSMain}. The quantity

\begin{equation}
{\cal{A}} = \frac{\sqrt{\Omega_M}}{z_{LRG}} \left [
\frac{z_{LRG}}{E(z_{LRG})} y^2(z_{LRG}) \right ]^{1/3} \label{eq:
defapar}
\end{equation}
is related to the position of acoustic peak where $z_{LRG} = 0.35$
is the effective redshift of the above sample. The parameter
${\cal{A}}$ depends  on the dimensionless coordinate distance (and
thus on the integrated expansion rate),  on $\Omega_M$ and $E(z)$.
This dependence removes some of the degeneracies intrinsic in
distance fitting methods.

Due to this reason, it is particularly interesting to include
${\cal{A}}$ as a further constraint on the model parameters using
its measured value  $ {\cal{A}} = 0.469 {\pm} 0.017  $
\cite{Eis05}. Note that, although similar to the usual $\chi^2$
introduced in statistics, the reduced $\chi^2$ (i.e., the ratio
between the $\chi^2$ and the number of degrees of freedom) is not
forced to be 1 for the best fit model because of the presence of
the priors on ${\cal{R}}$ and ${\cal{A}}$ and since the
uncertainties $\sigma_i$ are not Gaussian distributed, but take
care of both statistical errors and systematic uncertainties. With
the definition (\ref{eq: deflike1}) of the likelihood function,
the best fit model parameters are those that maximize
${\cal{L}}({\bf p})$.

In order to implement the above sketched method, much attention,
has been devoted to standard candles, i.e. astrophysical objects
whose absolute magnitude $M$ is known (or may be exactly
predicted) {\it a priori} so that a measurement of the apparent
magnitude $m$ immediately gives the distance modulus $\mu = m -
M$. Specifically, the distance to the object,  estimated in Mpc,
is\,:

\begin{equation}
\mu(z) = 5 \log{D_L(z)/Mpc} + 25 \label{eq: distmod}
\end{equation}
with $D_L(z)$ the luminosity distance  and $z$ the redshift of the
object. The number 25 depends on the distance modulus calculated
in Mpc. The relation between $\mu$ and $z$ is what is referred to
as Hubble diagram and it is an open window on the cosmography of
the Universe. Furthermore, the Hubble diagram is a powerful
cosmological test since the luminosity distance is determined by
the expansion rate as\,:

\begin{equation}
D_L(z) = \frac{c}{H_0} (1 + z) \int_{0}^{z}{\frac{d z'}{E(z')}}
\label{eq: dl}
\end{equation}
with $E(z)$ defined above. Being the Hubble diagram related to the
luminosity distance and being $D_L$ determined by the expansion
rate $H(z)$, it is clear why it may be used as an efficient tool
to test cosmological models and constrain their parameters.

To this aim, however, it is mandatory that the relation $\mu =
\mu(z)$ is measured up to high enough redshift since, for low $z$,
$D_L$ reduces to a linear function of the redshift (thus
recovering the Hubble law) whatever the background cosmological
model is. This necessity claims for standard candles that are
bright enough to be visible at such high redshift that the Hubble
diagram may discriminate among different rival theories. SNeIa
are, up to now, the objects that best match these requirements.

It is thus not surprising that the first evidences of an
accelerating Universe came from the SNeIa Hubble diagram  and
dedicated survey (like the SNAP satellite \cite{SNAP}) have been
planned in order to increase the number of SNeIa observed and the
redshift range probed.

A reliable compilation of SNeIa is the {\it Gold} dataset released
by Riess et al. \cite{Riess04}. The authors have compiled a
catalog containing 157 SNeIa with $z$ in the range $(0.01, 1.70)$
and visual absorption $A_V < 0.5$. The distance modulus of each
object has been evaluated by using a set of calibrated methods so
that the sample is homogenous in the sense that all the SNeIa have
been re-analyzed using the same technique in such a way that the
resulting Hubble diagram is indeed reliable and accurate. Given a
cosmological model assigned by a set of parameters ${\bf p} =
(p_1, \ldots, p_n)$, the luminosity distance may be evaluated with
Eq.(\ref{eq: dl}) and the predicted Hubble diagram contrasted with
the observed SNeIa one. Constraints on the model parameters may
then be extracted by mean of a $\chi^2$\,-\,based analysis
defining, in this case,  the above $\chi^2$ as\,:

\begin{equation}
\chi_{SNeIa}^2 = \sum_{i = 1}^{N_{SNeIa}}{\left [ \frac{\mu(z_i,
{\bf p}) - \mu_{obs}(z_i)}{\sigma_i} \right ]^2} \label{eq:
chisneia}
\end{equation}
where $\sigma_i$ is the error on the distance modulus at redshift
$z_i$ and the sum is over the $N_{SNeIa}$ SNeIa observed. It is
worth stressing that the uncertainty on  measurements also takes
into account  errors on the redshifts and they are not Gaussian
distributed.

As a consequence, the reduced $\chi^2$ (i.e., $\chi_{SNeIa}^2$
divided by the number of degrees of freedom) for the best fit
model is not forced to be close to unity. Nonetheless, different
models may still be compared on the basis of the $\chi^2$ value\,:
the lower is $\chi_{SNeIa}^2$, the better the model fits the SNeIa
Hubble diagram.

The method outlined  is a simple and quite efficient way to test
whether a given model is a viable candidate to describe the late
time evolution of the Universe. Nonetheless, it is affected by
some degeneracies that could be only partially broken by
increasing the sample size and extending the probed redshift
range. A straightforward example may help in elucidating this
point. Let us consider the flat concordance cosmological model
with matter and cosmological constant. It is\,:

\begin{displaymath}
E^2(z) = \Omega_M (1+z)^3 + (1 - \Omega_M)
\end{displaymath}
so that $\chi_{SNeIa}^2$ will only depend on the Hubble constant
$H_0$ and the matter density parameter $\Omega_M$. Actually, we
could split the matter term in a baryonic and a non-baryonic part
denoting with $\Omega_b$ the baryon density parameter. Since both
baryons and non baryonic dark matter scales as $(1 + z)^3$, $E(z)$
and thus the luminosity distance will depend only on the total
matter density parameter and we could never constrain $\Omega_b$
by fitting the SNeIa Hubble diagram. Similar degeneracies could
also happen with other cosmological models thus stressing the need
for complementary probes to be combined with the SNeIa data. For a
review, see the contribution by Bob Nichols in this volume.

To this aim,  a recently proposed test, based on the gas mass
fraction in galaxy clusters, can be considered. We briefly outline
here the method referring the interested reader to the literature
for further details \cite{fgasbib,fgasapp}. Both theoretical
arguments and numerical simulations predict that the baryonic mass
fraction in the largest relaxed galaxy clusters should be
invariant with the redshift (see, e.g., Ref.\,\cite{ENF98}).

However, this will only appear to be the case when the reference
cosmology in making the baryonic mass fraction measurements
matches the true underlying cosmology. From the observational
point of view, it is worth noticing that the baryonic content in
galaxy clusters is dominated by the hot X\,-\,ray emitting
intra-cluster gas so that what is actually measured is the gas
mass fraction $f_{gas}$ and it is this quantity that should be
invariant with the redshift within the caveat quoted above.
Moreover, it is expected that the baryonic fraction in clusters
equals the universal ratio $\Omega_b/\Omega_M$ so that $f_{gas}$
should indeed be given by $b {\times} \Omega_b/\Omega_M$ where the
multiplicative factor $b$ is motivated by simulations that suggest
that the gas fraction is slightly lower than the universal ratio
because of processes that convert part of the gas into stars or
eject it outside the cluster itself.

Following Ref.\,\cite{fgasdata}, we adopt the SCDM model (i.e., a
flat Universe with $\Omega_M = 1$ and $h = 0.5$, being $h$ the
Hubble constant in units of $100 \ {\rm km \ s^{-1} \ Mpc^{-1}}$)
as reference cosmology in making the measurements so that the
theoretical expectation for the apparent variation of $f_{gas}$
with the redshift is \cite{fgasdata}\,:

\begin{equation}
f_{gas}(z) = \frac{b \Omega_b}{(1 + 0.19 \sqrt{h}) \Omega_M} \left
[ \frac{D_A^{SCDM}(z)}{D_A^{mod}(z)} \right ]^{1.5} \label{eq:
fgas}
\end{equation}
where $D_A^{SCDM}$ and $D_A^{mod}$ is the angular diameter
distance for the SCDM and the model to be tested respectively.
$D_A(z)$ may be evaluated from the luminosity distance $D_L(z)$
as\,:

\begin{equation}
D_A(z) = (1 + z)^{-2} D_L(z) \label{eq: da}
\end{equation}
with $D_L(z)$ given by Eq.(\ref{eq: dl}) above.

In\cite{fgasdata}, it has been extensively analyzed the set of
simulations in Ref.\,\cite{ENF98} to get $b = 0.824 {\pm} 0.089$.
For values in the $1 \sigma$ range quoted above, the main results
are independent on $b$. It is worth noticing that, while the
angular diameter distance depends on $E(z)$ and thus on $h$ and
$\Omega_M$, the prefactor in Eq.(\ref{eq: fgas}) makes $f_{gas}$
explicitly depending on $\Omega_b/\Omega_M$ so that a direct
estimate of $\Omega_b$ is (in principle) possible. Actually,  for
the models which we are going to consider, the quantity that is
constrained by the data is the ratio $\Omega_b/\Omega_M$ rather
than $\Omega_b$ itself.

To simultaneously take into account both the fit to the SNeIa
Hubble diagram and the test on the $f_{gas}$ data, it is
convenient to perform a likelihood analysis defining the following
likelihood function\,:

\begin{equation}
{\cal{L}}({\bf p}) \propto \exp{\left [ - \frac{\chi^2({\bf
p})}{2} \right ]} \label{eq: deflike}
\end{equation}
with\,:

\begin{equation}
\chi^2 = \chi_{SNeIa}^2 + \chi_{gas}^2 + \left ( \frac{h -
0.72}{0.08} \right )^2 + \left ( \frac{\Omega_b/\Omega_M -
0.16}{0.06} \right )^2 \label{eq: defchi}
\end{equation}
where it is possible to define\,:

\begin{equation}
\chi_{gas}^2 = \sum_{i = 1}^{N_{gas}}{\left [ \frac{f_{gas}(z_i,
{\bf p}) - f_{gas}^{obs}(z_i)}{\sigma_{gi}} \right ]^2} \label{eq:
chigas}
\end{equation}
being $f_{gas}^{obs}(z_i)$ the measured gas fraction in a galaxy
clusters at redshift $z_i$ with an error $\sigma_{gi}$ and the sum
is over the $N_{gas}$ clusters considered. In order to avoid
possible systematic errors in the $f_{gas}$ measurement, it is
desirable that the cluster is both highly luminous (so that the
S/N ratio is high) and relaxed so that both merging processes and
cooling flows are absent. A catalog  of 26 large relaxed clusters,
with a  measurement of both the gas mass fraction $f_{gas}$ and
the redshift $z$ is in \cite{fgasdata}. These data can be used to
perform a suitable likelihood analysis.

Note that, in Eq.(\ref{eq: defchi}), we have explicitly introduced
two Gaussian priors to better constrain the model parameters.
First, there is a prior on the Hubble constant $h$ determined by
the results of the HST Key project \cite{HSTKey} from an accurate
calibration of a set of different local distance estimators.
Further, we impose a constraint on the ratio $\Omega_b/\Omega_M$
by considering the estimates of $\Omega_b h^2$ and $\Omega_M h^2$
obtained by Tegmark et al. \cite{Teg03} from a combined fit to the
SNeIa Hubble diagram, the CMBR anisotropy spectrum measured by
WMAP and the matter power spectrum extracted from over 200000
galaxies observed by the SDSS collaboration. It is worth noticing
that, while our prior on $h$ is the same as that used by many
authors when applying the $f_{gas}$ test \cite{fgasapp,fgasdata},
it is common to put a second prior on $\Omega_b$ rather than
$\Omega_b/\Omega_M$. Actually, this choice can be motivated by the
peculiar features of the models which one is going to consider.

With the definition (\ref{eq: deflike}) of the likelihood
function, the best fit model parameters are those that maximize
${\cal{L}}({\bf p})$. However, to constrain a given parameter
$p_i$, one resorts to the marginalized likelihood function defined
as\,:

\begin{equation}
{\cal{L}}_{p_i}(p_i) \propto \int{dp_1 \ldots \int{dp_{i - 1}
\int{dp_{i + 1} \ldots \int{dp_n {\cal{L}}({\bf p})}}}} \label{eq:
defmarglike}
\end{equation}
that is normalized at unity at maximum. The $1 \sigma$ confidence
regions are determined by $\delta \chi^2 = \chi^2 - \chi_0^2 = 1$,
while the condition $\delta \chi^2  = 4$ delimited the $2 \sigma$
confidence regions. It is worth stressing that $\delta\chi^2=1$
for 1-dimensional likelihoods. Here, $\chi_0^2$ is the value of
the $\chi^2$ for the best fit model. Projections of the likelihood
function allow to show eventual correlations among the model
parameters.

Using the method sketched above, the classes of models which we
are going to study can be constrained and selected by
observations. However, most of the tests recently used to
constrain cosmological parameters (such as the SNeIa Hubble
diagram and the angular size\,-\,redshift) are essentially
distance\,-\,based methods. The proposal of Dalal et al.
\cite{DAJM01} to use the lookback time to high redshift objects is
thus particularly interesting since it relies on a completely
different observable. The lookback time is defined as the
difference between the present day age of the Universe and its age
at redshift $z$ and may be computed as\,:

\begin{equation}
t_L(z, {\bf p}) = t_H \int_{0}^{z}{\frac{dz'}{(1 + z') E(z', {\bf p})}}
\label{eq: deftl}
\end{equation}
where $t_H = 1/H_0 = 9.78 h^{-1} \ {\rm Gyr}$ is the Hubble  time,
and, as above, $E(z, {\bf p})$ is the dimensionless Hubble
parameter, where the set of parameters characterizing the
cosmological model, $\{{\bf p}\}$, can be taken into account. It
is worth noticing that, by definition, the lookback time is not
sensible to the present day age of the Universe $t_0$ so that it
could be possible that a model fits well the data on the lookback
time, but nonetheless it  predicts a wrong value for $t_0$. This
latter parameter can be evaluated from Eq.(\ref{eq: deftl}) by
changing the upper integration limit from $z$ to infinity. This
shows that it is a different quantity indeed  since it depends on
the full evolution of the Universe and not only on how the
Universe evolves from the redshift $z$ to now. That is why this
quantity can be explicitly introduced as a further constraint.
However, it is possible to show from the observations that
$t_L(z)$ converges to $t_0$ already at low $z$ and then the method
can be considered reliable.

As an example, let us now sketch how to use the lookback time and
the age of the Universe to test a given cosmological model. To
this end, let us consider an object $i$ at redshift $z$ and denote
by $t_i(z)$ its age defined as the difference between the age of
the Universe when the object was born, i.e. at the formation
redshift $z_F$, and the one at $z$. It is\,:

\begin{eqnarray}
t_i(z) & = & \displaystyle{\int_{z}^{\infty}{\frac{dz'}{(1 + z')
E(z', {\bf p})}} - \int_{z_F}^{\infty}{\frac{dz'}{(1 + z') E(z', {\bf p})}}} \nonumber \\
~ & = & \displaystyle{\int_{z}^{z_F}{\frac{dz'}{(1 + z') E(z',
{\bf p})}}}
 =  t_L(z_F) - t_L(z) \ . \label{eq: titl}
\end{eqnarray}
where, in the last row, we have used the definition (\ref {eq:
deftl})  of the lookback time. Suppose now we have $N$ objects and
we have  been able to estimate the age $t_i$ of the object at
redshift $z_i$ for $i = 1, 2, \ldots, N$. Using the previous
relation, we can estimate the lookback time $t_{L}^{obs}(z_i)$
as\,:

\begin{eqnarray}
t_{L}^{obs}(z_i) & = & t_L(z_F) - t_i(z) \nonumber \\
~ & = & [t_0^{obs} - t_i(z)] - [t_0^{obs} - t_L(z_F)] \nonumber \\
~ & = & t_{0}^{obs} - t_i(z) - \Delta f \ , \label{eq: deftlobs}
\end{eqnarray}
where $t_{0}^{obs}$ is the today estimated age of  the Universe
and a {\it delay factor} can be defined as\,:

\begin{equation}
\Delta f = t_0^{obs} - t_L(z_F) \ .
\end{equation}
The delay factor is introduced to take into account  our ignorance
of the formation redshift $z_F$ of the object. Actually, what can
be measured is the age $t_i(z)$ of the object at redshift $z$. To
estimate $z_F$, one should use Eq.(\ref{eq: titl}) assuming a
background cosmological model. Since our aim is to determine what
is the background cosmological model, it is clear that we cannot
infer $z_F$ from the measured age so that this quantity is
completely undetermined.

It is worth stressing that, in principle, $\Delta f$ should be
different for each object in the sample unless there is a
theoretical reason to assume the same redshift at the formation of
all the objects. If this is indeed the case (as we will assume
later), then it is computationally convenient to consider $\Delta
f$ rather than $z_F$ as the unknown parameter to be determined
from the data. Again a likelihood function can be defined as\,:

\begin{equation}
{\cal{L}}_{lt}({\bf p}, \Delta f) \propto
\exp{[-\chi^{2}_{lt}({\bf p}, \Delta f)/2]} \label{eq: deflikelt}
\end{equation}
with\,:

\begin{equation}
\chi^{2}_{lt} = \displaystyle{\frac{1}{N - N_p + 1}} \left \{
\left [ \frac{t_{0}^{theor}({\bf p}) -
t_{0}^{obs}}{\sigma_{t_{0}^{obs}}} \right ]^2
 + \sum_{i = 1}^{N}{\left [ \frac{t_{L}^{theor}(z_i, {\bf p}) -
t_{L}^{obs}(z_i)}{\sqrt{\sigma_{i}^2 + \sigma_{t}^{2}}} \right
]^2} \right \} \label{eq: defchia}
\end{equation}
where $N_p$ is the number of parameters of the model,  $\sigma_t$
is the uncertainty on $t_{0}^{obs}$, $\sigma_i$ the one on
$t_{L}^{obs}(z_i)$ and the superscript {\it theor} denotes the
predicted values of a given quantity. Note that the delay factor
enters the definition of $\chi^2_{lt}$ since it determines
$t_{L}^{obs}(z_i)$ from $t_i(z)$ in virtue of Eq.(\ref{eq:
deftlobs}), but the theoretical lookback time does not depend on
$\Delta f$.

In principle, such a method should work efficiently to
discriminate among the various dark energy models. Actually, this
is not exactly the case due to the paucity of the available data
which leads to large uncertainties on the estimated parameters. In
order to partially alleviate this problem, it is convenient to add
further constraints on the models by using  Gaussian
priors\footnote{The need for  priors to reduce the parameter
uncertainties is often advocated for  cosmological tests. For
instance, in Ref.\,\cite{LA00} a strong prior on $\Omega_M$ is
introduced to  constrain the dark energy equation of state. It is
likely, that extending the dataset to higher redshifts and
reducing the uncertainties on the age estimate will allow to avoid
resorting to priors.} on the Hubble constant, i.e. redefining the
likelihood function as\,:

\begin{equation}
{\cal{L}}({\bf p}) \propto {\cal{L}}_{lt}({\bf p}) \exp{\left [
-\frac{1}{2} \left ( \frac{h - h^{obs}}{\sigma_h}  \right )^2
\right ]} \propto \exp{[- \chi^2({\bf p})/2]} \label{eq: deflikea}
\end{equation}
where we have absorbed $\Delta f$ in the set of parameters ${\bf
p}$ and have defined\,:
\begin{equation}
\chi^2 = \chi^{2}_{lt} + \left ( \frac{h - h^{obs}}{\sigma_h} \right )^2
\label{eq: newchi}
\end{equation}
with $h^{obs}$ the estimated value of $h$ and  $\sigma_h$ its
uncertainty. The HST Key project results \cite{Freedman} can be
used setting $(h, \sigma_h) = (0.72, 0.08)$. Note that this
estimate is independent of the cosmological model since it has
been obtained from local distance ladder methods. The best fit
model parameters ${\bf p}$ may  be obtained by maximizing
${\cal{L}}({\bf p})$ which is equivalent to minimize the $\chi^2$
defined in Eq.(\ref{eq: newchi}).

It is worth stressing again that such a function should not be
considered as a {\it statistical $\chi^2$} in the sense that it is
not forced to be of order 1 for the best fit model to consider a
fit as successful. Actually, such an interpretation is not
possible since the errors on the measured quantities (both $t_i$
and $t_0$) are not Gaussian distributed and, moreover, there are
uncontrolled systematic uncertainties that may also dominate the
error budget.

Nonetheless, a qualitative comparison among different models may
be obtained by comparing the values of this pseudo $\chi^2$ even
if this should not be considered as a definitive evidence against
a given model. Having more than one parameter, one obtains the
best fit  value of each single parameter $p_i$ as the value which
maximizes the marginalized likelihood for that parameter defined
in Eq.(\ref{eq: defmarglike}). After having normalized  the
marginalized likelihood to 1  at maximum, one computes the $1
\sigma$ and $2 \sigma$ confidence limits (CL) on that parameter by
solving ${\cal{L}}_{p_i} = \exp{(-0.5)}$ and ${\cal{L}}_{p_i} =
\exp{(-2)}$ respectively. In summary, taking into account the
above procedures for distance and time measurements, one can
reasonably constrain a given cosmological model. In any case, the
main and obvious issue is to have at disposal sufficient and good
quality data sets.

\subsection{Samples of data to constrain models: the case of LSS for lookback time}

In order to apply the method outlined above, we need a  set of
distant objects whose distances and ages can be somehow estimated.
As an example for the lookback time method, let us consider the
clusters of galaxies which seem to be ideal candidates since they
can be detected up to high redshift and their redshift, at
formation epoch \footnote{It is worth stressing that, in
literature, the cluster formation redshift is defined as the
redshift at which the last episode of star formation happened. In
this sense, we should modify our definition of $\Delta f$ by
adding a constant term which takes care of how long is the star
formation process and what is the time elapsed from the beginning
of the Universe to the birth of the first cluster of galaxies. For
this reason, it is still possible to consider the delay factor to
be the same for all clusters, but it is not possible to infer
$z_F$ from the fitted value of $\Delta f$ because we do not know
the detail of
 star formation history. This approach is
particular useful since it allows  to overcome the problem to
consider lower limits of the Universe age at $z$ rather than  the
actual values.} is almost the same for all the clusters.
Furthermore, it is relatively easy to estimate their age from
photometric data only. To this end, the color of their component
galaxies,  in particular the reddest ones,  is needed.

Actually, the stellar populations of the reddest galaxies become
redder and redder as they evolve. It is just a matter, then, to
assume a stellar population synthesis model, and to look at how
old the latest episode of star formation should be happened in the
galaxy past to produce colors as red as the observed ones. This is
what is referred to as {\it color age}. The main limitation of the
method relies in the stellar population synthesis model, and on a
few (unknown) ingredients (among which the metallicity and the
star formation rate).

The choice of the evolutionary model is a key step  in the
estimate of the color age and the main source of uncertainty
\cite{Worthey}. An alternative and more robust route to cluster
age is to consider the color scatter (see \cite{Bower} for an
early application of this approach). The argument, qualitatively,
goes in this way\,: if galaxies have an extreme similarity in
their color and nothing is conspiring to make the color scatter
surreptitiously small, then the latest episode of star formation
should happen in the galaxy far past, otherwise the observed color
scatter would be larger.

Quantitatively, the scatter in color should thus be equal to  the
derivative of color with time multiplied the scatter of star
formation times. The first quantity may be predicted using
population synthesis models and turns out to be almost the same
for all the evolutionary models thus significantly reducing the
systematic uncertainty. We will refer to the age estimated by this
method as {\it scatter age}. The dataset we need to apply the
method may  be obtained using the following procedure. First, for
a given redshift $z_i$, we collect the colors of the reddest
galaxies in a cluster at that redshift and then use one of the two
methods outlined above to determine the color or the scatter age
of the cluster. If more than one cluster is available at that
redshift, we average the results from different clusters in order
to reduce systematic errors. Having thus obtained $t_i(z_i)$, we
then use Eq.(\ref{eq: deftlobs}) to estimate the value of the
lookback time at that redshift.

Actually, what we measure is $t_{L}^{obs}(z_i) + d\Delta f$ that
is the quantity that enters the definition (\ref{eq: defchia}) of
$\chi^2_{lt}$ and then the likelihood function. To estimate the
color age, following \cite{Andreon4}, it is possible to choose,
among the various available stellar population synthesis models,
the Kodama and Arimoto one \cite{Kodama1}, which, unlike other
models, allows a chemical evolution neglected elsewhere. This
gives us three points on the diagram $z$ vs. $t_{L}^{obs}$
obtained by applying the method to a set of six clusters at three
different redshifts as detailed in Table\,1.

Using a large sample of low redshift SDSS clusters, it is possible
to evaluate the scatter age for clusters age at $z = 0.10$ and $z
= 0.25$ \cite{Andreon1}.  Blakeslee et al. \cite{Blakeslee}
applied the same method to a single, high redshift $(z = 1.27)$
cluster. Collecting the data using both  the color age and the
scatter age, we end up with a sample of $\sim 160$ clusters at six
redshifts (listed in Table\,1) which probe the redshift range
$(0.10, 1.27)$. This nicely overlaps the one probed by SNeIa
Hubble diagram so that a comparison among our results and those
from SNeIa is possible. We assume a $\sigma = 1 \ {\rm Gyr}$ as
uncertainty on the cluster age, no matter what is the method used
to get that estimate.

Note that this is a very conservative choice. Actually, if the
error on the age were so large, the color\,-\,magnitude relation
for reddest cluster galaxies should have a large scatter that is
not observed. We have, however, chosen such a large error to take
qualitatively into account the systematic uncertainties related to
the choice of the evolutionary model.

\begin{table}
\begin{center}
\begin{tabular}{|c|c|c|c|c|c|c|c|}
\hline
 \multicolumn{4}{|c|}{Color age} & \multicolumn{4}{|c|}{Scatter age} \\
\hline $z$  & N & Age (Gyr) & Ref & $z$  & N & Age (Gyr) & Ref \\
\hline\hline
0.60 & 1 & 4.53 & \cite{Andreon4} & 0.10 & 55 & 10.65 & \cite{Andreon1} \\
0.70 & 3 & 3.93 & \cite{Andreon4} & 0.25 & 103 & 8.89 & \cite{Andreon1} \\
0.80 & 2 & 3.41 & \cite{Andreon4} & 1.27 &   1 & 1.60 & \cite{Blakeslee} \\
\hline
\end{tabular}
\end{center}
\caption{Main properties of the cluster sample used  for  the
analysis. The data in the left part of the Table refers to
clusters whose age has been estimated from the color of the
reddest galaxies (color age), while that of clusters in the right
part has been obtained by the color scatter (scatter age). For
each data point, we give the redshift $z$, the number $N$ of
clusters used, the age estimate and the relevant reference.}
\end{table}

Finally, we need an estimate of $t_{0}^{obs}$ to apply the method.
Following Rebolo et al. \cite{VSA}, one can choose $(t_{0}^{obs},
\sigma_t) = (14.4, 1.4) \ {\rm Gyr}$ as obtained by a combined
analysis of the WMAP and VSA data on the  CMBR anisotropy spectrum
and SDSS galaxy clustering. Actually, this estimate is model
dependent since Rebolo et al. \cite{VSA} implicitly assumes that
the $\Lambda$CDM model is the correct one. However, this value is
in perfect agreement with $t_{0}^{obs} = 12.6^{+3.4}_{-2.4} \ {\rm
Gyr}$ determined from globular clusters age \cite{krauss} and
$t_{0}^{obs} > 12.5 \pm 3.5 \ {\rm Gyr}$ from radioisotopes
studies \cite{Cayrel}. For this reason, one is confident that no
systematic error is induced on the adopted  method using the
Rebolo et al. estimate for $t_{0}^{obs}$ even when testing
cosmological models other than the $\Lambda$CDM one.

\subsection{Dark Energy as a curvature effect}

The methods outlined above allow to constrain dark energy models
without considering the nature of dark constituents. In
\cite{mimick}, it is shown that the most popular quintessence
(dark energy) models can be reproduced, in principle, only
considering "curvature effects" i.e. only generalizing the theory
of gravity to some  $f(R)$ which is not supposed to be simply
linear in $R$. From our point of view, this approach seems
"economic" and "conservative" and does not claim for unknown
fundamental ingredients,  up to now not detected, in the cosmic
fluid\footnote{Following the Occam razor prescription:
\textit{"Entia non sunt multiplicanda praeter necessitatem."}}. As
it is clear, from Eq.(\ref{h4}), the curvature stress\,-\,energy
tensor formally plays the role of a further source term in the
field equations so that its effect is the same as that of an
effective fluid of purely geometric origin. Let us rewrite it here
for convenience:
\begin{equation} \label{c6}
T^{curv}_{\alpha\beta}\,=\,\frac{1}{f'(R)}\left\{\frac{1}{2}g_{\alpha\beta}\left[f(R)-Rf'(R)\right]
+f'(R)^{;\mu\nu}(g_{\alpha\mu}g_{\beta\nu}-g_{\alpha\beta}g_{\mu\nu})
\right\}\,.
\end{equation}
Our aim is to show that such a quantity provides all the
ingredients we need to tackle with the dark side of the Universe.
In fact, depending on the scales, such a curvature fluid can play,
in principle,  the role of dark matter and dark energy. To be more
precise, also the coupling $1/f'(R)$ in front of the matter stress
energy tensor, see Eqs.(\ref{h4}), plays  a fundamental role in
the dynamics since it  affects, in principle, all the physical
processes (e.g. the nucleo- synthesis) and the observable
(luminous, clustered, baryonic) quantities. This means that the
whole problem of the dark side of the Universe could be addressed
considering a comprehensive theory where the interplay between the
geometry and the matter has to be reconsidered assuming non-linear
contributions and non-minimal couplings in curvature invariants.

From the cosmological point of view, in the standard framework of
a spatially flat homogenous and isotropic Universe, the
cosmological dynamics is determined by its energy budget through
the Friedmann equations. In particular, the cosmic acceleration is
achieved when the r.h.s. of the acceleration equation remains
positive. Specifically the Friedmann equation, in physical units,
is

\begin{equation}
\frac{\ddot{a}}{a} = - \frac{1}{6} \left ( \rho_{tot} + 3 p_{tot}
\right ) \ . \label{eq: fried21}
\end{equation}
The subscript $tot$ denotes the sum of the curvature fluid and the
matter contribution to the energy density and pressure. From the
above relation, the acceleration condition, for a dust dominated
model, leads to\,:
\begin{equation}
\rho_{curv} + \rho_m + 3p_{curv} < 0 \rightarrow w_{curv} < -
\frac{\rho_{tot}}{3 \rho_{curv}} \label{eq: condition}
\end{equation}
so that a key role is played by the effective quantities\,:
\begin{equation}
\rho_{curv} = \frac{1}{f'(R)} \left \{ \frac{1}{2} \left [ f(R)  -
R f'(R) \right ] - 3 H \dot{R} f''(R) \right \} \ , \label{eq:
rhocurv}
\end{equation}
and
\begin{equation}
w_{curv} = -1 + \frac{\ddot{R} f''(R) + \dot{R} \left [ \dot{R}
f'''(R) - H f''(R) \right ]} {\left [ f(R) - R f'(R) \right ]/2 -
3 H \dot{R} f''(R)} \,, \label{eq: wcurv}
\end{equation}
deduced from Eq.(\ref{c6}). As a first simple choice, one may
neglect ordinary matter and assume a power\,-\,law form $f(R) =
f_0 R^n$, with $n$ a real number, which represents a
straightforward generalization of  Einstein GR in the limit $n=1$.
One can find power\,-\,law solutions for $a(t)$ providing a
satisfactory fit to the SNeIa data and a good agreement with the
estimated age of the Universe in the range $1.366 < n < 1.376$
\cite{lookback,curvfit}.
\begin{figure}
\centering \resizebox{8.5cm}{!}{\includegraphics{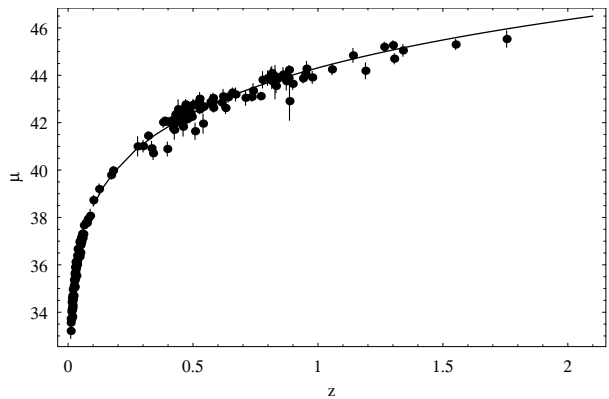}}
\caption{Best fit curve to the SNeIa Hubble diagram for the power
law Lagrangian model. Only data of ``Gold" sample of SNeIa have
been used.} \label{fig: snefitpl}
\end{figure}
On the other side, one can develop the same analysis in presence
of the ordinary matter component, although in such a case, one has
to solve numerically the field equations. Then, it is still
possible to confront the Hubble flow described by such a model
with the Hubble diagram of SNeIa using the above mentioned
methods.
\begin{figure}
\centering\resizebox{7.5cm}{!}{\includegraphics{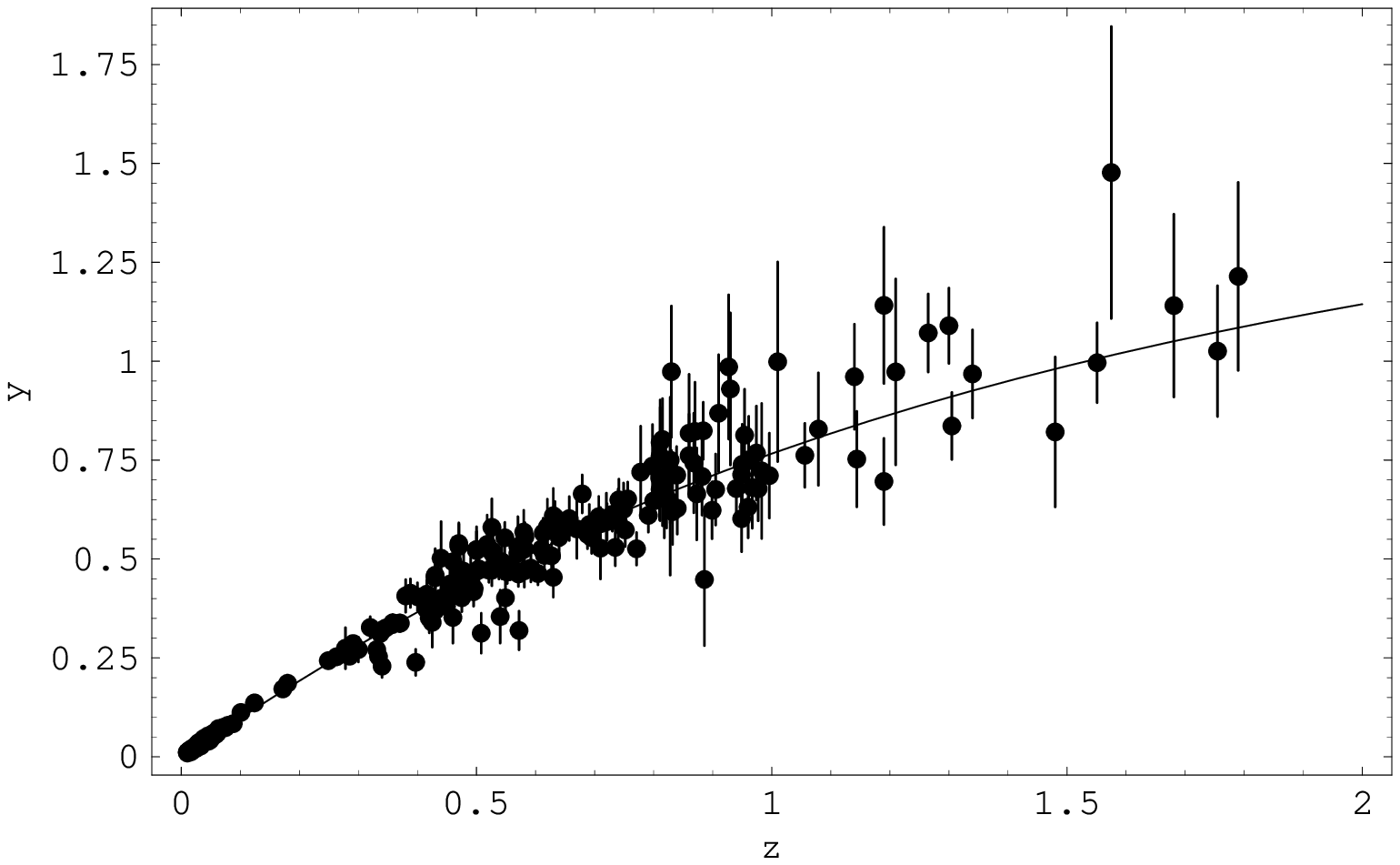}}
\centering\resizebox{8cm}{!}{\includegraphics{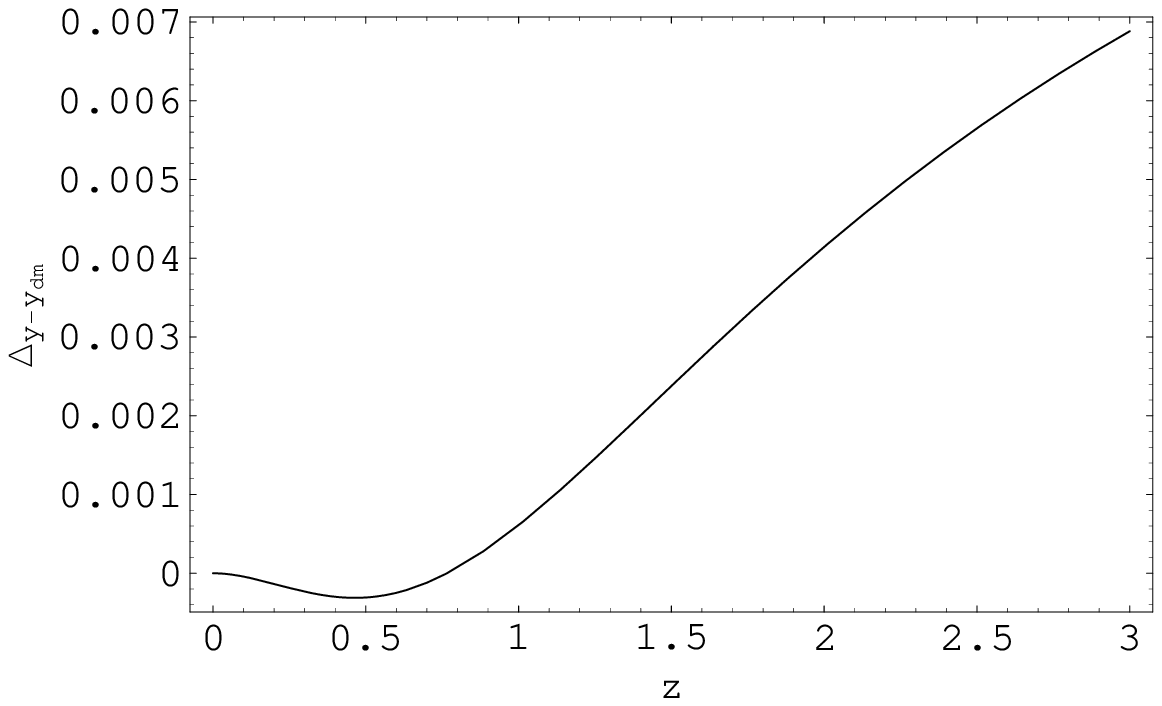}}
\caption{The Hubble diagram of 20 radio galaxies together with the
``Gold" sample of SNeIa, in term of the redshift as suggested in
\cite{daly}. The best fit curve refers to the $R^n$\,-\,gravity
model without dark matter (left), while in the {\it right} panel
it is shown the difference between the luminosity distances
calculated without dark matter and in presence of this component
in term of redshift. It is evident that the two behaviors are
quite indistinguishable. \label{fig: SNeIa}}
\end{figure}
The data fit turns out to be significant (see Fig. \ref{fig:
SNeIa}) improving the $\chi^2$ value and it fixes the best fit
value at $n\,=\,3.46$ when it is accounted only the baryon
contribute $\Omega_b \approx 0.04$ (according with BBN
prescriptions). It has to be remarked that considering dark matter
does not modify the result of the fit, as it is evident from Fig.
\ref{fig: SNeIa}, in some sense positively supporting the
assumption of no need for dark matter in this model. A part the
simplicity of the  power law model, the theoretical implications
of the best fit values found for $n$ are telling us that dynamics
related to cosmological constant (whose theoretical shortcomings
are well known) could be seriously addressed by finding a reliable
$f(R)$ gravity model (see also \cite{starob2}).

From the evolution of the Hubble parameter in term of redshift,
one can even calculate the age of Universe. In Fig. \ref{fig:
age}, it is sketched the age of the Universe as a function of the
correlation between the deceleration parameter $q_0$ and the model
parameter $n$. The best fit value $n\,=\,3.46$ provides
$t_{univ}\approx 12.41$ Gyr.
\begin{figure}
\centering\resizebox{7.5cm}{!}{\includegraphics{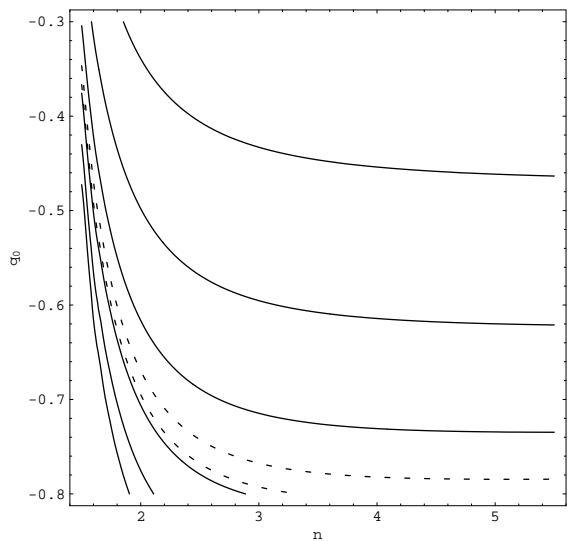}}
\caption{Contour plot in the plane ($q_0\,,\ n$) describing the
Universe age as induced by $R^n$\,-\,gravity model without dark
matter. The contours refer to age ranging from 11 Gyr to 16 Gyr
from up to down. The dashed curves define the $1-\sigma$ region
relative to the best fit Universe age suggested by the last WMAP
release ($13.73_{-0.17}^{+0.13}$ Gyr) in the case of $\Lambda$-CDM
model \cite{wmap2}. At the best fit $n\simeq 3.5$ for SNeIa, the
measured $q_0\simeq -0.5$ gives a rather short age  (about $11.5$
Gyr) with respect to the WMAP constraint. This is an indication
that the $f(R)$ model has to be further improved. \label{fig:
age}}
\end{figure}

It is worth noticing that considering $f(R)\,=\,f_0\,R^n$ gravity
is only the simplest generalization of the Einstein theory. In
other words, it has to be considered that $R^n$\,-\,gravity
represents just a working hypothesis as there is no overconfidence
that such a model is the correct final gravity theory. In a sense,
we want only to suggest that several cosmological and
astrophysical results can be well interpreted in the realm of a
power law extended gravity model.

As matter of fact, this approach gives no rigidity about the value
of the power $n$, although it would be preferable to determine a
model capable of working at different scales.  Furthermore, we do
not expect to be able to reproduce the whole cosmological
phenomenology by means of a simple power law model, which has been
demonstrated to be not sufficiently versatile
\cite{cnot,tsu2,tsu3}.

For example, we can easily demonstrate that this model fails when
it is analyzed with respect to its capability of providing the
correct evolutionary conditions for the perturbation spectra of
matter overdensity. This point is typically addressed as one of
the most important issues which suggest the need for dark matter.
In fact, if one wants to discard this component, it is crucial to
match the experimental results related to the Large Scale
Structure of the Universe and the CMBR which show, respectively at
late time and at early time, the signature of the initial matter
spectrum.

As important remark, we notice that the quantum spectrum of
primordial perturbations, which provides the seeds of matter
perturbations, can be positively recovered in the framework of
$R^n$\,-\,gravity. In fact, $f(R)\,\propto \,R^2$ can represent a
viable model with respect to CMBR data and it is a good candidate
for cosmological Inflation (see \cite{hwang} and references
therein).

In order to develop the matter power spectrum suggested by this
model, we resort to the equation for the matter contrast obtained
in \cite{pengjie} in the case of fourth order gravity (see even
\cite{mukhanov} for a review on cosmological perturbations in
$f(R)$\,-\,theories). This equation can be deduced considering the
conformal Newtonian gauge for the perturbed metric
\cite{pengjie}\,:
\begin{equation}\label{metric-pert}
ds^2\,=\,(1+2\psi)dt^2\,-\,a^2(1+2\phi)\Sigma_{i\,=1}^3(dx^i)\,.
\end{equation}
where $\psi$ and $\phi$ are now gravitational perturbation
potentials. In GR, it is $\phi\,=\,-\psi$, since there is no
anisotropic stress; in ETGs, this relation breaks, in general, and
the $i\,\neq\,j$ components of field equations give new relations
between $\phi$ and $\psi$.

In particular, for $f(R)$ gravity, due to the non-vanishing
derivatives $f_{R;i;j}$ (with $i\,\neq\,j$), the $\phi\,-\,\psi$
relation becomes scale dependent. Instead of the perturbation
equation for the matter contrast $\delta$, we provide here its
evolution in term of the growth index ${\cal F}
\,=\,d\ln{\delta}/d\ln{a}$, which is the directly measured
quantity at $z\sim 0.15$\,:
\begin{equation} \label{growind}
{\cal F}'(a)-\frac{{\cal F}(a)^2}{a}+
\left[\frac{2}{a}+\frac{1}{a}E'(a)\right]{\cal
F}(a)-\frac{1-2Q}{2-3Q}\cdot\frac{3\Omega_m\,a^{-4}}{n\,E(a)^2\tilde{R}^{n-1}}\,=\,0\,,
\end{equation}
(the prime, in this case, means the derivative with respect to
$a$, $n$ is the model parameter, being $f(R)\propto R^n$),
$E(a)\,=\,H(a)/H_0$, $\tilde{R}$ is the dimensionless Ricci
scalar, and
\begin{equation}\label{Q}
Q\,=\,-\frac{2f_{RR}\,k^2}{f_R\,a^2}\,.
\end{equation}
For $n\,=\,1$ the previous expression gives the ordinary growth
index relation for the  Cosmological Standard  Model. It is clear,
from Eq.(\ref{growind}), that such a model suggests a scale
dependence of the growth index which is contained into the
corrective term $Q$ so that, when $Q\rightarrow0$, this dependence
can be reasonably neglected.

\begin{figure}
\centering{!}{\includegraphics{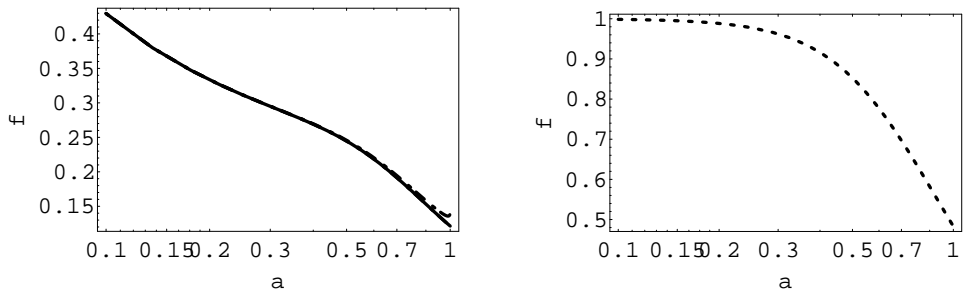}} \caption{Scale factor
evolution of the growth index \,: ({\it left}) modified gravity,
in the case $\Omega_m\,=\,\Omega_{bar}\,\sim 0.04$, for the SNeIa
best fit model with $n\,=\,3.46$, ({\it right}) the same evolution
in the case of a $\Lambda$CDM model. In the case of
$R^n$\,-\,gravity it is shown also the dependence on the scale
$k$. The three cases $k\,=\,0.01,\ 0.001,\ 0.0002$  have been
checked. Only the latter case shows a very small deviation from
the leading behavior. Clearly, the trend is that the growth law
saturates to ${\cal F}=1$ for higher redshifts (i.e. $a\sim 0.001$
to $0.01$). This behavior agrees with observations since we know
that comparing CMB anisotropies and LSS, we need roughly $\delta
\propto a$ between recombination and $z\sim 5$ to generate the
present LSS from the small fluctuations at recombination seen in
the CMB. \label{fig: grwf}}
\end{figure}

\begin{figure}
\centering\resizebox{13.0cm}{!}{\includegraphics{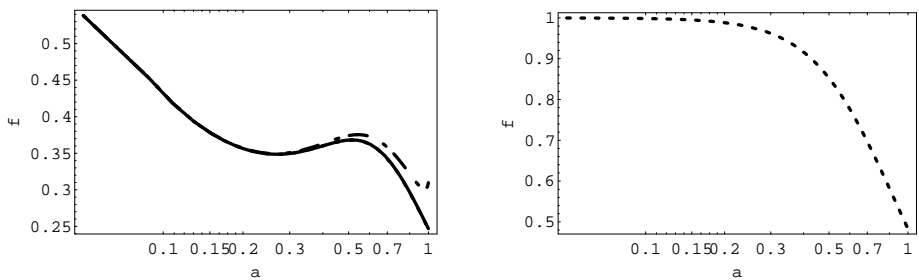}}
\caption{The evolution of the growth index in terms of the scale
factor when dark matter is included in the whole energy budget.
Again, the {\it left} plot shows the modified gravity evolution
for the SNeIa best fit model with $n\,=\,3.46$, while the {\it
right} one refers to $\Lambda$CDM model. \label{fig: grwfDM}}
\end{figure}

In the most general case, one can resort to the limit
$aH\,<\,k\,<\,10^{-2}h\,Mpc^{-1}$, where Eq.(\ref{growind}) is a
good approximation, and non-linear effects on the matter power
spectrum can be neglected. Studying numerically
Eq.(\ref{growind}), one obtains the growth index evolution in term
of the scale factor; for the sake of simplicity, we assume the
initial condition ${\cal F}(a_{ls})\,=\,1$ at the last scattering
surface as in the case of matter-like domination. The results are
summarized in Fig.(\ref{fig: grwf})\,-\,(\ref{fig: grwfDM}), where
we have displayed, in parallel, the growth index evolution in
$R^n$\,-\,gravity and in the $\Lambda$CDM model.
In the case of $\Omega_m\,=\,\Omega_{bar}\,\sim 0.04$, one can
observe a strong disagreement between the expected rate of the
growth index and the behavior induced by power law fourth order
gravity models.

This negative result is evidenced by the predicted value of ${\cal
F}(a_{z\,=\,0.15})$, which has been observationally estimated by
the analysis of the correlation function for 220000 galaxies in
2dFGRS dataset sample at the survey effective depth $z\,=\,0.15$.
The observational result suggests ${\cal F}\,=\,0.58\pm0.11$
\cite{lahav}, while our model gives ${\cal
F}(a_{z\,=\,0.15})\,\sim\,0.117\ (k\,=\,0.01),\ 0.117\
(k\,=\,0.001),\ 0.122\ (k\,=\,0.0002)$.

Although this result seems frustrating with respect to the
underlying idea to discard the dark matter component from the
cosmological dynamics, it does not give substantial improvement in
the case of $R^n$\,-\,gravity model plus dark matter. In fact, as
it is possible to observe from Fig.(\ref{fig: grwfDM}), even in
this case the growth index prediction is far to be in agreement
with the $\Lambda$CDM model and again, at the observational scale
$z\,=\,0.15$, there is not enough growth of perturbations to match
the observed Large Scale Structure. In such a case one obtains\,:
${\cal F}(a_{z\,=\,0.15})\,\sim\,0.29\ (k\,=\,0.01),\ 0.29\
(k\,=\,0.001),\ 0.31\ (k\,=\,0.0002)$, which are quite increased
with respect to the previous case but still very far from the
experimental estimate.

It is worth noticing that no significant different results are
obtained if one varies the power $n$. Of course in the case of
$n\rightarrow 1$, one recovers the standard behavior if a
cosmological constant contribution is added.
These results seem to suggest that an ETG model which considers a
simple power law of Ricci scalar, although cosmologically relevant
at late times, is not viable to describe the evolution of Universe
at all scales.

In other words such a scheme seems too simple to give account of
the whole cosmological phenomenology. In fact, in \cite{pengjie} a
gravity Lagrangian considering an exponential correction to the
Ricci scalar, $f(R)\,=\,R\,+\,A\exp(-B\,R)$ (with $A,\ B$ two
constants), gives a grow factor rate which is in agreement with
the observational results at least in the dark matter case. To
corroborate this point of view, one has to consider that when the
choice of $f(R)$ is performed starting from observational data
(pursuing an inverse approach) as in \cite{mimick}, the
reconstructed Lagrangian is a non\,-\,trivial polynomial in term
of the Ricci scalar, as we shall see below.

A result which directly suggests that the whole cosmological
phenomenology can be accounted only by a suitable non\,-\,trivial
function of the Ricci scalar rather than a simple power law
function. In this case, cosmological equations, coming from an
$f(R)$ action, can be reduced to a linear third order differential
equation for the function $f(R(z))$, where $z$ is the redshift.
The Hubble parameter $H(z)$ inferred from the data and the
relation between $z$ and $R$ can be used to finally work out
$f(R)$.

This scheme provides even another interesting result. Indeed, one
may consider the expression for $H(z)$ in a given dark energy
model as the input for the reconstruction of $f(R)$ and thus work
out a $f(R)$ theory giving rise to the same dynamics as the input
model.

This suggests the intriguing possibility to consider
observationally viable dark energy models (such as $\Lambda$CDM
and quintessence) only as effective parameterizations of the
curvature fluid \cite{mimick,cnot}. As matter of fact, the results
obtained with respect to the study of the matter power spectra in
the case of $R^n$\,-\,gravity do not invalidate the whole
approach, since they can be referred to the too simple form of the
model. Similar considerations can be developed for cosmological
solutions derived in Palatini approach (see \cite{CCF} for
details).

An important remark is in order at this point. If the power $n$ is
not a natural number, $R^n$ models could be not analytic for
$R\rightarrow 0$. In this case, the Minkowski space is not a
solution and, in general, the post-Minkowskian limit of the theory
could be bad defined. Actually this is not a true shortcoming if
we consider $R^n$-gravity as a toy model for  a (still unknown)
self-consistent and comprehensive theory  working at all scales.

\begin{figure}
\centering \resizebox{8.5cm}{!}{\includegraphics{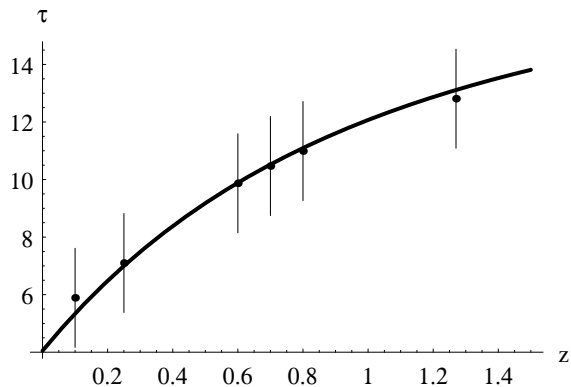}}
\hfill \caption{Comparison between predicted and observed values
of $\tau = t_L(z) + \Delta f$ for the best fit $\Lambda$CDM model.
Data in Table I have been used. } \label{fig: taulcdm}
\end{figure}

\begin{figure}
\centering \resizebox{8.5cm}{!}{\includegraphics{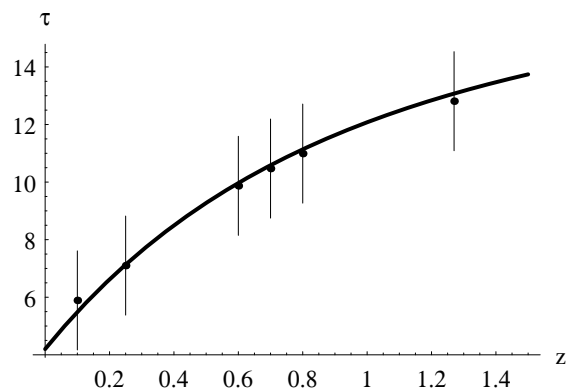}}
\hfill \caption{Comparison between predicted and observed  values
of $\tau = t_L(z) + \Delta f$ for the best fit  $f(R)$  power-law
model as in Fig.\ref{fig: SNeIa}. Data in Table I have been used.
Also for this test, it is evident the strict concordance with
$\Lambda$CDM model in Fig.\ref{fig: taulcdm}. } \label{fig:
tauhobbit}
\end{figure}

However, the discussion is not definitely closed since some
authors  support the point of view that no $f(R)$ theories with
$f=R+\alpha R^n$, $n\neq 1$ can evolve from a matter-dominated
epoch $a(t)\propto t^{2/3}$ to an accelerated phase \cite{tsu3}.
This result could be the end of such theories, if the phase space
analysis of cosmological solutions is not correctly faced.

In \cite{noi-phase}, and recently in \cite{ctd}, it is shown that
transient matter-dominated evolutions evolving toward accelerated
phases are actually possible and the lack of such solutions in
\cite{tsu3} depends on an incomplete parameterization of the phase
space.

In general,  by performing a conformal transformation on a generic
$f(R)$ gravity theory, it is possible to achieve, in the Einstein
frame, dust matter behaviors which are compatible with
observational prescriptions. In addition, by exploiting the
analogy between the two frames and between modified gravity and
scalar-tensor gravity, one can realize that physical results, in
the two conformally related frames, could be completely different.
In other words one can pass from a non\,-\,phantom phase behavior
(Einstein frame) to a phantom regime (Jordan frame) \cite{cno}.

Now, we can suppose to change completely the point of view. In
fact, we can rely directly with the Jordan frame and we can verify
if a dust matter regime is intrinsically compatible with modified
gravity.

As a first  example, one can cite the exact solution provided in
\cite{curvfit}, which has been deduced working only in the Jordan
Frame (FRW Universe). In particular, one is able to find a power
law regime for the scale factor whose rate is connected with the
power $n$ of the Lagrangian $f(R)\,=\,f_0 R^n$.

In other words, one has $a(t)\,=\,a_0 t^{\alpha}$ with
${\displaystyle \alpha\,=\,\frac{2n^2-3n+1}{2-n}}$. Such an exact
solution is found out when only baryonic matter is considered
\cite{jcap,mnras}. It is evident that such a solution allows to
obtain an ordinary matter behavior ($\alpha\,=\,2/3$) for given
values of the parameter $n$ (i.e. $n \sim -0.13,\ n \sim 1.29$).

Such solutions are nevertheless stable and no transition to
acceleration phase then occurs. In general, it is possible to show
that solutions of the type
\begin{equation}
\label{friedman-transient} a\,=\,a_0(t-t_0)^{\frac{2n}{3(1+w)}}\,,
\end{equation}
where  $w$ is the barotropic index of standard perfect fluid,
 arises as a transient phase, and this phase evolves into an
accelerated solution representing an attractor for the system
\cite{noi-phase}. In any case,  a single solution exactly
matching, in sequence, radiation, matter and accelerated phases is
unrealistic to be found out in  the framework of  simple
$f(R)$-power law theories. The discussion can be further extended
as follows.

Modified gravity can span a wide range of analytic functions of
the Ricci scalar where $f(R)\,=\,f_0R^n$ only represents  the
simplest choice. In general, one can reverse the perspective and
try to derive the form of  gravity Lagrangian directly from the
data or mimicking other cosmological models.

Such an approach has been developed in \cite{mimick}, and allows
to recover modified gravity Lagrangians by the Hubble flow
dynamics $H(z)$: in particular, it is possible to show that wide
classes of  dark energy models worked out in the Einstein frame
can be consistently reproduced by $f(R)$-gravity as quintessence
models with exponential potential \cite{mimick-scafi}.

Clearly the approach works also for  the case of coupled
quintessence scalar fields. In other words, the dynamics of
$H(z)$, considered in the Jordan frame, is reconstructed by
observational data considered in the Einstein frame then assuming
one of the two frames as the "physical frame" could be misleading.
Here we further develop this approach with the aim to show, in
general, the viability of $f(R)$ gravity to recover a
matter-dominated phase capable of evolving in a late accelerating
phase.

 From a formal point of view, the reconstruction of the gravity
Lagrangian from data is based on the relation which expresses the
Ricci scalar in terms of the Hubble parameter\,:
\begin{equation}
R = -6 \left ( \dot{H} + 2 H^2 +\frac{k}{a^2}\right )\,.
\label{eq: constr}
\end{equation}
Now, starting from the above the $f(R)$ field equations (\ref{h4})
one can reconstruct the form of $f(R)$ from the Hubble parameter
as a function of the redshift $z$ exploiting the relation
(\ref{eq: constr}) after this expression has been rewritten in
term of the redshift itself.

A key role in this discussion is played by the conservation
equation for the curvature and the matter fluids which, in the
case of dust matter, (i.e. $p_m = 0$) gives\,:
\begin{eqnarray}
\dot{\rho}_{curv} + 3 H (1 + w_{curv}) \rho_{curv} & = & - \frac{1}{f'(R)} (\dot{\rho}_m + 3 H \rho_m) \nonumber \\
 ~ & ~ & - \rho_m \frac{df'(R)}{dt} \ . \label{eq: cons}
\end{eqnarray}
In particular, one may assume that the matter energy density is
conserved\,:
\begin{equation}
\rho_m = \Omega_M \rho_{crit} a^{-3} = 3 H_0^2 \Omega_M (1 + z)^3
\label{eq: mattrho}
\end{equation}
with $z = 1/a - 1$ the redshift   (having set $a(t_0) = 1$),
$\Omega_M$ the matter density parameter (also here,  quantities
labelled with the subscript $0$ refers to present day ($z = 0$)
values). Eq.(\ref{eq: mattrho}) inserted into Eq.(\ref{eq: cons}),
allows to write a conservation equation for the effective
curvature fluid\,:
\begin{eqnarray}
\dot{\rho}_{curv} + 3 H (1 + w_{curv}) \rho_{curv} & = & 3 H_0^2 \Omega_M (1 + z)^3  \nonumber \\
 ~ & ~ & \times \ \frac{\dot{R} f''(R)}{\left [ f'(R) \right ]^2} \ . \label{eq: curvcons}
\end{eqnarray}
Actually, since the continuity   equation and the field equations
are not independent  \cite{mimick}, one can reduce  to the
following single equation
\begin{eqnarray}
\dot{H} & = & -\frac{1}{2 f'(R)} \left \{ 3 H_0^2 \Omega_M (1 + z)^3 + \ddot{R} f''(R)+ \right . \nonumber \\
 ~ & ~ & \left . + \dot{R} \left [ \dot{R} f'''(R) - H f''(R) \right ] \right \} \,, \label{eq: presingleeq}
\end{eqnarray}
where all quantities can be expressed in term of  redshift by
means of the relation ${\displaystyle \frac{d}{dt} = - (1 + z) H
\frac{d}{dz}}$. In particular, for a flat FRW metric, one has\,:
\begin{equation}
R = -6 \left [ 2 H^2 - (1 + z) H \frac{dH}{dz} \right ]
\,,\label{eq: rvsh}
\end{equation}
\begin{equation}
f'(R) = \left ( \frac{dR}{dz} \right )^{-1} \frac{df}{dz} \ ,
\label{eq: fp}
\end{equation}
\begin{equation}
f''(R) = \left ( \frac{dR}{dz} \right )^{-2} \frac{d^2f}{dz^2} -
\left ( \frac{dR}{dz} \right )^{-3} \frac{d^2R}{dz^2}
\frac{df}{dz} \ , \label{eq: fpp}
\end{equation}
\begin{eqnarray}
f'''(R) & = & \left ( \frac{dR}{dz} \right )^{-3}
\frac{d^3f}{dz^3}
+ 3 \left ( \frac{dR}{dz} \right )^{-5} \left ( \frac{d^2R}{dz^2} \right )^2 \frac{df}{dz} +\nonumber \\
 ~ & - & \left ( \frac{dR}{dz} \right )^{-4} \left ( 3 \frac{d^2R}{dz^2} \frac{d^2f}{dz^2} + \frac{d^3R}{dz^3}
\frac{df}{dz} \right ) \ . \label{eq: f3p}
\end{eqnarray}
Now, we have all the ingredients to reconstruct the shape of
$f(R)$ by data or, in general,  by the definition of a suitable
$H(z)$ viable with respect to observational results. In
particular,  we can show that a standard matter regime (necessary
to cluster large scale structure) can arise, in this scheme,
before the accelerating phase arises as, for example, in the so
called {\it quiessence} model.

A quiessence model is based on an ordinary matter fluid plus a
cosmological component whose equation of state $w$ is constant but
can scatter from  $w\,=\,-1$. This approach represents the easiest
generalization of the cosmological constant model, and it has been
successfully tested against the SNeIa Hubble diagram and the CMBR
anisotropy spectrum so that it allows to severely constraint the
barotropic index $w$ \cite{EstW}.

It is worth noticing that these constraints extend into the region
$w < -1$, therefore models (phantom models) violating the weak
energy condition are allowed. From the cosmological dynamics
viewpoint, such a model, by definition, has to display an
evolutionary rate of expansion which moves from the standard
matter regime to the accelerated behavior in relation to the value
of $w$. In particular, this quantity parameterizes  the transition
point to the accelerated epoch.

Actually, if it is possible to find out a $f(R)$-gravity model
compatible with the evolution of the Hubble parameter of the
quiessence model, this result suggests that modified gravity  is
compatible with a phase of standard matter domination. To be
precise, let us consider the Hubble flow defined by this model,
where, as above\,:
\begin{equation}
H(z) = H_0 \sqrt{\Omega_M (1 + z)^3 + \Omega_X (1 + z)^{3 (1 +
w)}} \label{eq: hqcdm}
\end{equation}
with $\Omega_X = (1 - \Omega_M)$ and $w$ the constant parameter
defining the dark energy barotropic index. This definition of the
Hubble parameter implies:
\begin{equation}
R = - 3 H_0^2 \left [ \Omega_M (1 + z)^3 + \Omega_X (1 - 3 w) (1 +
z)^{3 (1 + w)} \right ] \,. \label{eq: rvszqcdm}
\end{equation}
The ansatz in Eq.(\ref{eq: hqcdm}) allows to obtain from
Eq.(\ref{eq: presingleeq}) a differential relation for $f(R(z))$
which can be solved numerically by choosing suitable boundary
 conditions. In particular we choose\,:
\begin{equation}
\left ( \frac{df}{dz} \right )_{z = 0} = \left ( \frac{dR}{dz}
\right )_{z = 0} \ , \label{eq: fpzero}
\end{equation}
\begin{equation}
\left ( \frac{d^2f}{dz^2} \right )_{z = 0} = \left (
\frac{d^2R}{dz^2} \right )_{z = 0} \ . \label{eq: fppzero}
\end{equation}
\begin{equation}
f(z = 0) = f(R_0) = 6 H_0^2 (1 - \Omega_M) + R_0 \,. \label{eq:
fzero}
\end{equation}
A comment  is in order here.  We have derived the present day
values of $df/dz$ and $d^2f/dz^2$ by imposing the consistency of
the reconstructed $f(R)$ theory with {\it local} Solar System
tests. One could wonder whether tests on local scales could be
used to set the boundary conditions for a cosmological problem. It
is easy to see that this is indeed meaningful.

Actually, the isotropy and homogeneity of the Universe ensure that
the present day value of a whatever cosmological quantity does not
depend on where the observer is. As a consequence,  hypothetical
observers living in the Andromeda galaxy and testing gravity in
his planetary system should get the same results. As such, the
present day values of $df/dz$ and $d^2f/dz^2$ adopted by these
hypothetical observers are the same as those we have used, based
on our Solar System experiments. Therefore, there is no systematic
error induced by our method of setting the boundary conditions.

Once one has obtained  the numerical solution for $f(z)$,
inverting again numerically Eq.(\ref{eq: rvszqcdm}), we may obtain
$z = z(R)$ and finally get $f(R)$  for several values of $w$.

It turns out that $f(R)$ is the same for different models for low
values of $R$ and hence of $z$. This is a consequence of the well
known degeneracy among different quiessence models at low $z$
that, in the standard analysis, leads to large uncertainties on
$w$.  This is reflected in the shape of the reconstructed $f(R)$
that is almost $w$\,-\,independent in this redshift range.

An analytic representation of the reconstructed fourth order
gravity model, can be obtained considering that the following
empirical function

\begin{equation}
\ln{(-f)} = l_1 \left [ \ln{(-R)} \right ]^{l_2} \left [ 1 +
\ln{(-R)} \right ]^{l_3} + l_4 \label{eq: fit}
\end{equation}
approximates very well the numerical  solution, provided that the
parameters $(l_1, l_2, l_3, l_4)$ are suitably chosen for a given
value of $w$. For instance, for $w = -1$ (the cosmological
constant) it is\,:
\begin{displaymath}
(l_1, l_2, l_3, l_4) = (2.6693, 0.5950, 0.0719, -3.0099) \ .
\end{displaymath}
At this point, one can wonder if it is possible to improve such a
result considering even the radiation, although energetically
negligible. Rather than inserting radiation in the (\ref{eq:
hqcdm}), a more general approach in this sense is to consider the
Hubble parameter descending from a unified model like those
discussed  in \cite{Hobbit}. In such a scheme one takes into
account energy density which scales as\,:
\begin{equation}
\rho(z) = A \ \left ( 1 + \frac{1 + z}{1 + z_s} \right )^{\beta -
\alpha} \ \left [ 1 + \left ( \frac{1 + z}{1 + z_b} \right
)^{\alpha} \right ] \label{eq: rhoz}
\end{equation}
having defined\,:
\begin{equation}\label{eq: defzb}
z_s = 1/s - 1 \ ,\ \ \ \ \ \ \ \ z_b = 1/b - 1 \ .
\end{equation}
This model, with the choice $(\alpha, \beta) = (3, 4)$,  is able
to mimic a Universe undergoing first a radiation dominated era
(for $z \gg z_s$), then a matter dominated phase (for $z_b \ll z
\ll z_s$) and finally approaching a de\,Sitter phase with constant
energy.

In other words, it works  in the way we are asking for. In such a
case, the Hubble parameter can be written, in natural units, as
$H\,=\,\sqrt{\frac{\rho(z)}{3}}$ and one can perform the same
calculation as in the quiessence case.

As a final result, it is  again possible to find out a suitable
$f(R)$-gravity model which,  for numerical reasons, it is
preferable to interpolate as $f(R)/R$\,:
\begin{displaymath}\label{fr-hobbit}
\frac{f(R)}{R}\,=\,1.02\times\frac{R}{R_0}\left[1+\left(-0.04\times(\frac{R}{R_0})^{0.31}
\right.\right.
\end{displaymath}
\begin{equation}
\left.\left. +0.69\times(\frac{R}{R_0})^{-0.53}\right)\times
\ln({\frac{R}{R_0}})\right] \,,
\end{equation}
where $R_0$ is a normalization constant. This result once more
confutes issues addressing modified gravity as incompatible with
structure formation prescriptions. In fact, also in this case, it
is straightforward to show that a phase of ordinary matter
(radiation and dust) domination  can be obtained and it is
followed by an accelerated phase.

Furthermore, several recent studies are pointing out that large
scale structure and CMBR anisotropy spectrum are compatible with
$f(R)$ gravity as discussed in details in \cite{song,tsu4} for the
metric approach and in \cite{koivisto} for the Palatini approach.

In particular, in \cite{song}, it is shown that several classes of
$f(R)$ theories can tune the large-angle CMB anisotropy,  the
shape of the linear matter power spectrum, and qualitatively
change the correlations between the CMB and galaxy surveys. All
these phenomena are accessible with current and future data and
will soon provide stringent tests for such theories at
cosmological scales \cite{fosalba}.

\subsection{The stochastic background of Gravitational Waves "tuned" by $f(R)$ Gravity}

As we have seen, a  pragmatic point of view could be to
``reconstruct'' the suitable theory of gravity starting from data.
The main issues of this ``inverse '' approach is matching
consistently observations at different scales and taking into
account wide classes of gravitational theories where ``ad hoc''
hypotheses are avoided. In principle, as discussed in the previous
section,  the most popular dark energy cosmological models can be
achieved by considering $f(R)$ gravity without considering unknown
ingredients. The main issue to achieve such a goal is to have at
disposal suitable datasets at every redshift. In particular, this
philosophy can be taken into account also for the cosmological
stochastic background of gravitational waves (GW) which, together
with CMBR, would carry, if detected, a huge amount of information
on the early stages of the Universe evolution \cite{CC}. Here, we
want to show that cosmological information coming from
cosmological stochastic background of GWs could constitute a
benchmark for cosmological models coming from ETGs, in particular
for $f(R)$.

As well known, GWs are perturbations $h_{\mu\nu}$ of the metric
$g_{\mu\nu}$ which transform as 3-tensors.  The GW-equations in
the transverse-traceless gauge are
\begin{equation}
\square h_{i}^{j}=0\label{eq: 1}\,.
\end{equation}
  Latin indexes run from 1
to 3. Our task is now to derive the analog of Eqs.\ (\ref{eq: 1})
for a generic $f(R)$. As we have seen from conformal
transformation, the extra degrees of
 freedom related to  higher order gravity can be recast into a
 scalar field being
\begin{equation}
\widetilde{g}_{\mu\nu}=e^{2\phi}g_{\mu\nu}\qquad \mbox{with}
\qquad e^{2\phi}=f'(R)\,.\label{eq:3}
\end{equation}
and
\begin{equation}
\widetilde{R}=e^{-2\phi}\left(R-6\square\phi-6\phi_{;\delta}\phi^{;\delta}\right)\,.\label{eq:6}\end{equation}
 The GW-equation is now
\begin{equation}
\widetilde{\square}\tilde{h}_{i}^{j}=0\label{eq:7}
\end{equation}
where
\begin{equation}
\widetilde{\square}=e^{-2\phi}\left(\square+2\phi^{;\lambda}\nabla_{;\lambda}\right)\label{eq:9}\,.\end{equation}
Since no scalar perturbation couples to the tensor part of
gravitational waves, we have
\begin{equation}
\widetilde{h}_{i}^{j}=\widetilde{g}^{lj}\delta\widetilde{g}_{il}=e^{-2\phi}g^{lj}e^{2\phi}\delta
g_{il}=h_{i}^{j}\label{eq:8}
\end{equation}
which means that $h_{i}^{j}$ is a conformal invariant. As a
consequence, the plane-wave amplitudes
$h_{i}^{j}(t)=h(t)e_{i}^{j}\exp(ik_{m}x^{m}),$ where $e_{i}^{j}$
is the polarization tensor, are the same in both metrics. This
fact will assume a key role in the following discussion.

In a FRW background,  Eq.(\ref{eq:7}) becomes
\begin{equation}
\ddot{h}+\left(3H+2\dot{\phi}\right)\dot{h}+k^{2}a^{-2}h=0\label{eq:10}
\end{equation}
being  $a(t)$ the scale factor,  $k$ the wave number and $h$ the
GW amplitude. Solutions are combinations of Bessel's functions.
Several mechanisms can be considered for the production of
cosmological GWs.  In principle, we could seek for contributions
due to every high-energy  process in the early phases of the
Universe.

In the case of inflation, GW-stochastic background is strictly
related to dynamics of cosmological model. This is the case we are
considering here. In particular, one can assume that the main
contribution to the stochastic background comes from the
amplification of vacuum fluctuations at the transition between the
inflationary phase and the radiation  era. However,  we can assume
that the GWs generated as zero-point fluctuations during the
inflation undergo adiabatically damped oscillations $(\sim 1/a)$
until they reach the Hubble radius $H^{-1}$. This is the particle
horizon for the growth of perturbations. Besides, any previous
fluctuation is smoothed away by the inflationary expansion. The
GWs freeze out for $a/k\gg H^{-1}$ and reenter the $H^{-1}$ radius
after the reheating. The reenter in the Friedmann era depends on
the scale of the GW. After the reenter, GWs can be constrained by
the  Sachs-Wolfe effect on the temperature anisotropy
$\bigtriangleup T/T$ at the decoupling. More precisely, such
fluctuations are degenerated with scalar fluctuations, but GWs
can, in principle, be measured via B-polarization of the CMB. The
measurement is very hard to be performed, but many experiments in
this direction are presently planned. In any case, $\bigtriangleup
T/T$ can always be used to derive constraints.

If $\phi$ acts as the inflaton, we have $\dot{\phi}\ll H$ during
the inflation. Adopting the conformal time $d\eta=dt/a$, Eq.\
(\ref{eq:10}) reads
\begin{equation}
h''+2\frac{\chi'}{\chi}h'+k^{2}h=0\label{eq:16}
\end{equation}
where $\chi=ae^{\phi}$. The derivation is  now with respect to
$\eta$.  Inside the  radius $H^{-1}$, we have $k\eta\gg 1.$
Considering the absence of gravitons in the initial vacuum state,
we have only negative-frequency modes and then the solution of
(\ref{eq:16}) is
\begin{equation}
h=k^{1/2}\sqrt{2/\pi}\frac{1}{aH}C\exp(-ik\eta)\,.\label{eq:18}
\end{equation}
 $C$ is the amplitude parameter. At the first horizon crossing
$(aH=k)$ the averaged amplitude
$A_{h}^k=(k/2\pi)^{3/2}\left|h\right|$ of the perturbation is
\begin{equation}
A_{h}^k=\frac{1}{2\pi^{2}}C\,.\label{eq:19}
\end{equation}

\begin{figure}
\begin{tabular}{|c|c|}
\hline
\includegraphics[scale=0.7]{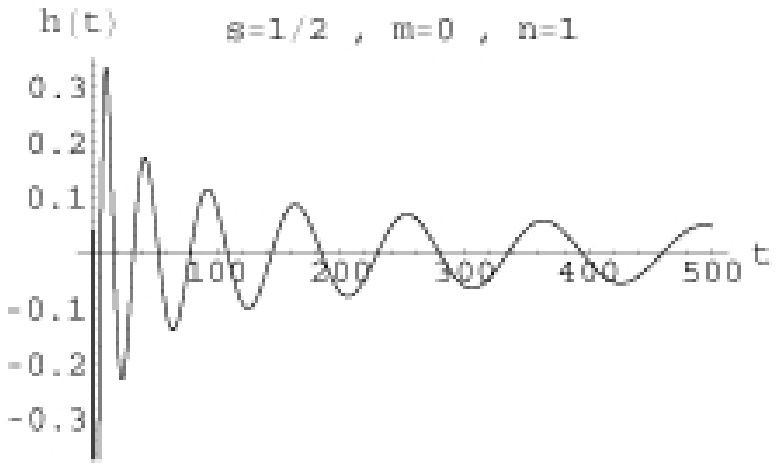}&
\includegraphics[scale=0.7]{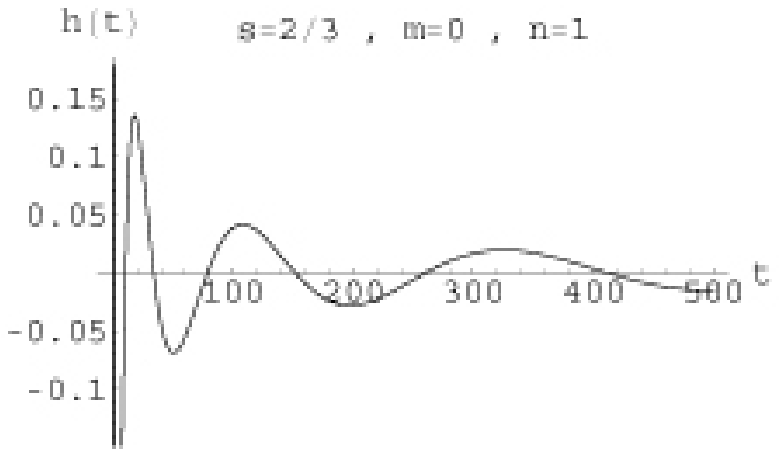}\tabularnewline
\hline
\includegraphics[scale=0.7]{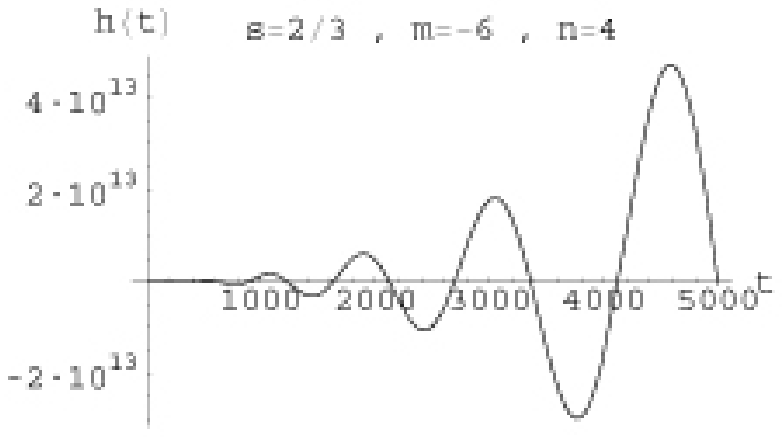}&
\includegraphics[scale=0.7]{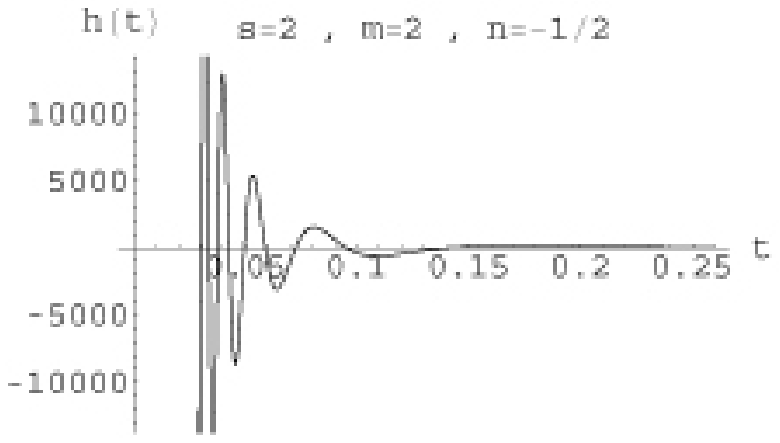}\tabularnewline
\hline
\includegraphics[scale=0.7]{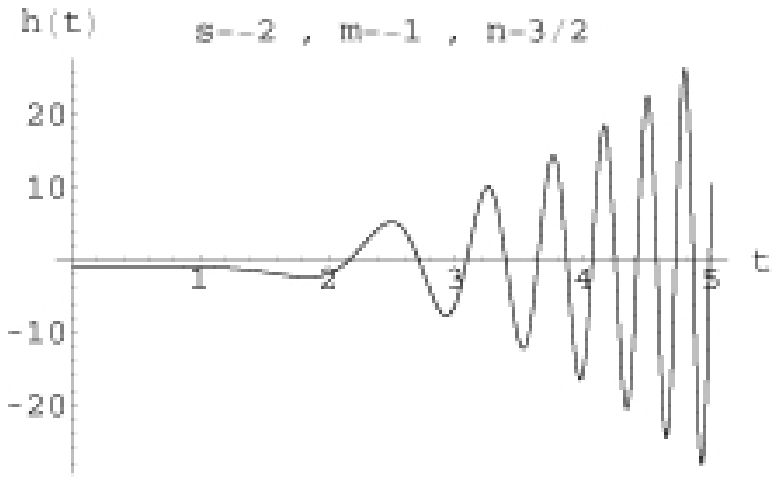}&
\includegraphics[scale=0.7]{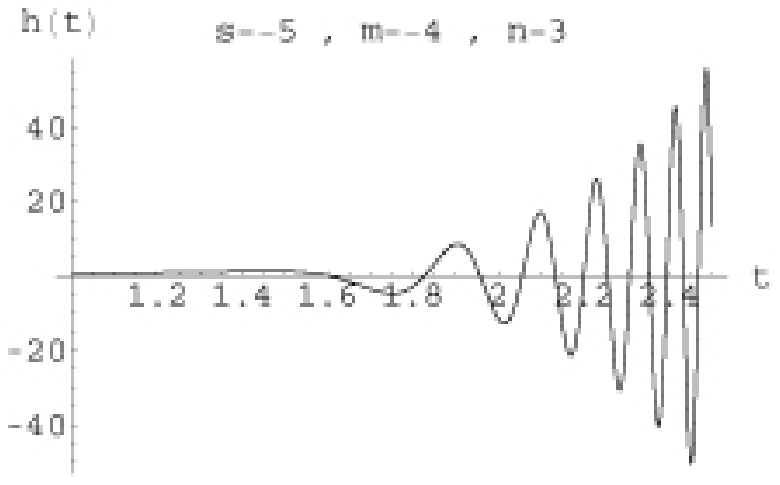}\tabularnewline
\hline
\end{tabular}
\caption {Evolution of the GW amplitude for some power-law
behaviors of $a(t)\sim t^s$, $\phi\sim t^m$ and $f(R)\sim R^n$.
The scales of time and amplitude  strictly depend on the
cosmological background giving a "signature" for the model.}
\label{fig:1}
\end{figure}

When the scale $a/k$ becomes larger than the Hubble radius
$H^{-1}$, the growing  mode of evolution is constant, i.e. it is
frozen.   It can be shown that $\bigtriangleup T/T\lesssim
A_{h}^k$, as an upper limit to $A_{h}^k$, since other effects can
contribute to the background anisotropy. From this consideration,
it is clear that the only relevant quantity is the initial
amplitude $C$ in Eq.\ (\ref{eq:18}), which is conserved until the
reenter. Such an amplitude depends  on the fundamental mechanism
generating perturbations. Inflation gives rise to processes
capable of producing perturbations as zero-point energy
fluctuations. Such a mechanism depends on the gravitational
interaction and then $(\bigtriangleup T/T)$ could constitute a
further constraint to select a suitable theory of gravity.
Considering a single graviton in the form of a monochromatic wave,
its zero-point amplitude is derived through the commutation
relations:
\begin{equation}
\left[h(t,x),\,\pi_{h}(t,y)\right]=i\delta^{3}(x-y)\label{eq:20}
\end{equation}
calculated at a fixed time $t$, where the amplitude $h$ is the
field and $\pi_{h}$ is the conjugate momentum operator. Writing
the Lagrangian for $h$
\begin{equation}
\widetilde{\mathcal{L}}=\frac{1}{2}\sqrt{-\widetilde{g}}\widetilde{g}^{\mu\nu}h_{;\mu}h{}_{;\nu}\label{eq:21}
\end{equation}
in the conformal FRW metric $\widetilde{g}_{\mu\nu}$, where the
amplitude $h$ is conformally invariant, we obtain
\begin{equation}
\pi_{h}=\frac{\partial\widetilde{\mathcal{L}}}{\partial\dot{h}}=e^{2\phi}a^{3}\dot{h}\label{eq:22}\end{equation}

Eq.\ (\ref{eq:20}) becomes
\begin{equation}
\left[h(t,x),\,\dot{h}(y,y)\right]=i\frac{\delta^{3}(x-y)}{a^{3}e^{2\phi}}\label{eq:23}
\end{equation}
and the fields $h$ and $\dot{h}$ can be expanded in terms of
creation and annihilation operators
\begin{equation}
h(t,x)=\frac{1}{(2\pi)^{3/2}}\int
d^{3}k\left[h(t)e^{-ikx}+h^{*}(t)e^{+ikx}\right],\label{eq:24}
\end{equation}
\begin{equation}
\dot{h}(t,x)=\frac{1}{(2\pi)^{3/2}}\int
d^{3}k\left[\dot{h}(t)e^{-ikx}+\dot{h}^{*}(t)e^{+ikx}\right].\label{eq:25}
\end{equation}

The commutation relations in conformal time are
\begin{equation}
\left[hh'^{*}-h^{*}h'\right]=\frac{i(2\pi)^{3}}{a^{3}e^{2\phi}}\,.\label{eq:26}
\end{equation}
From (\ref{eq:18}) and (\ref{eq:19}), we obtain
$C=\sqrt{2}\pi^{2}He^{-\phi}$, where $H$ and $\phi$ are calculated
at the first horizon-crossing and, being $e^{2\phi}=f'(R)$, the
remarkable relation
\begin{equation}
A_{h}^k=\frac{H}{\sqrt{2f'(R)}}\,, \label{eq:27}
\end{equation}
holds for a generic $f(R)$ theory at a given $k$.  Clearly the
amplitude of GWs produced during inflation depends on the theory
of gravity which, if different from GR, gives extra degrees of
freedom.  On the other hand, the Sachs-Wolfe effect could
constitute a test for gravity at early epochs. This probe could
give further constraints on the GW-stochastic background, if ETGs
are independently probed at other scales.

In summary,  the amplitudes of tensor GWs are conformally
invariant and their evolution depends on the cosmological
background. Such a background is tuned by a conformal scalar field
which is not present in the standard GR. Assuming that primordial
vacuum fluctuations produce stochastic GWs, beside scalar
perturbations, kinematical distortions and so on, the initial
amplitude of these ones is a function of the $f(R)$-theory of
gravity and then the stochastic background can be, in a certain
sense ``tuned'' by the theory. Viceversa, data coming from the
Sachs-Wolfe effect could contribute to select a suitable $f(R)$
theory which can be consistently  matched with other observations.
However, further and accurate studies  are needed in order to test
the relation between Sachs-Wolfe effect and $f(R)$ gravity. This
goal could be achieved very soon through the forthcoming space
(LISA) and ground-based (VIRGO, LIGO) interferometers.

\section{Applications to galactic dynamics}

The results obtained at  cosmological scales motivates further
applications of ETGs, in particular of $f(R)$ theories. In
general, one is wondering whether ETG models, working as dark
energy models, can also play a  role  to explain the dark matter
phenomenology at scales of galaxies and clusters of galaxies.

Several studies have been pursued in this direction \cite{mond}
but the main goal remains that to seek a unified model capable of
explain dynamics at every scale without introducing {\it ad hoc}
components.

\subsection{Dark matter as a curvature effect: the case of flat
rotation curves of LSB galaxies}

It is well known that, in the low energy limit, higher order
gravity implies modified gravitational potentials
\cite{schmidt,modpot}. By considering the case of a pointlike mass
$m$ and solving the vacuum field equations for a
Schwarzschild\,-\,like metric \cite{noipla,mnras}, one gets as
exact solution from a theory $f(R)=f_0 R^n$, the modified
gravitational potential\,:
\begin{equation}
\Phi(r) = - \frac{G m}{2r} \left [ 1 + \left ( \frac{r}{r_c}
\right )^{\beta} \right ] \label{eq: pointphi}
\end{equation}
where
\begin{equation}
\beta = \frac{12n^2 - 7n - 1 - \sqrt{36n^4 + 12n^3 - 83n^2 + 50n +
1}}{6n^2 - 4n + 2} \label{eq: bnfinal}
\end{equation}
which corrects the ordinary Newtonian potential by a power\,-\,law
term. As we will see, it has to be $\beta >0$ and then  $n>0$. In
particular, the best fit value will be $\beta\simeq 0.8$ and then
$n=3.2$. Standard units have been considered here. In particular,
this correction sets in on scales larger than $r_c$ which value
depends essentially on the mass of the system \cite{mnras}. This
quantity deserves some discussion. As shown in \cite{mnras}, it is
derived from the initial conditions of the models and it
correlates with the core masses of the LSB galaxies which we have
taken into account. In some sense, it is a sort of further {\it
gravitational radius}, beside the standard Schwarzschild radius,
which rules the central mass of the galaxies, and then it is
different for different systems. It is interesting to note that,
given a generic $2n$-order theory of gravity, it is possible to
find out $n$ characteristic radii \cite{schmidt} and it is
intriguing to suspect that they could likely rule the structure
and the stability of the astrophysical self-gravitating structures
\cite{betty}. But this is a working hypothesis which has to be
firmly demonstrated.

The corrected potential (\ref{eq: pointphi}) reduces to the
standard $\Phi \propto 1/r$ for $n=1$ as it can be seen from the
relation (\ref{eq: bnfinal}). The generalization of Eq.(\ref{eq:
pointphi}) to extended systems is straightforward. We simply
divide the system in infinitesimal mass elements and sum up the
potentials generated by each single element. In the continuum
limit, we replace the sum with an integral over the mass density
of the system taking care of eventual symmetries of the mass
distribution \cite{mnras}. Once the gravitational potential has
been computed, one may evaluate the rotation curve $v_c^2(r)$ and
compare it with the data. For extended systems, one has typically
to resort to numerical techniques, but the main effect may be
illustrated by the rotation curve for the pointlike case\,:
\begin{equation}
v_c^2(r) = \frac{G m}{2r} \left [ 1 + (1 - \beta) \left (
\frac{r}{r_c} \right )^{\beta} \right ] \ . \label{eq: vcpoint}
\end{equation}
Compared with the Newtonian result $v_c^2 = G m/r$, the corrected
rotation curve is modified by the addition of the second term in
the r.h.s. of Eq.(\ref{eq: vcpoint}). For $0 <\, \beta \,< 1$, the
corrected rotation curve is higher than the Newtonian one. Since
measurements of spiral galaxies rotation curves signal a circular
velocity higher than those which are predicted on the basis of the
observed luminous mass and the Newtonian potential, the above
result suggests the possibility that such a modified gravitational
potential may fill the gap between theory and observations without
the need of additional dark matter. It is worth noticing that the
corrected rotation curve is asymptotically vanishing as in the
Newtonian case, while it is usually claimed that observed rotation
curves are flat (i.e., asymptotically constant). Actually,
observations do not probe $v_c$ up to infinity, but only show that
the rotation curve is flat within the measurement uncertainties up
to the last measured point. This fact by no way excludes the
possibility that $v_c$ goes to zero at infinity.

\begin{table}
\begin{center}
\begin{tabular}{|c|c|c|c|c|c|c|}
\hline
Id & $D$ & $\mu_0$ & $r_d$ & $r_{HI}$ & $M_{HI}$ & Type \\
\hline UGC 1230 & 51 & 22.6 & 4.5 & 101 & 58.0 & Sm\\
UGC 1281 & 5.5 & 22.7 & 1.7 & 206 & 3.2 & Sdm\\ UGC 3137 & 18.4 &
23.2 & 2.0 & 297 & 43.6 & Sbc \\ UGC 3371 & 12.8 & 23.3 & 3.1 & 188 & 12.2 & Im \\
UGC 4173 & 16.8 & 24.3 & 4.5 & 178 & 21.2 & Im \\ UGC 4325 & 10.1
& 21.6 & 1.6 & 142 & 7.5 & SAm \\ NGC 2366 & 3.4 & 22.6 & 1.5 &
439 & 7.3 & IB(s)m \\ IC 2233 & 10.5 & 22.5 & 2.3 & 193 & 13.6 &
SBd \\ NGC 3274 & 6.7 & 20.2 & 0.5 & 225 & 6.6 & SABd \\ NGC 4395
& 3.5 & 22.2 & 2.3 & 527 & 9.7 & SAm \\ NGC 4455 & 6.8 & 20.8 &
0.7 & 192 & 5.4 & SBd \\ NGC 5023 & 4.8 & 20.9 & 0.8 & 256 & 3.5 &
Scd \\ DDO 185 & 5.1 & 23.2 & 1.2 & 136 & 1.6 & IBm \\ DDO 189 &
12.6 & 22.6 & 1.2 & 167 & 10.5 & Im \\ UGC 10310 & 15.6 & 22.0 &
1.9 & 130 & 12.6 & SBm \\ \hline
\end{tabular}
\caption{Properties of sample galaxies. Explanation of the
columns\,: name of the galaxy, distance in Mpc; disk central
surface brightness in the ${\cal R}$ band (corrected for galactic
extinction); disk scalelength in kpc; radius at which the gas
surface density equals $1 \ {\rm M_{\odot}/pc^2}$ in arcsec; total
HI gas mass in $10^8 \ {\rm M_{\odot}}$; Hubble type as reported
in the NED database.}
\end{center}
\end{table}

In order to observationally check the above result, one can take
into account  samples of low surface brightness (LSB) galaxies
with well measured HI + H$\alpha$ rotation curves extending far
beyond the visible edge of the system. LSB galaxies are known to
be ideal candidates to test dark matter models since, because of
their high gas content, the rotation curves can be well measured
and corrected for possible systematic errors by comparing
21\,-\,cm HI line emission with optical H$\alpha$ and ${\rm
[NII]}$ data. Moreover, they are supposed to be dark matter
dominated so that fitting their rotation curves without this
elusive component could be a strong evidence in favor of any
successful alternative theory of gravity. The considered sample
(Table II) contains 15 LSB galaxies with data on  the rotation
curve, the surface mass density of the gas component and ${\cal
R}$\,-\, photometric band, disk photometry extracted from a larger
sample selected by de Blok \& Bosma \cite{dbb02}. We assume the
stars are distributed in a thin and circularly symmetric disk with
surface density $\Sigma(r) = \Upsilon_\star I_0 exp{(-r/r_d)}$
where the central surface luminosity $I_0$ and the disk
scalelength $r_d$ are obtained from fitting to the stellar
photometry. The gas surface density has been obtained by
interpolating the data over the range probed by HI measurements
and extrapolated outside this range.

\begin{table*}
\begin{center}
\begin{tabular}{|c|c|c|c|c|c|c|}
\hline
Id & $\beta$ & $\log{r_c}$ & $f_g$ & $\Upsilon_{\star}$ & $\chi^2/dof$ & $\sigma_{rms}$ \\
\hline
UGC 1230 & 0.83 $\pm$ 0.02 & -0.39 $\pm$ 0.09 & 0.15 $\pm$ 0.01 & 15.9 $\pm$ 0.5 & 2.97/8 & 0.96 \\
UGC 1281 & 0.38 $\pm$ 0.01 & -3.93 $\pm$ 0.80 & 0.65 $\pm$ 0.08 & 0.64 $\pm$ 0.33 & 3.48/21 & 1.05 \\
UGC 3137 & 0.72 $\pm$ 0.03 & -1.86 $\pm$ 0.06 & 0.65 $\pm$ 0.02 & 9.8 $\pm$ 0.9 & 48.1/26 & 1.81 \\
UGC 3371 & 0.78 $\pm$ 0.05 & -1.85 $\pm$ 0.01 & 0.41 $\pm$ 0.01 & 3.3 $\pm$ 0.2 & 0.48/15 & 1.30 \\
UGC 4173 & 0.94 $\pm$ 0.02 & -0.97 $\pm$ 0.22 & 0.34 $\pm$ 0.01 & 9.37 $\pm$ 0.04 & 0.12/10 & 0.52 \\
UGC 4325 & 0.79 $\pm$ 0.07 & -2.85 $\pm$ 0.44 & 0.70 $\pm$ 0.02 & 0.50 $\pm$ 0.05 & 0.09/13 & 1.19 \\
NGC 2366 & 0.96 $\pm$ 0.14 & -0.58 $\pm$ 0.42 & 0.64 $\pm$ 0.01 & 14.5 $\pm$ 0.9 & 28.6/25 & 1.10 \\
IC 2233 & 0.42 $\pm$ 0.01 & -3.50 $\pm$ 0.05 & 0.64 $\pm$ 0.01 & 1.29 $\pm$ 0.06 & 6.1/22 & 2.10\\
NGC 3274 & 0.71 $\pm$ 0.03 & -2.30 $\pm$ 0.19 & 0.55 $\pm$ 0.03 & 2.3 $\pm$ 0.3 & 17.6/20 & 2.7 \\
NGC 4395 & 0.13 $\pm$ 0.02 & -3.68 $\pm$ 0.31 & 0.14 $\pm$ 0.01 & 7.6 $\pm$ 0.3 & 37.7/52 & 1.40 \\
NGC 4455 & 0.87 $\pm$ 0.05 & -2.32 $\pm$ 0.07 & 0.83 $\pm$ 0.01 & 0.42 $\pm$ 0.04 & 3.3/17 & 1.12 \\
NGC 5023 & 0.81 $\pm$ 0.02 & -2.54 $\pm$ 0.05 & 0.53 $\pm$ 0.02 & 0.91 $\pm$ 0.06 & 8.9/30 & 2.50 \\
DDO 185 & 0.92 $\pm$ 0.10 & -2.75 $\pm$ 0.35 & 0.90 $\pm$ 0.03 & 0.21 $\pm$ 0.07 & 5.03/5 & 0.81 \\
DDO 189 & 0.54 $\pm$ 0.08 & -2.40 $\pm$ 0.61 & 0.63 $\pm$ 0.04 & 4.2 $\pm$ 0.7 & 0.44/8 & 1.08 \\
UGC 10310 & 0.72 $\pm$ 0.04 & -1.87 $\pm$ 0.04 & 0.59 $\pm$ 0.02 & 1.39 $\pm$ 0.04 & 2.90/13 & 1.02 \\
\hline
\end{tabular}
\end{center}
\caption{Best fit values of the model parameters from minimizing
$\chi^2(\beta, \log{r_c}, f_g)$.  The values of
$\Upsilon_{\star}$, the $\chi^2/dof$ are reported for  the best
fit parameters (with $dof = N - 3$ and $N$ the number of
datapoints) and the root mean square $\sigma_{rms}$ of the fit
residuals. Errors on the fitting parameters and the $M/L$ ratio
are estimated through the jacknife method hence do not take into
account parameter degeneracies \cite{mnras}.\label{tab}}
\end{table*}

When fitting to the theoretical rotation curve, there are three
quantities to be determined, namely the stellar
mass\,-\,to\,-\,light (M/L) ratio, $\Upsilon_{\star}$ and the
theory parameters $(\beta, r_c)$. It is worth stressing that,
while fit results for different galaxies should give the same
$\beta$, $r_c$ is related to one of the integration constants of
the field equations. As such, it is not a universal quantity and
its value must be set on a galaxy\,-\,by\,-\,galaxy basis.
\begin{figure}
\centering\resizebox{7.5cm}{!}{\includegraphics{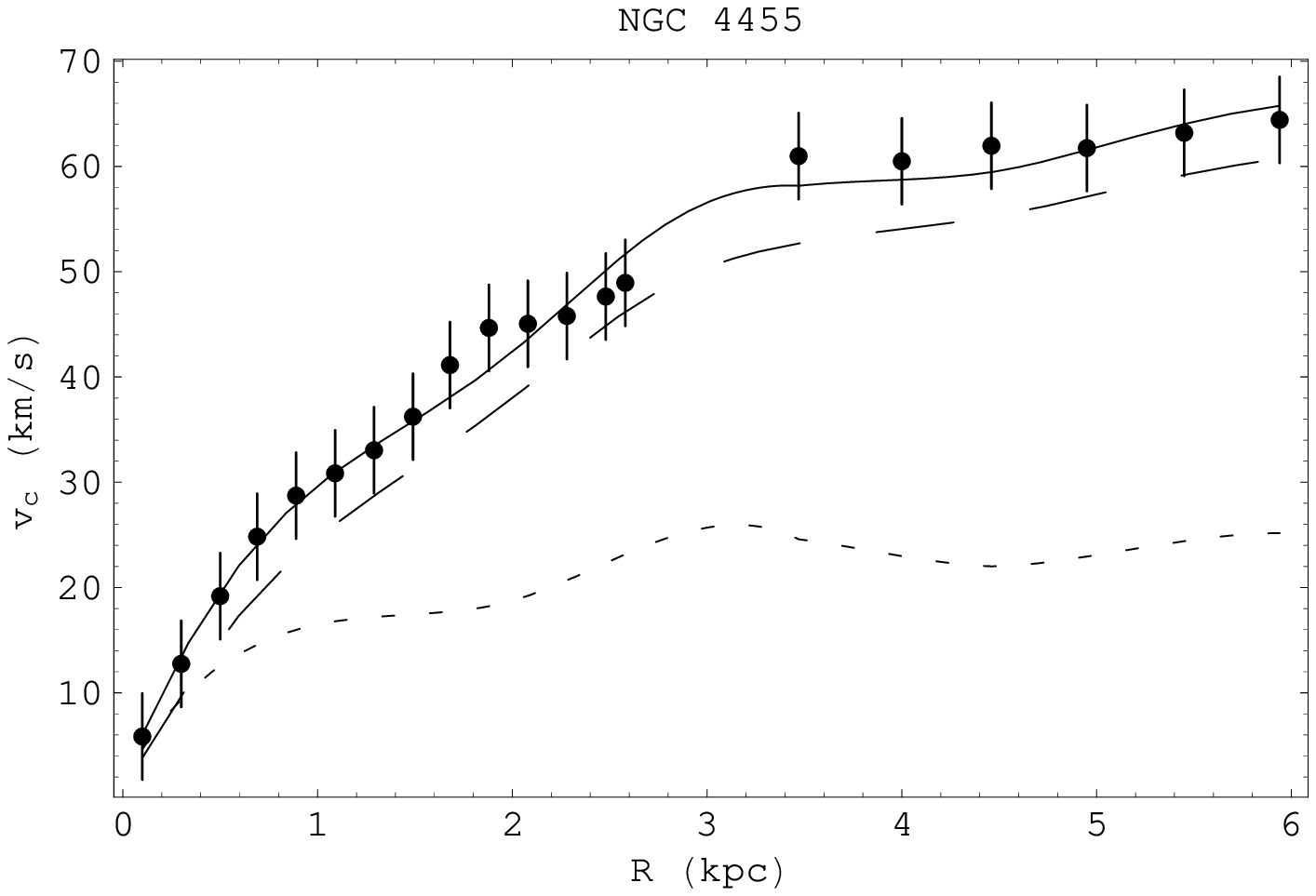}}
\centering\resizebox{7.5cm}{!}{\includegraphics{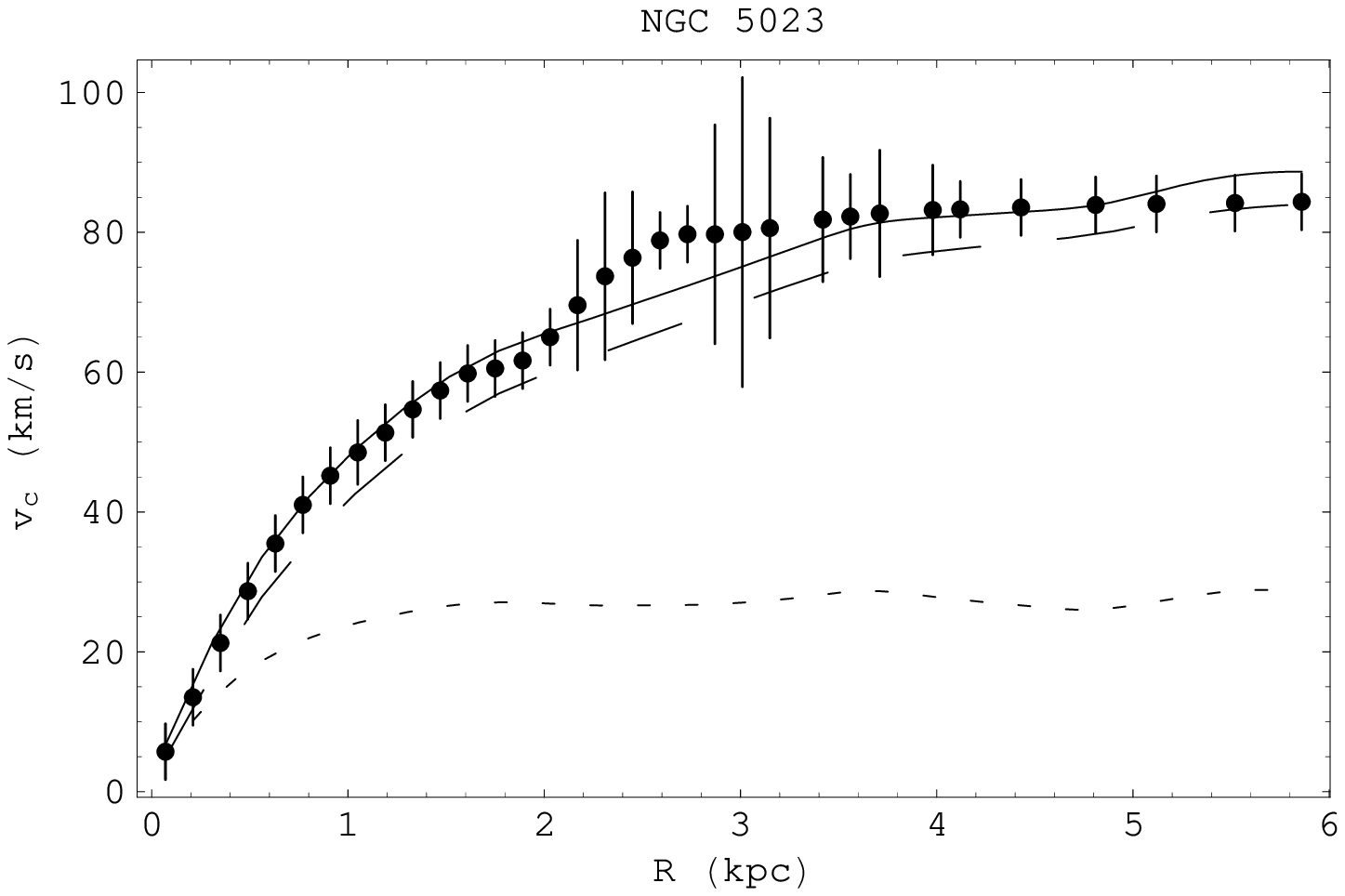}}
\caption{Best fit theoretical rotation curve superimposed to the
data for the LSB galaxy NGC 4455 (left) and NGC 5023 (right).
These two cases are considered to better show the effect of the
correction to the Newtonian gravitational potential. We report the
total rotation curve $v_c(r)$ (solid line), the Newtonian one
(short dashed) and the corrected term (long dashed).\label{fig:
lsb1}}
\end{figure}
However, it is expected that galaxies having similar properties in
terms of mass distribution have similar values of $r_c$ so that
the scatter in $r_c$ must reflect somewhat the scatter in the
terminal circular velocities. In order to match the model with the
data, we perform a likelihood analysis  for each galaxy, using, as
fitting parameters $\beta$, $\log{r_c}$ (with $r_c$ in kpc) and
the gas mass fraction\footnote{This is related to the $M/L$ ratio
as $\Upsilon_{\star} = [(1 - f_g) M_{g}]/(f_g L_d)$ with $M_g =
1.4 M_{HI}$ the gas (HI + He) mass, $M_d = \Upsilon_{\star} L_d$
and $L_d = 2 \pi I_0 r_d^2$ the disk total mass and luminosity.}
$f_g$. As it is evident considering the results from the different
fits summarized in Table \ref{tab}, the experimental data are
successfully fitted by the model. In particular,  the best fit
range of $\beta$ $(\beta=0.80\pm 0.08)$, corresponding to $R^n$
gravity with $2.3 < n <5.3$ (best fit value $n\,=\,3.2$), seems
well overlaps the above mentioned range of $n$ fitting SNeIa
Hubble diagram.

\begin{figure*}
\begin{tabular}{c c c}

\includegraphics[width=5cm]{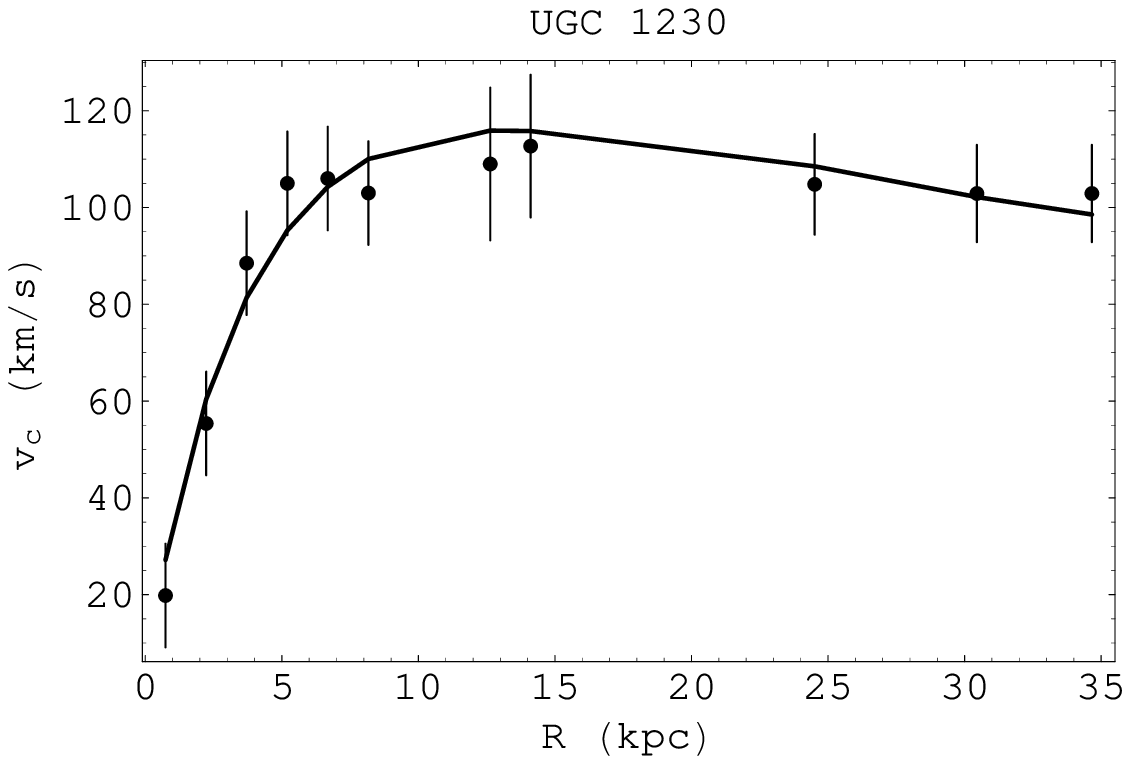}&
\includegraphics[width=5cm]{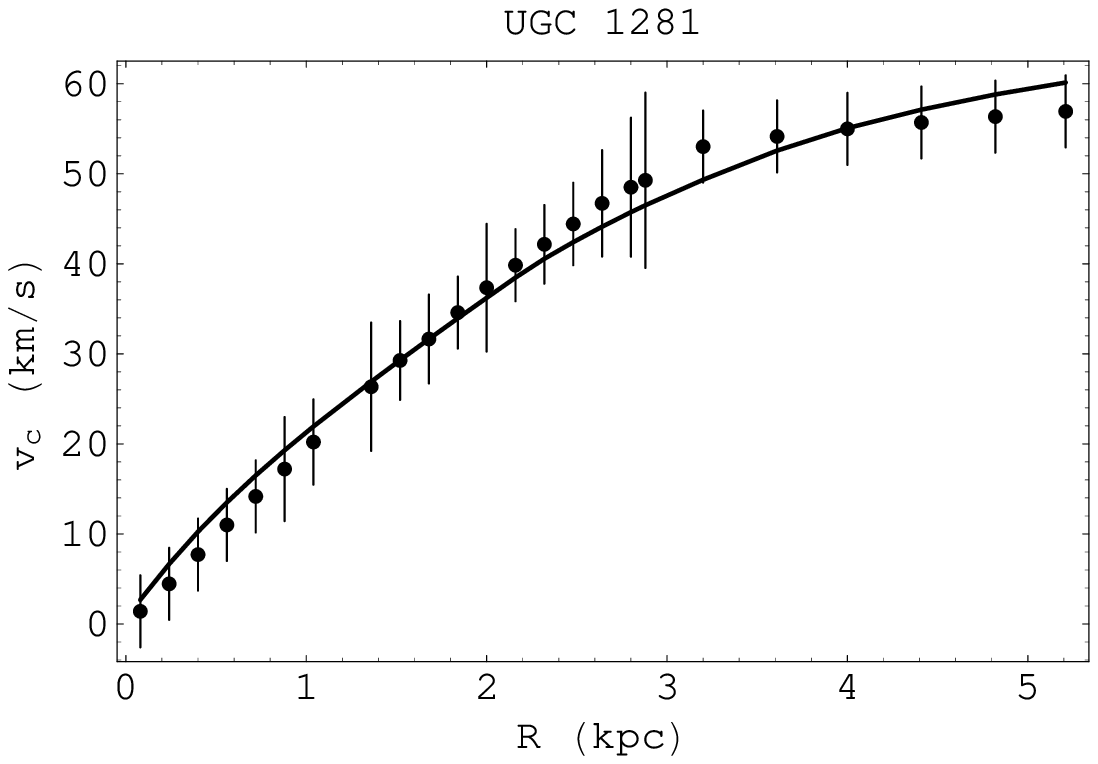} &
\includegraphics[width=5cm]{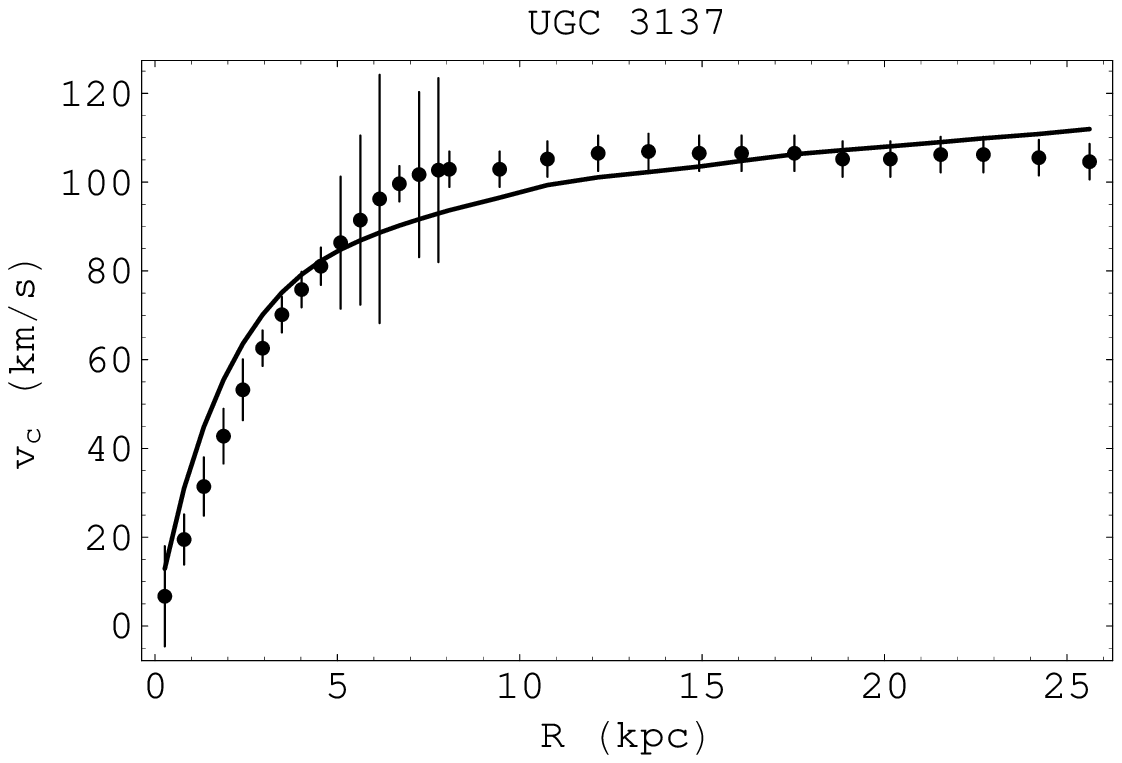} \\

\includegraphics[width=5cm]{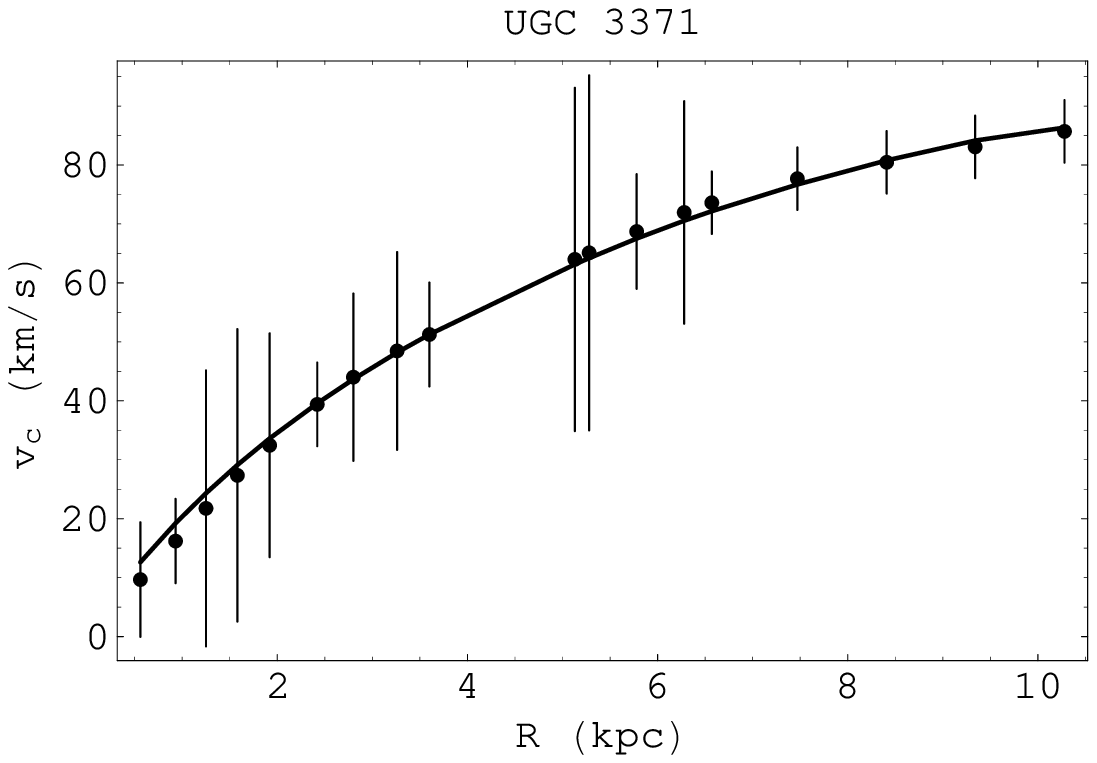}&
\includegraphics[width=5cm]{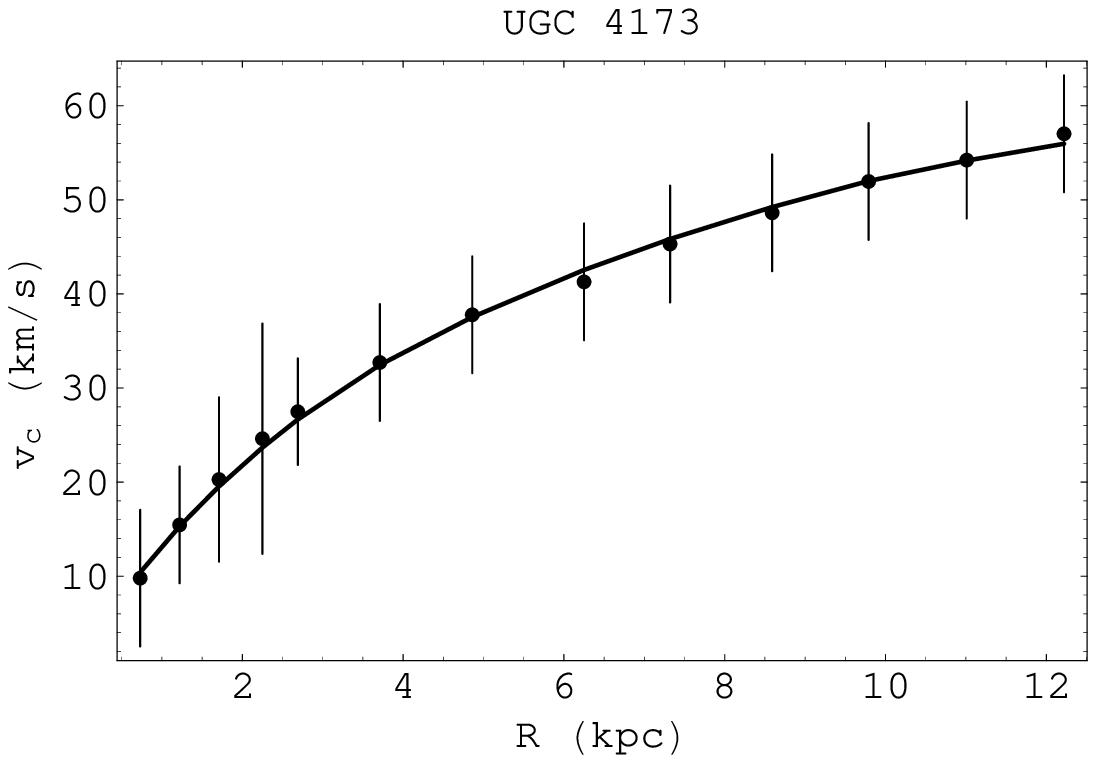}&
\includegraphics[width=5cm]{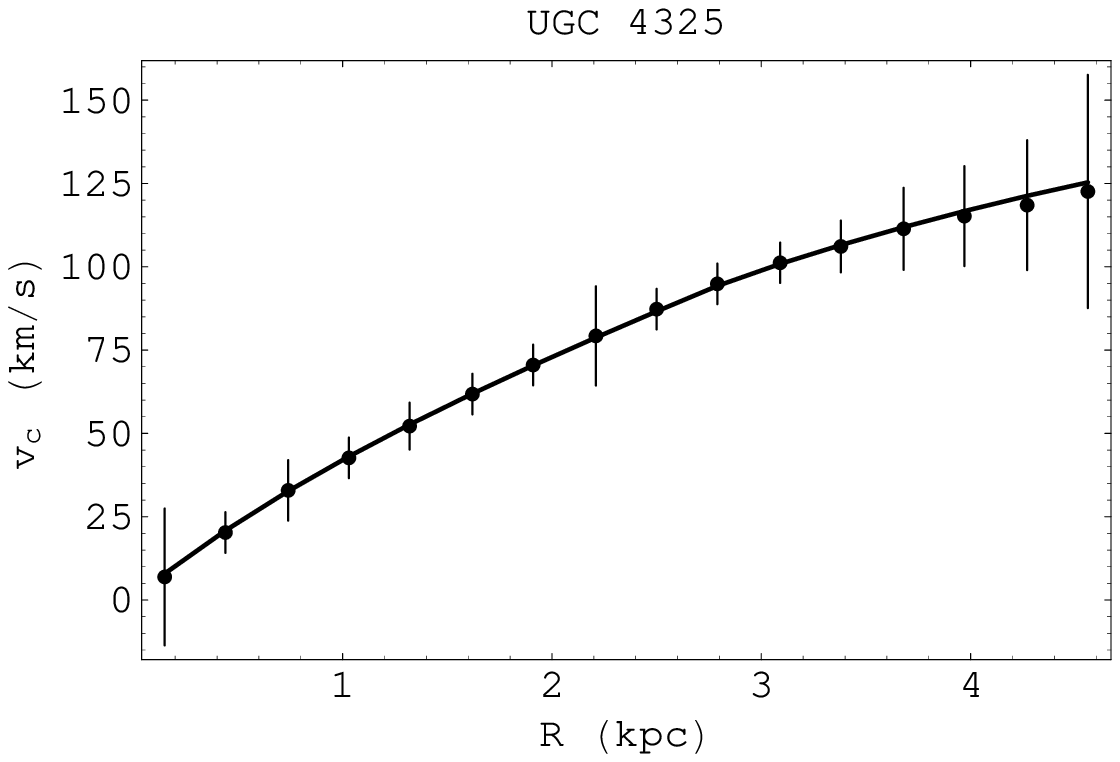} \\

\includegraphics[width=5cm]{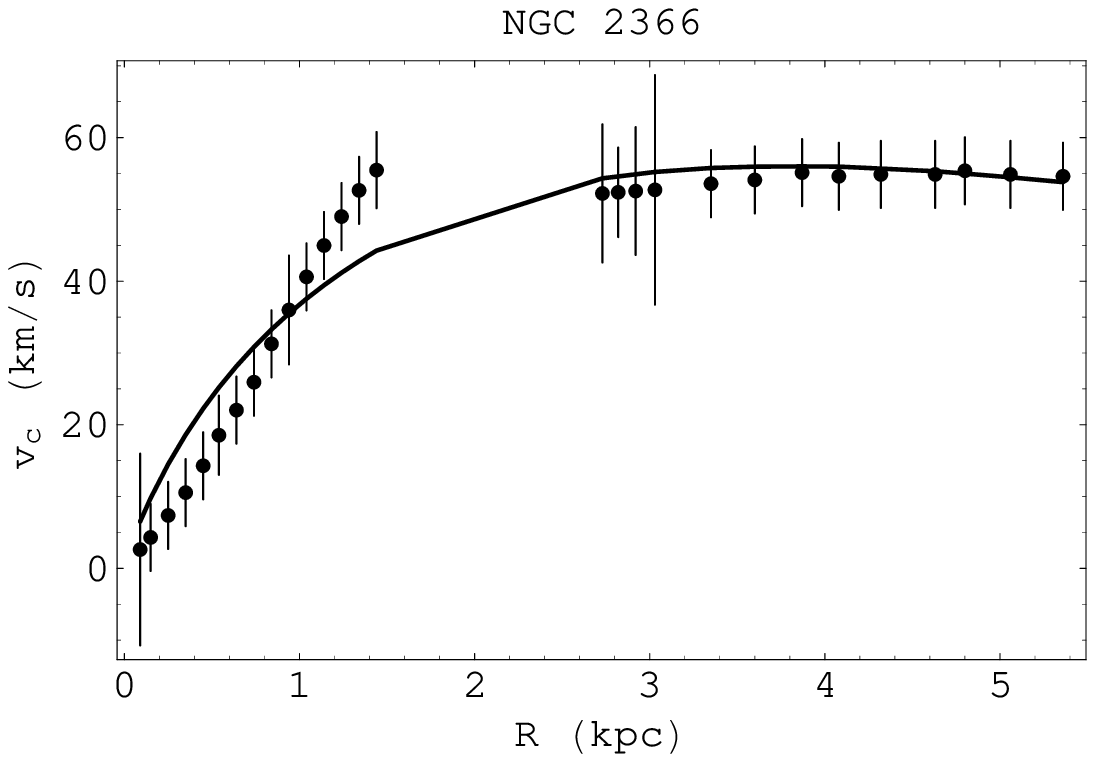}&
\includegraphics[width=5cm]{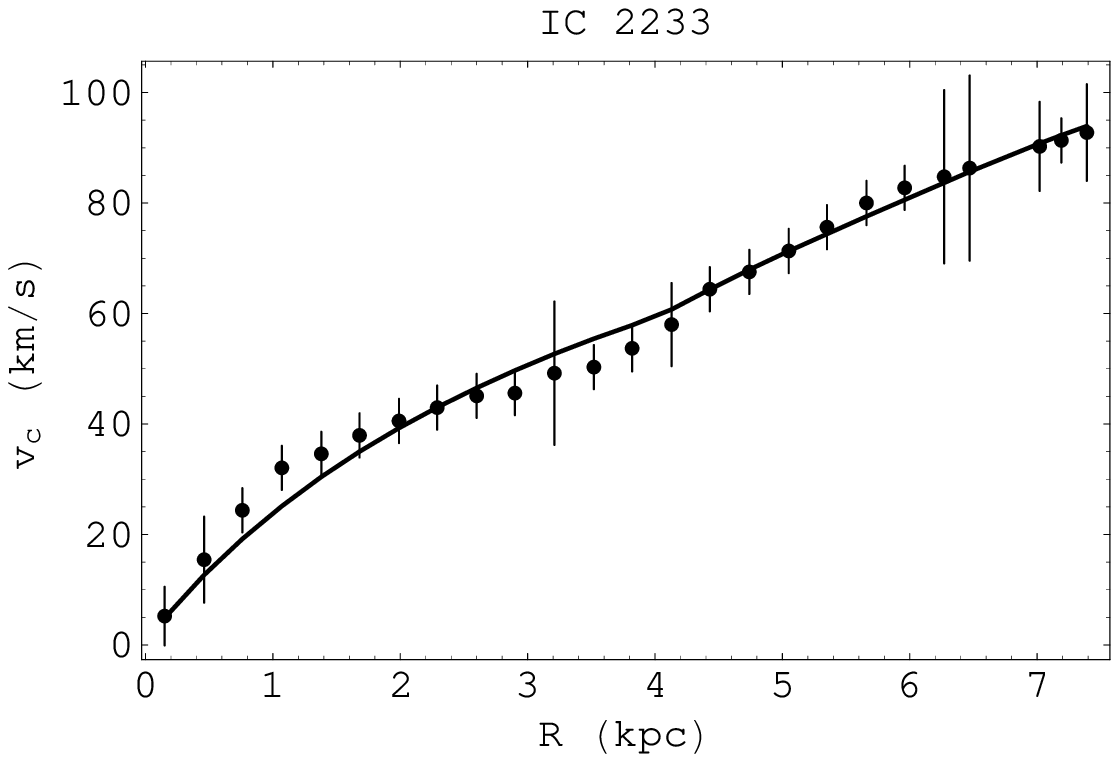}&
\includegraphics[width=5cm]{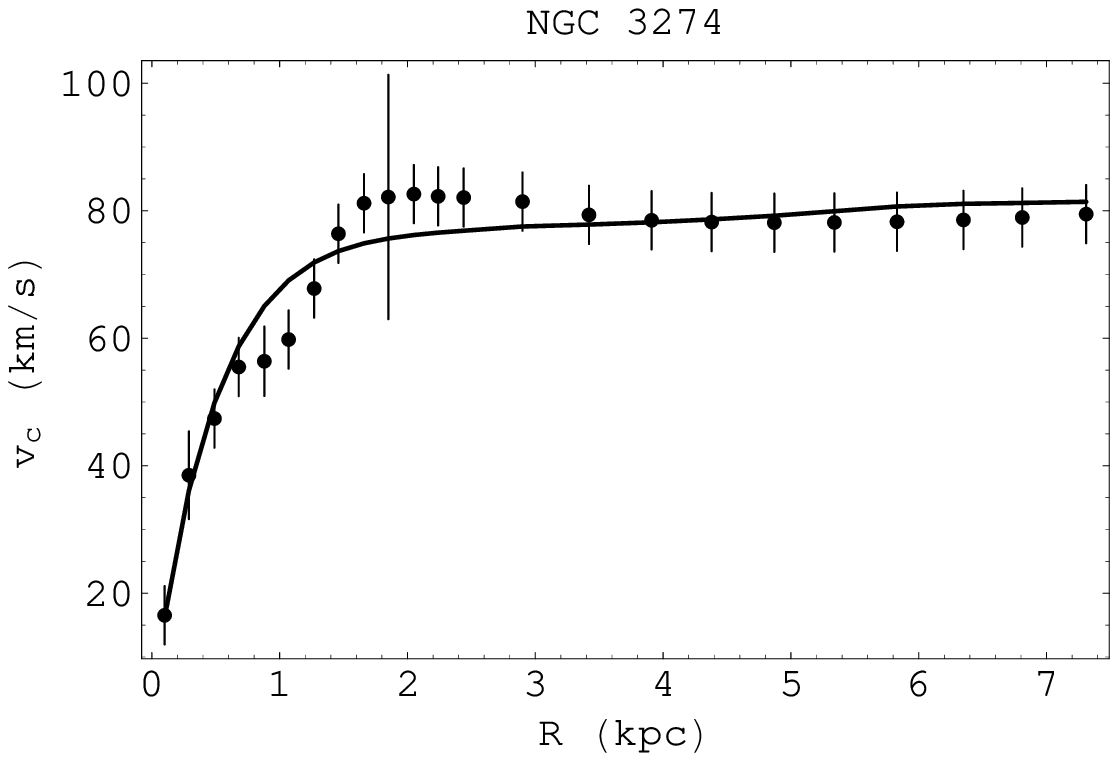}\\

\includegraphics[width=5cm]{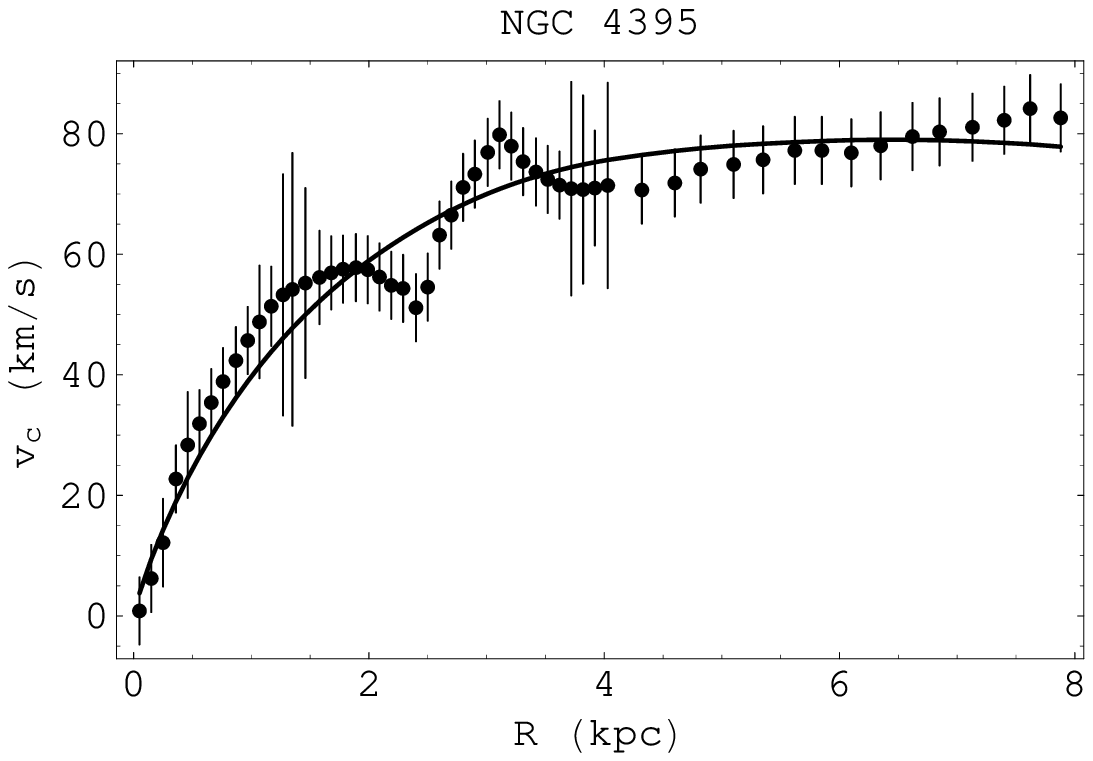}&
\includegraphics[width=5cm]{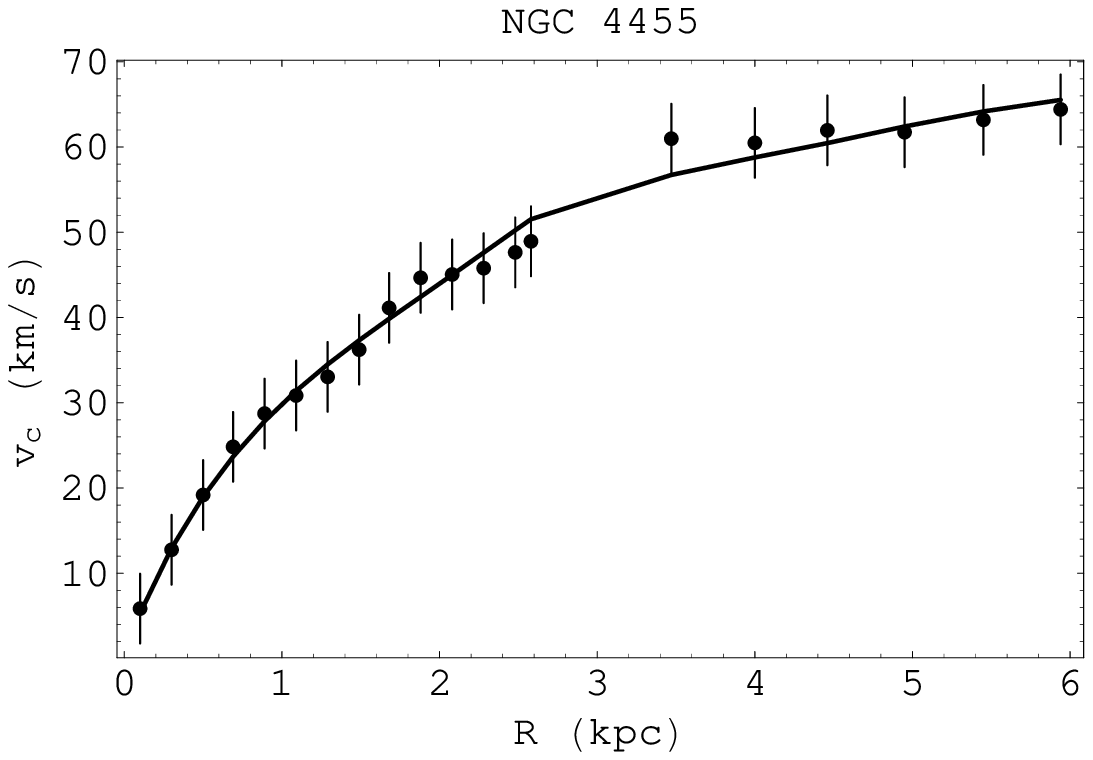}&
\includegraphics[width=5cm]{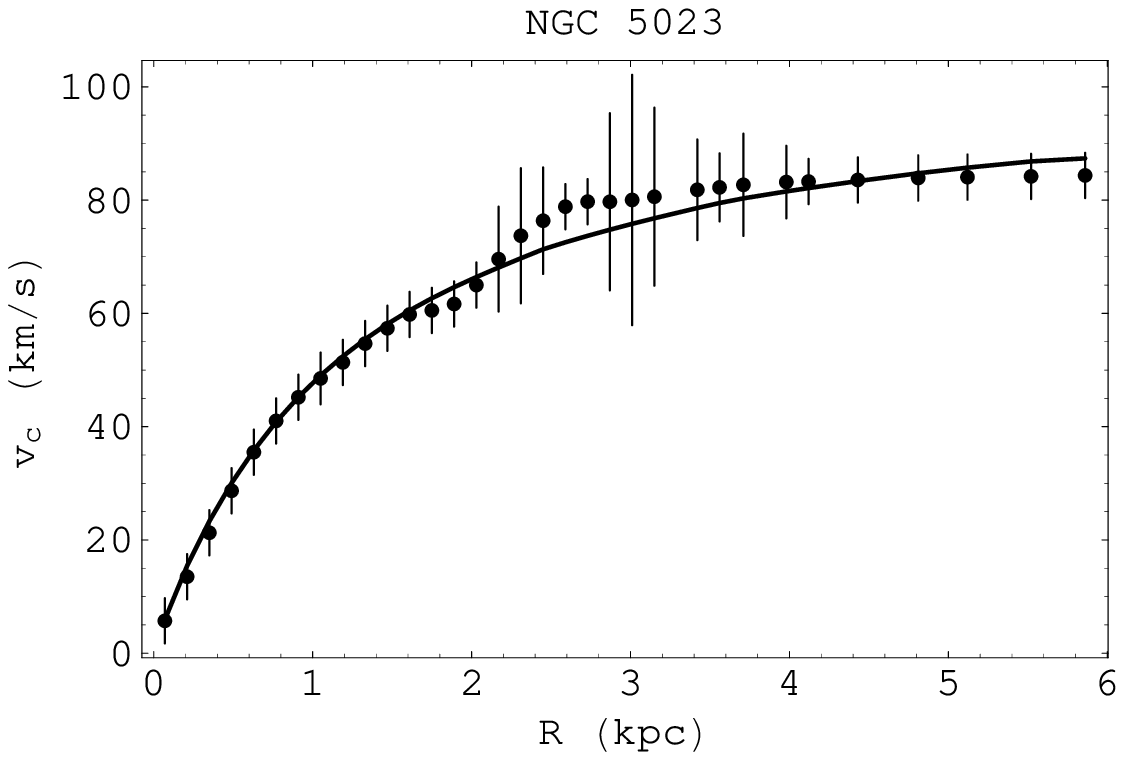}\\

\includegraphics[width=5cm]{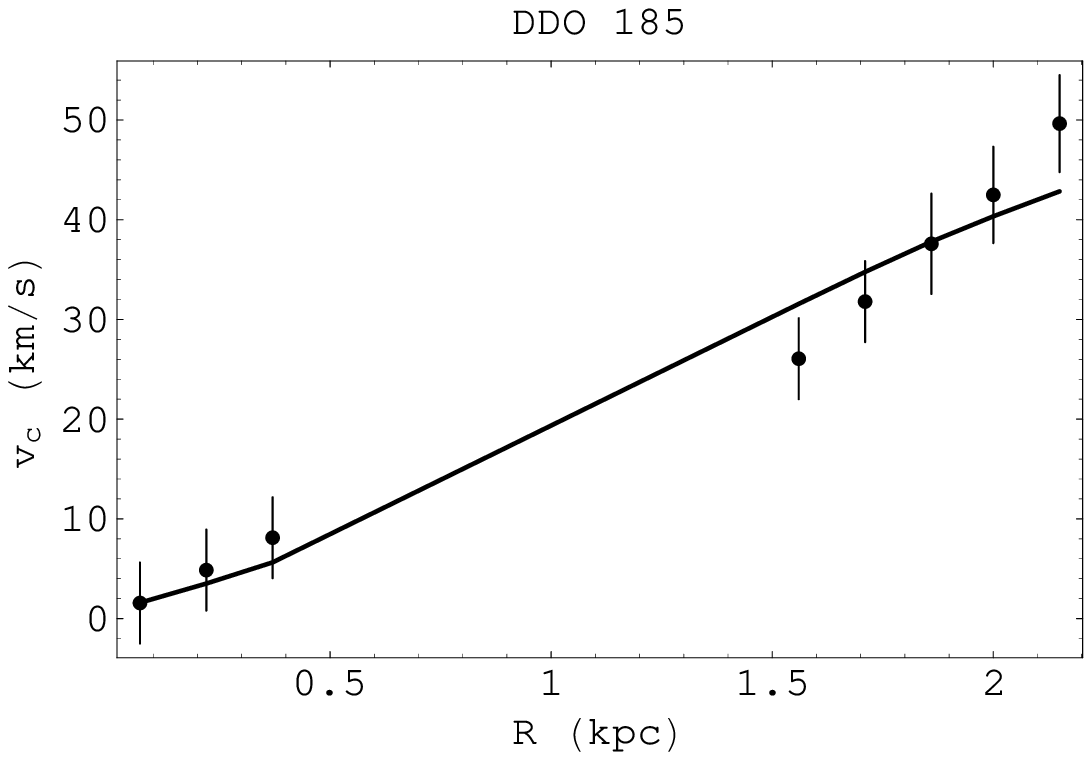}&
\includegraphics[width=5cm]{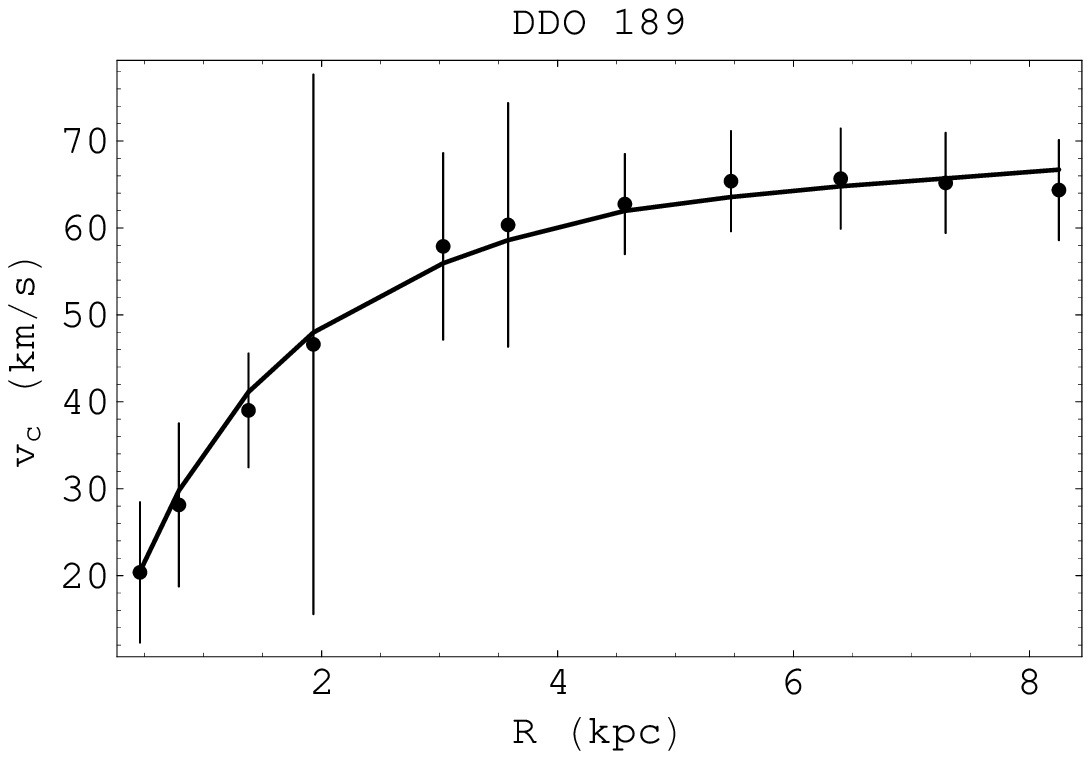}&
\includegraphics[width=5cm]{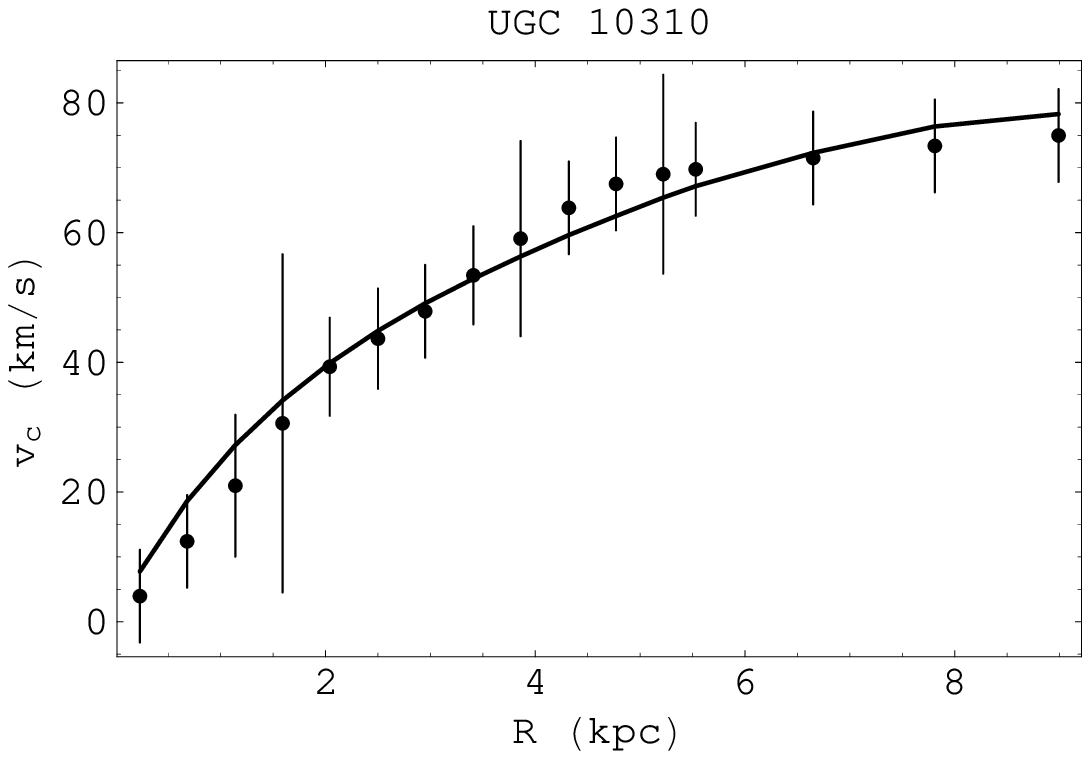}\\

\end{tabular}
\caption{Best fit curves superimposed to the data for the total
sample of 15 LSB galaxies considered.} \label{fig: bf}
\end{figure*}

However, these are only preliminary results which do not
completely solve the problem of dark matter in galaxies by models
coming from ETGs and do not fit the growth of structures. In any
case, further evidences on the same line of thinking  are coming
from other samples of galaxies (where also high surface brightness
galaxies are considered) \cite{salucci}, or from galaxy clusters,
where the dark matter range is completely different \cite{betty}.

\subsection{Dark Matter Halos inspired by $f(R)$\,-\,gravity}

At this point, it is worth wondering whether a link may be found
between $f(R)$ gravity and the standard approach based on dark
matter haloes since both theories fit equally well the same data.
The {\it trait\,-\,de\,-\,union} between these two different
schemes can be found in the modified gravitational potential which
induces a correction to the rotation curve in a similar manner as
a dark matter halo does. As a matter of fact, it is possible to
define an {\it effective dark matter halo} by imposing that its
rotation curve equals the correction term to the Newtonian curve
induced by $f(R)$ gravity. Mathematically, one has to split the
total rotation curve derived from $f(R)$ gravity as $v_c^2(r) =
v_{c, N}^2(r) + v_{c, corr}^2(r)$ where the second term is the
correction. Considering, for simplicity a spherical halo where  a
thin exponential disk is embedded, one can write the total
rotation curve as $v_c^2(r) = v_{c, disk}^2(r) + v_{c, DM}^2(r)$
with $v_{c, disk}^2(r)$ the Newtonian disk rotation curve and
$v_{c, DM}^2(r) = G M_{DM}(r)/r$ the dark matter one, $M_{DM}(r)$
being its mass distribution. Equating the two expressions, we
get\,:
\begin{equation}
M_{DM}(\eta) =
M_{vir}\left(\frac{\eta}{\eta_{vir}}\right)\frac{2^{\beta-5}\eta^{-\beta}_c(1-\beta)\eta^{\frac{\beta-5}{2}}{\cal
I}_0(\eta)-{\cal V}_d(\eta)}
{2^{\beta-5}\eta^{-\beta}_c(1-\beta)\eta^{\frac{\beta-5}{2}}{\cal
I}_0(\eta_{vir})-{\cal V}_d(\eta_{vir})}\ . \label{eq: mdm}
\end{equation}
with $\eta = r/r_d$, $\Sigma_0 = \Upsilon_{\star} i_0$, ${\cal
V}_d(\eta)\,=\,I_0(\eta/2)K_0(\eta/2)\times
I_1(\eta/2)K_1(\eta/2)$ and\footnote{Here $I_l$ and $K_l$, with
$l\,=\,1,2$ are the Bessel functions of first and second type.}\,:
\begin{equation} {\cal{I}}_0(\eta, \beta) =
\int_{0}^{\infty}{{\cal{F}}_0(\eta, \eta', \beta) k^{3 - \beta}
\eta'^{\frac{\beta - 1}{2}} {\rm e}^{- \eta'} d\eta'} \label{eq:
deficorr}
\end{equation}
with ${\cal{F}}_0$ only depending on the geometry of the system
and ``$vir$" indicating virial quantities. Eq.(\ref{eq: mdm})
defines the mass profile of an effective spherically symmetric
dark matter halo whose ordinary rotation curve provides the part
of the corrected disk rotation curve due to the addition of the
curvature corrective term to the gravitational potential. It is
evident that, from an observational viewpoint, there is no way to
discriminate between this dark halo model and a $f(R)$ power-law
gravity model. Having assumed spherical symmetry for the mass
distribution, it is straightforward to compute the mass density
for the effective dark halo as $\rho_{DM}(r) = (1/4 \pi r^2)
dM_{DM}/dr$. The most interesting features of the density profile
are its asymptotic behaviors that may be quantified by the
logarithmic slope $\alpha_{DM} = d\ln{\rho_{DM}}/d\ln{r}$ which
can be  numerically computed as function of $\eta$ for fixed
values of $\beta$ (or $n$). As expected, $\alpha_{DM}$ depends
explicitly on $\beta$, while $(r_c, \Sigma_0, r_d)$ enter
indirectly through $\eta_{vir}$. The asymptotic values at the
center and at infinity denoted as $\alpha_0$ and $\alpha_{\infty}$
result particularly interesting. It turns out that $\alpha_0$
almost vanishes so that in the innermost regions the density is
approximately constant. Indeed, $\alpha_0 = 0$ is the value
corresponding to models having an inner core such as the cored
isothermal sphere \cite{BT87} and the Burkert model \cite{burk}.
Moreover, it is well known that galactic rotation curves are
typically best fitted by cored dark halo models (see, e.g.,
\cite{GS04} and references therein). On the other hand, the outer
asymptotic slope is between $-3$ and $-2$, that are values typical
of most dark halo models in literature. In particular, for $\beta
= 0.80$ one finds $(\alpha_0, \alpha_{\infty}) = (-0.002, -2.41)$,
which are quite similar to the value for the Burkert model $(0,
-3)$. It is worth noticing that the Burkert model has been {\it
empirically} proposed to provide a good fit to the LSB and dwarf
galaxies rotation curves. The values of $(\alpha_0,
\alpha_{\infty})$ we find for our best fit effective dark halo
therefore suggest a possible theoretical motivation for the
Burkert\,-\,like models. Due to the construction, the properties
of the effective dark matter halo are closely related to the disk
one. As such, we do expect some correlation between the dark halo
and the disk parameters. To this aim, exploiting the relation
between the virial mass and the disk parameters, one can obtain a
relation for the Newtonian virial velocity $V_{vir} = G
M_{vir}/R_{vir}$\,:

\begin{equation}
M_d \propto \frac{(3/4 \pi \delta_{th} \Omega_m
\rho_{crit})^{\frac{1 - \beta}{4}} r_d^{\frac{1 + \beta}{2}}
\eta_c^{\beta}}{2^{\beta - 6}
 (1 - \beta) G^{\frac{5 - \beta}{4}}} \frac{V_{vir}^{\frac{5 -
\beta}{2}}}{{\cal{I}}_0(V_{vir}, \beta)} \label{eq: btfvir} \ .
\end{equation}
One can numerically check that Eq.(\ref{eq: btfvir}) may be well
approximated as $M_d \propto V_{vir}^{a}$ which has the same
formal structure as the baryonic Tully\,-\,Fisher (BTF) relation
$M_b \propto V_{flat}^a$ with $M_b$ the total (gas + stars)
baryonic mass and $V_{flat}$ the circular velocity on the flat
part of the observed rotation curve. In order to test whether the
BTF can be explained thanks to the effective dark matter halo we
are proposing, we should look for a relation between $V_{vir}$ and
$V_{flat}$. This is not analytically possible since the estimate
of $V_{flat}$ depends on the peculiarities of the observed
rotation curve such as how  far it extends and the uncertainties
on the outermost points. Therefore, for given values of the disk
parameters, it is possible to simulate theoretical rotation curves
for some values of $r_c$ and measure $V_{flat}$ finally choosing
the fiducial value for $r_c$ which gives a value of $V_{flat}$ as
similar as possible to the measured one. Inserting the relation
thus found between $V_{flat}$ and $V_{vir}$ into Eq.(\ref{eq:
btfvir}) and averaging over different simulations, one finally
gets\,:
\begin{equation}
\log{M_b} = (2.88 \pm 0.04) \log{V_{flat}} + (4.14 \pm 0.09)
\label{eq: btfour}
\end{equation}
while a direct fit to the observed data gives \cite{ssm}\,:
\begin{equation}
\log{M_b} = (2.98 \pm 0.29) \log{V_{flat}} + (3.37 \pm 0.13) \ .
\label{eq: btfssm}
\end{equation}
The slope of the predicted and observed BTF are in good agreement
 leading further support to  the $f(R)$ gravity model. The zeropoint is
markedly different with the predicted one being significantly
larger than the observed one, but it is worth stressing, however,
that both relations fit the data with similar scatter. A
discrepancy in the zeropoint may be due to the approximate
treatment of the effective halo which does not take into account
the gas component. Neglecting this term, one should increase the
effective halo mass and hence $V_{vir}$ which affects the relation
with $V_{flat}$ leading to a higher than observed zeropoint.
Indeed, the larger is $M_g/M_d$, the more the point deviate from
our predicted BTF thus confirming our hypothesis. Given this
caveat, we may therefore conclude with confidence that $f(R)$
gravity offers a theoretical foundation even for the empirically
found BTF relation.

All these results converge toward  the picture that data coming
from observations at galactic, extragalactic and cosmological
scales could be seriously framed  in ETGs without considering huge
amounts of dark energy and dark matter.

\section{Discussion and Conclusions}

Extended Theories of Gravity  can be considered as the natural
extension of General Relativity. Also if  they are not the final
theory of gravity at fundamental level (i.e. quantum gravity),
they could be a useful approach to address several shortcomings of
GR. In fact, also at Solar System scales, where GR has been
strongly confirmed, some conundrums come out as the indications of
an apparent, anomalous, long-range acceleration revealed from the
data analysis of Pioneer 10/11, Galileo, and Ulysses spacecrafts.
Such results are difficult to be framed in the standard theory of
GR and in its low energy limit \cite{anderson}.

 Furthermore, at
galactic scales, huge bulks of dark matter are needed to provide
realistic models matching with observations. In this case,
retaining GR and its low energy limit, implies the introduction of
an actually unknown ingredient (a huge amount of missing matter).

We face a similar situation even at larger scales: clusters of
galaxies are gravitationally stable and bound only if large
amounts of dark matter are supposed in their potential wells.

Finally, an unknown form of dark energy is required to explain the
observed accelerated expansion of cosmic fluid. Summarizing,
almost $95\%$ of matter-energy content of the Universe is unknown
while we can experimentally probe only gravity and ordinary
(baryonic  and radiation)  matter.

Considering another point of view, anomalous acceleration (Solar
System), dark matter (galaxies,  galaxy clusters and clustered
structures in general), dark energy (cosmology) could be nothing
else but the indications that  gravity is an interaction depending
on the scale and the assumption of a linear Lagrangian density in
the Ricci scalar $R$, as the Hilbert-Einstein action, could be too
simple for a comprehensive picture  at any scale.

Due to these facts, several motivations suggest to generalize GR
by considering gravitational actions where generic functions of
curvature invariants and scalar fields are present. This viewpoint
is physically motivated by several unification schemes and by
field quantization on curved spacetime \cite{birrell}.
Furthermore,   it is well known that revisions of GR  can solve
shortcomings at early cosmological epochs (giving rise to suitable
inflationary behaviors \cite{starobinsky,la}) and explain the
today observed accelerated behavior \cite{curvature,odinoj}. These
results can be achieved in metric and Palatini approaches
\cite{noi-review,francaviglia_1,palatiniCosmo, multamaki}.

In addition, reversing the problem, one can reconstruct the form
of the gravity Lagrangian by observational data of cosmological
relevance through a "back scattering" procedure \cite{mimick}.

All these facts suggest that the theory should be more general
than the linear Hilbert-Einstein one implying that extended
gravity could be a suitable approach to solve GR shortcomings
without introducing mysterious ingredients as dark energy and dark
matter which seem without explanation at fundamental level.
However, changing gravitational side could be nothing else but a
matter of taste since final probes for dark energy and dark matter
could come out from the forthcoming experiments as LHC.

Furthermore, in recent papers, some authors  have confronted this
kind of theories even with the PPN prescriptions considering both
metric and Palatini approaches. The results seem controversial
since in some cases \cite{olmo} it is argued that GR is always
valid  and there is no room for other theories while other studies
\cite{ppn-noi,mpla,allemandi-ruggiero} find that recent
experiments as Cassini and Lunar Laser Ranging allow the
possibility that ETGs could be  taken into account. In particular,
it is possible to define generalized PPN-parameters  and several
ETGs could result compatible with experiments in Solar System
\cite{will,mpla,damour}.

In principle, any  analytic ETGs can be compared with the
Hilbert-Einstein Lagrangian  provided suitable values of the
coefficients.  This consideration suggests to take into account,
as physical theories, functions of the Ricci scalar which slightly
deviates from GR, i.e. $f(R)\,=\,f_0R^{(1+\epsilon)}$ with
$\epsilon$ a small parameter which indicates how much the theory
deviates from GR and then approximate as
\begin{equation}\label{fReps} f_0|R|^{(1+\epsilon)}\simeq
f_0|R|\biggl(1+\epsilon\ln|R|+\frac{\epsilon^2\ln^2|R|}{2}+\dots\biggr)\,.
\end{equation}

Actually, the PPN - Eddington parameters $\beta$ and $\gamma$ may
represent the key parameters  to discriminate among relativistic
theories of gravity. In particular, these quantities should be
significatively tested at Solar System scales by forthcoming
experiments like LATOR \cite{nordtvedt} while the  today available
releases are far, in our opinion, to be conclusive in this sense,
as a rapid inspection of Table IV  suggests. In other words, ETGs
cannot be {\it a priori} excluded also at Solar System scales.

\begin{table}[ht]
\centering
\begin{tabular}{|l|c|}
\hline\hline
  Mercury Perihelion Shift& $|2\gamma-\beta-1|<3\times10^{-3}$ \\\hline
 Lunar Laser Ranging &  $4\beta-\gamma-3\,=\,-(0.7\pm 1)\times{10^{-3}}$ \\\hline
 Very Long Baseline Interf. &  $|\gamma -1|\,=\,4\times10^{-4}$ \\\hline
 Cassini Spacecraft &  $\gamma-1\,=\,(2.1\pm 2.3)\times10^{-5}$ \\\hline\hline
\end{tabular}
\caption{\small \label{ppn} A schematic resume of recent
experimental constraints on the PPN-parameters. They are the
perihelion shift of Mercury \cite{mercury}, the Lunar Laser
Ranging \cite{lls}, the upper limit coming from the Very Long
Baseline Interferometry \cite{VLBI} and the results obtained by
the estimate of the Cassini spacecraft delay into the radio waves
transmission near the Solar conjunction \cite{cassini}.}
\end{table}

In this paper, we have outlined what one should intend for ETGs in
the metric and in the Palatini approach. In particular, we have
discussed the higher-order  and the scalar-tensor theories of
gravity showing the relations between them and their connection to
GR via the conformal transformations.

In the so called Einstein frame, any ETG can be reduced to the
Hilbert-Einstein action plus one or more than one scalar field(s).
The physical meaning of conformal transformations can be
particularly devised in the Palatini approach, as discussed in
Sec.IV. After, we have discussed some cosmological and
astrophysical applications of ETGs.

Although the results outlined are referred to the simplest class
of  ETGs, power law $f(R)$, they  could represent an interesting
paradigm. Assuming both metric and Palatini  approach,  it is
possible to investigate the viability of $f(R)$ cosmological
models. The expansion rate $H = \dot{a}/a$ may be analytically
expressed as a function of the redshift $z$, so that it is
possible to contrast the model predictions against the
observations. In particular, the SNeIa Hubble diagram,  the gas
mass fraction in relaxed galaxy clusters, the lookback time to
galaxy clusters, and  radio galaxies can be used to constrain
cosmological parameters by distance and time-based methods.

Also if such  models are, up to now, not completely satisfactory
to match all the observations, they allow to recover accelerated
behavior of Hubble fluid without any unknown form of dark energy.
However, the issue of structure formation has to be seriously
faced in order to understand if such toy models could give rise to
a self-consistent alternative theory to GR.

Furthermore,  it is possible to "tune" the stochastic background
of GWs and this occurrence could constitute a further cosmological
test capable of confirming or ruling out ETGs once data from
interferometers, like VIRGO, LIGO and LISA, will be available.

In addition, the modification of the gravitational potential
arising as a natural effect in the framework of ETGs can represent
a fundamental tool to interpret the  rotation curves of spiral
galaxies. Besides, if one considers the model parameters settled
by the fit over the observational data on rotation curves, it is
possible to construct a phenomenological analogous of dark matter
halo whose shape is similar to the one of the so called Burkert
model. Since Burkert's model has been empirically introduced to
give account for the dark matter distribution in the case of LSB
and dwarf galaxies, this result could represent an interesting
achievement since it provides a theoretical foundation to such a
model.

By investigating the relation between dark halo and the galaxy
disk parameters,  a relation between $M_d$ and $V_{flat}$,
reproducing the baryonic Tully\,-\,Fisher, can be deduced. In
fact, exploiting the relation between the virial mass and the disk
parameters, one obtains a relation for the virial velocity which
can be satisfactory approximated as $M_d \propto V_{vir}^{a}$.
Even such a result seems  intriguing since it provides  a
theoretical interpretation for a phenomenological relation.

As a matter of fact, although not definitive, these
phenomenological issues  can represent a viable approach for
future, more exhaustive investigations of ETGs. In particular,
they support the quest for a unified view of the dark side of the
Universe. In summary, these results seem to motivate a careful
search for a fundamental theory of gravity capable of explaining
the full cosmic dynamics by the only  "ingredients" which we can
directly and firmly experience, namely the background gravity, the
baryonic matter, the radiation and  also the neutrinos
\cite{vitiello}.

\section*{Acknowledgments}
The authors wish to thank G. Allemandi, A. Borowiec,  V.F.
Cardone, S. Carloni, S. Nojiri, S.D. Odintsov, A. Stabile and A.
Troisi for discussions, suggestions and  results presented in this
paper.

\end{document}